\documentclass[amsmath,amssymb,aps,prl,reprint,superscriptaddress]{revtex4-1}
\usepackage{graphicx}% Include figure files
\usepackage{xcolor}
\usepackage[colorlinks=true,urlcolor=blue,citecolor=blue,linkcolor=blue]{hyperref}
\usepackage{physics}
\usepackage{siunitx}
\usepackage{xspace}
\usepackage{bm}
\usepackage{comment}

\newcommand{\bk}{\textbf{k}}
\newcommand{\bb}{\textbf{b}}

\newcommand{\bq}{\textbf{q}}

\newcommand{\br}{\textbf{r}}

\newcommand{\brr}{\textbf{R}}
\newcommand{\moire}{moir\'e\xspace}
\newcommand{\Moire}{Moir\'e\xspace}

\usepackage[normalem]{ulem}

\definecolor{JK-color}{named}{green}
\definecolor{AJ-color}{named}{blue}
\definecolor{ACG-color}{rgb}{0.97,0.57,0.11}
\definecolor{GB-color}{RGB}{128,0,128}

\definecolor{JK-color2}{named}{green}
\definecolor{AJ-color2}{named}{red}
\definecolor{ACG-color2}{rgb}{0.87,0.47,0.01}
\definecolor{GB-color2}{RGB}{128,0,128}

\begin{document}

% Use the \preprint command to place your local institutional report
% number in the upper righthand corner of the title page in preprint mode.
% Multiple \preprint commands are allowed.
% Use the 'preprintnumbers' class option to override journal defaults
% to display numbers if necessary
%\preprint{}

%Title of paper
\title{Non-local interactions and supersolidity of \moire excitons}
% repeat the \author .. \affiliation  etc. as needed
% \email, \thanks, \homepage, \altaffiliation all apply to the current
% author. Explanatory text should go in the []'s, actual e-mail
% address or url should go in the {}'s for \email and \homepage.
% Please use the appropriate macro foreach each type of information

% \affiliation command applies to all authors since the last
% \affiliation command. The \affiliation command should follow the
% other information
% \affiliation can be followed by \email, \homepage, \thanks as well.

%\email[]{Your e-mail address}
%\homepage[]{Your web page}
%\thanks{}
%\altaffiliation{}

\newcommand{\affiliationAarhus}{Center for Complex Quantum Systems, Department of Physics and Astronomy, Aarhus University, Ny Munkegade, DK-8000 Aarhus C, Denmark}
\newcommand{\affiliationAalto}{Department of Applied Physics, Aalto University, P.O.Box 15100, 00076 Aalto, Finland}
\newcommand{\affiliationCambridge}{T.C.M. Group, Cavendish Laboratory, University of Cambridge, JJ Thomson Avenue, Cambridge, CB3 0HE, U.K}
\newcommand{\affiliationChina}{Shenzhen Institute for Quantum Science and Engineering and Department of Physics, Southern University of Science and Technology, Shenzhen 518055, China}

\author{Aleksi Julku}
\affiliation{\affiliationAarhus}
%\author{Arturo Camacho-Guardian}
%\affiliation{\affiliationCambridge}
%\author{Georg M. Bruun}
%\affiliation{\affiliationAarhus}
%\affiliation{\affiliationChina}

%Collaboration name if desired (requires use of superscriptaddress
%option in \documentclass). \noaffiliation is required (may also be
%used with the \author command).
%\collaboration can be followed by \email, \homepage, \thanks as well.
%\collaboration{}
%\noaffiliation

\date{\today}

\begin{abstract}
Heterobilayer transition metal dichalcogenide (TMDC) \moire systems provide an ideal framework to investigate strongly correlated physics. Here we theoretically study bosonic many-body phases of excitons in \moire TMDCs. By using two \moire models and cluster mean-field theory, we reveal that, due to non-local Coulomb interactions, \moire excitons can feature exotic supersolid phases, i.e. superfluids of broken translational invariance, and correlated insulating states. The correlated phases exist at experimentally accessible temperatures and are tunable via the twist angle and exciton density.

\end{abstract}

% insert suggested PACS numbers in braces on next line
% insert suggested keywords - APS authors don't need to do this
%\keywords{}

%\maketitle must follow title, authors, abstract, \pacs, and \keywords
\maketitle

%\hat{x}

%%%%%%%%%%%%%%%%%%%%%%%%% 

%\section{Introduction} 
\textit{Introduction---} Rapid advances in nanofabrication techniques have allowed for experimental realizations of multilayer van der Waals \moire heterostructures, where lattice mismatch or a twist angle between monolayers, results in a tunable long-wavelength potential for electrons~\cite{Balents2020,Kennes2021,Andrei2021}.  \Moire potential leads to localized electrons, reducing their kinetic energy and thus enhancing the role of interactions. \Moire systems, therefore, serve as versatile platforms to study strongly correlated electronic systems.  Prominent interaction-driven phases observed in \moire systems include superconductivity in twisted bi- and multilayer graphene~\cite{Cao2018,Yankowitz2019,Chen2019} and correlated electronic states, such as Wigner crystals, stripes and Mott insulators in bilayer transition metal dichalcogenides (TMDC)~\cite{Wang2020,Huang2021,Shimazaki2020,Xu2020,Tang2020,Campbell2020,Liu2021,Jin2021,Miao2021}.

\Moire TMDCs are also an ideal platform for revealing many-body effects of bosons. Namely, in TMDC monolayers, excitons -- bound electron-hole pairs -- can be optically created. Correspondingly, the \moire potential of electrons leads to formation of \moire excitons~\cite{Tran2019,Alexeev2019,Ruiz-Tijerina2019,Shimazaki2020,Yu2017,Seyler2019,Jin2019,Yu2017,Jiang2021,Montblanch2021,Huang2022,Wang2022}. %\Moire TDMCs then allow studying strongly correlated bosons and, as excitons can interact with electrons, also Bose-Fermi mixtures. Consequently, interactions between \moire excitons and electrons have been exploited in recent experiments to probe correlated electronic states by measuring optical spectra of \moire excitons~\cite{Shimazaki2020,Tang2020,Campbell2020,Liu2021b,Jin2021,Miao2021}. 
While most of the research  have focused on probing \moire electrons with excitons~\cite{Shimazaki2020,Tang2020,Campbell2020,Liu2021b,Jin2021,Miao2021}, less  attention has been given to possible bosonic many-body phases of \moire excitons. As the \moire potential allows the confinement of bosons to triangular or honeycomb lattice geometries~\cite{Yu2017}, and as repulsive Coulomb interactions between \moire excitons can be very strong compared to their kinetic energy, it is tempting to expect that \moire excitons form Mott insulating phases. This was indeed predicted in a recent theoretical study~\cite{Gotting2022} for a large range of tunable parameters. Moreover, the possibility to reach superfluidity of \moire exictons has been also speculated~\cite{Gotting2022,Lagoin2021}.

Due to the strong on-site interaction, weaker non-local interactions between excitons is often ignored. However, at sufficiently low densities the on-site interaction can, by the virtue of hard-core boson constraint, be discarded so the non-local interactions become the dominant interaction channel. In this work, we theoretically study possible many-body phases of \moire excitons by taking into account non-local interactions. We employ two different continuum models~\cite{Ruiz-Tijerina2019,Wu2018a}, from which we derive the effective tight-binding models for \moire excitons and compute non-local exciton-exciton interactions. By using cluster mean field (CMF) theory~\cite{Luhmann2013,Yamamoto2012,Yamamoto2012b,Malakar2020,Hassan2007}, we show that,  in addition to possible conventional Mott and superfluid states, \moire excitons can also exhibit more exotic many-body states, namely correlated insulating and supefluid phases of broken translational invarince. The latter is widely known as the \textit{supersolid} phase. We show that these states are accessible using reasonable twist angles and exciton densities, and experimentally accessible temperatures. %Our work paves the way for understanding the interaction-driven many-body phases of \moire excitons.

%The emergence of supersolidity arises due to interplay between the kinetic energy and dipolar non-local interactions between \moire excitons. 

\begin{figure}
  \centering
    \includegraphics[width=1.0\columnwidth]{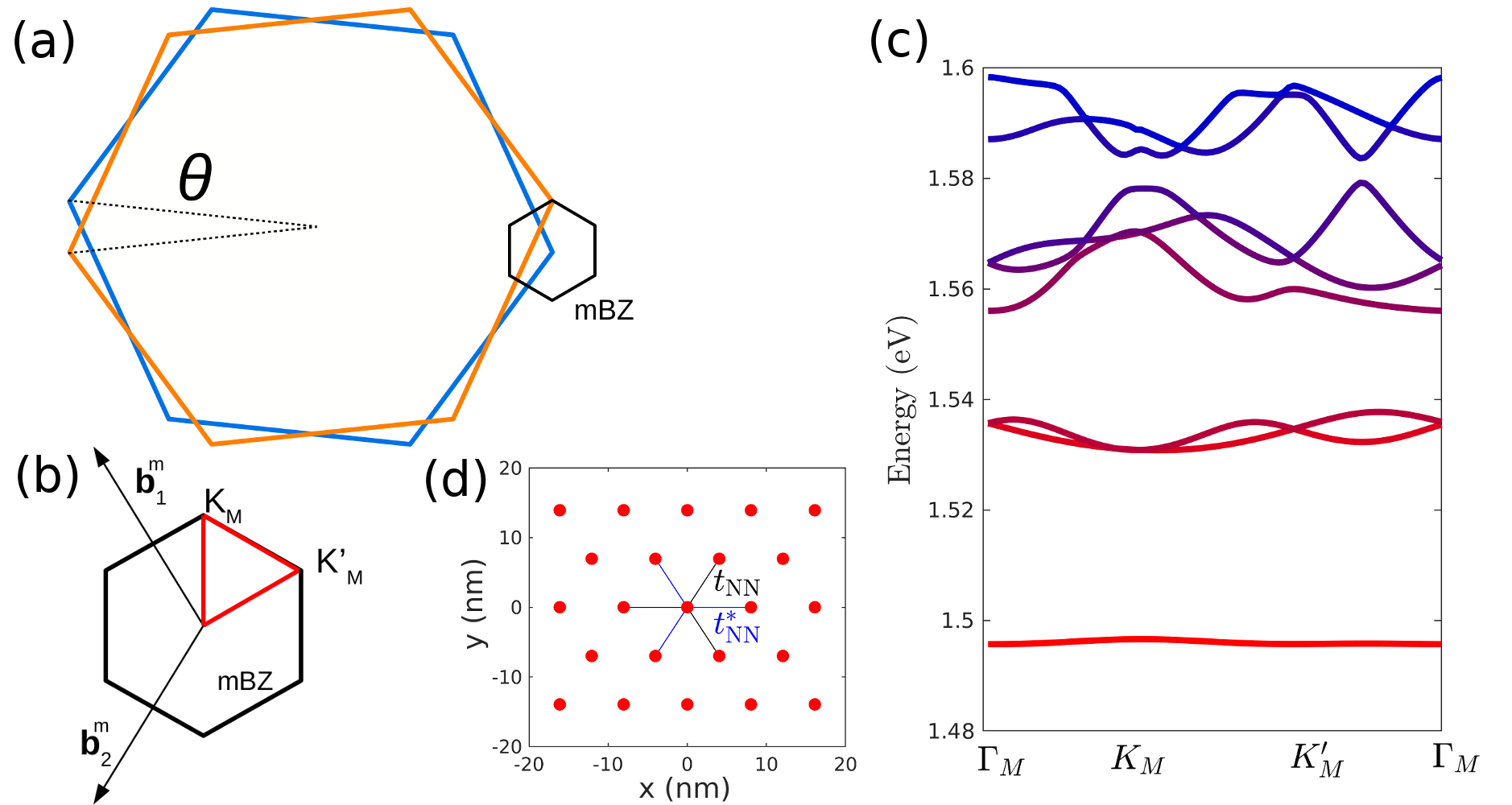}
    \caption{(a) Schematic of the \moire system in the momentum space. Blue (orange) hexagon depicts the BZ of layer 1 (2) and black hexagon presents the \moire BZ (mBZ) in the $K$-valley. (b) $K$-valley mBZ. The momentum space is spanned by vectors $\bb^m_1$ and $\bb^m_2$. The red path gives the x-axis of panel (c). (c) \Moire exciton $K$-valley spin-down band structure at $\theta = 0.5^\circ$ obtained for MoSe/WS$_2$ with $H_H$. (d) Corresponding triangular tight-binding model for the lowest band \moire excitons, characterized by the NN hopping $t_{\text{NN}}$ and \moire periodicity $a_m$.}
   \label{Fig:1}
\end{figure}

%%%%%%%%%%%%%%%%%%%%%%%%%%%%%%%%%%%%%%%%%%%%%%%%%%%%%%%%%%%%%%%%%%%%%%%%%%%

\textit{Hamiltonian---} We consider \moire excitons of a TMDC heterobilayer system of layers 1 and 2. TMDC monolayers have a hexagonal lattice structure and direct band gaps at the corners of their hexagonal Brillouin zone (BZ), namely in the $K$- and $K'$-valleys~\cite{Wang2018,Mueller2018,Yu2015}. Small lattice constant mismatch or twist angle $\theta$ between layers causes the interlayer electron tunneling to hybridize the low-energy states of two layers in the $K$- and $K'$-valleys~\cite{Wang2017,Ruiz-Tijerina2019}. This gives a rise to \moire flat bands of excitons, long \moire periodicity $a_m$ and reduced \moire BZ (mBZ), see Figs.~\ref{Fig:1}(a)-(c). As excitons can be created valley-selectively~\cite{Xiao2012,Cao2012,Zeng2012,Mak2012b,Wang2018,Schaibley2016,Zhang2019}, we from now on consider only the $K$-valley excitons and small twist angles $\theta \sim 0-4^\circ$.

We study the many-body properties of \moire excitons by employing two different one-particle continuum Hamiltonians. The first one, which we call hybridized \moire exciton model and denote $H_{H}$, has been used successfully in Refs.~\cite{Alexeev2019,Ruiz-Tijerina2019} to study the hybridization of intra- and interlayer excitons in \moire structures. The model treats the interlayer electron tunneling $t(\bk,\bk')$ from momentum $\bk'$ to $\bk$ in the microscopic level as $t(\bk,\bk')\propto  \delta_{\bk,\bk'} + \delta_{\bk - \bk', \bb^m_{1}} + \delta_{\bk - \bk', \bb^m_{2}}$ [$\bb^m_i$ are the \moire reciprocal vectors, see Fig.~\ref{Fig:1}(b)], which leads to the emergence of \moire excitons (see Supplementary Material (SM)~\cite{SM} for further details).

The second model, which we denote $H_E$, treats the \moire effects with a slowly-varying effective potential $\Delta(\br)$ ($\br$ being the spatial coordinate) i.e. the single-particle Hamiltonian for excitons is simply $H_E = -\nabla^2/2m + \Delta(\br)$, where $m$ is the exciton mass. We call this as an effective potential model, and it has been widely used to study \moire electrons~\cite{Wu2019a,Wu2018,Pan2020,Morales-Duran2021} and  excitons~\cite{Yu2017,Wu2017,Gotting2022,Wu2018a,Tran2019}. Recent first-principles studies~\cite{Naik2022} have argued that continuum models are sufficient to capture the nature of the lowest energy \moire excitons, the primary focus of this work.  Following the experimental works of~\cite{Alexeev2019,Tran2019}, we apply $H_H$ ($H_E$)  to study a hybrid \moire excitons (\moire interlayer excitons) in a MoSe$_2$/WS$_2$ (MoSe$_2$/WSe$_2$) heterobilayer system.

Both $H_H$ and $H_H$ yield one-particle Hamiltonians in the form $H_0 = \sum_{n\bk\in \text{mBZ}} \epsilon_{\bk n} \gamma_{\bk n}^\dag \gamma_{\bk n}$, where $n$ is the band index and $\gamma_{\bk n}$ annihilates a \moire exciton at momentum $\bk$ and energy $\epsilon_{\bk n}$~\cite{SM}. For both the models, the lowest energy band $\epsilon_{\bk 1}$ at small $\theta$ is extremely flat and well isolated from higher bands by a large band gap (see Fig.~\ref{Fig:1}(c) and SM~\cite{SM}). Subsequently, the Wannier functions of the lowest \moire exciton band form a triangular lattice characterized by the nearest-neighbour hopping $t_{\text{NN}}$~\cite{SM}, see Fig.~\ref{Fig:1}(d). We thus write the effective tight-binding Hamiltonian of the excitons in the lowest \moire band as
\begin{align}
\label{mainHam}
H = \sum_{<i,j>} t_{ij} x^\dag_i x_j + \sum_i U_0 x^\dag_i x^\dag_i x_i x_i + \sum_{i,j}U_{ij} x^\dag_i x^\dag_j x_j x_i   
\end{align}
where $x_i$ annihilates a \moire exciton in lattice site $i$, $t_{ij}$ describes hopping from site $i$ to $j$, $U_0$ is the repulsive on-site interaction, $U_{ij}$ denotes the non-local interactions between sites $i$ and $j$, and the sum over the hopping terms is limited to nearest-neighbouring sites. Interaction terms arise due to Coulomb interactions between excitons and the values of $t_{ij}$, $U_0$ and $U_{ij}$ depend on the chosen continuum model. The hopping values are obtained as the Fourier transform of the energies of the lowest \moire band, i.e. $t_{ij} = \frac{1}{N}\sum_{\bk \in \textrm{mBZ}} \epsilon_{\bk 1} e^{-i \bk \cdot (\brr_i -\brr_j)}$. Here, $N$ is the number of \moire unit cells, and $\brr_i$ denote the locations of \moire lattice sites. 

%%%%%%%%%%%%%%%%%%%%%%%%%%%%%%%%%%%%%%%%%%%%%%%%%%%%%%%%%%%%%%%%%%%%%%%%%%%

\begin{figure}
  \centering
    \includegraphics[width=1.0\columnwidth]{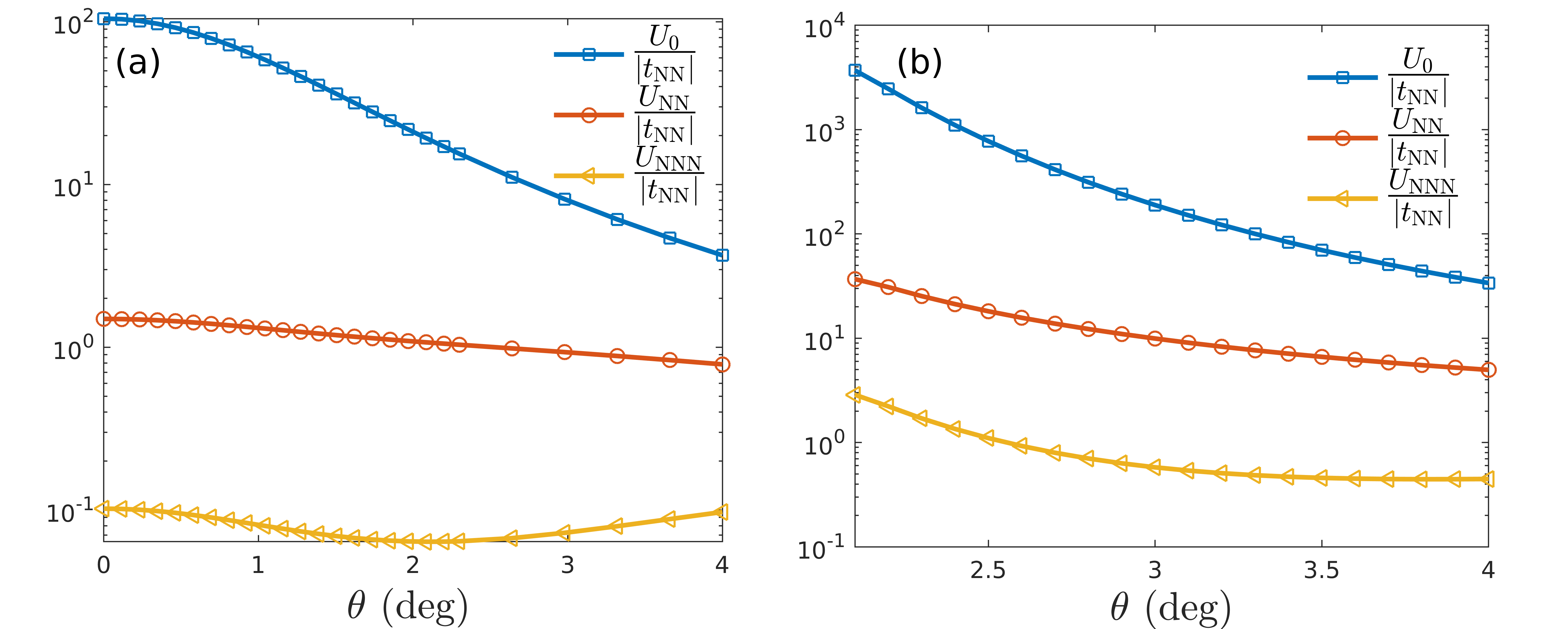}
    \caption{(a)-(b) $U_0$, $U_{\mathrm{NN}}$ and $U_{\mathrm{NNN}}$ with respect to $|t_{\mathrm{NN}}|$ as a function of $\theta$ obtained with $H_H$ and $H_E$ continuum models, respectively.}
   \label{Fig:2}
\end{figure}

%%%%%%%%%%%%%%%%%%%%%%%%%%%%%%%%%%%%%%%%%%%%%%%%%%%%%%%%%%%%%%%%%%%%%%%%%%%

Deriving interaction terms $U_0$ and $U_{ij}$ is more involved and depends on the chosen model. We detail how to do this for $H_H$ in the next section. In Fig.~\ref{Fig:2} we present $U_0$, nearest-neighbour (NN) and next-nearest-neighbour (NNN) interactions, $U_{\mathrm{NN}}$ and $U_{\mathrm{NNN}}$, with respect to $|t_{\mathrm{NN}}|$ as a function of $\theta$ for the two continuum models. In both the cases, $U_0$ is the dominant energy scale, being roughly one to two orders of magnitude larger than  $U_{\text{NN}}$. Furthermore, $U_{\text{NN}}$ is comparable to $t_{\text{NN}}$ in case of $H_H$ and much larger than $t_{\text{NN}}$  when using $H_E$.  We also see that in case of $H_E$, NNN interaction is comparable to $t_{\text{NN}}$ and cannot be ignored. We thus keep the interactions up to the NN (NNN) terms when using $H_H$ ($H_E$). In case of \moire electrons, non-local interactions have been predicted to lead a rich landscape of many-body phases~\cite{Wu2018,Pan2020,Pan2020b,Pan2020c,Pan2021,Morales-Duran2021}.

\textit{Exciton-exciton interactions---} We derive now the interaction terms of Eq.~\eqref{mainHam} in case of $H_H$ (derivation for $H_E$ is given in SM~\cite{SM}). In this model, the \moire excitons consist of superpositions of intra- and interlayer excitons labelled as $| \text{X} \rangle$, $| \text{X}' \rangle$, $| \text{IX} \rangle$ and $| \text{IX}' \rangle$. Intralayer excitons $| \text{X} \rangle$ ($| \text{X}' \rangle$) and holes of interlayer excitons $| \text{IX} \rangle$ ($|\text{IX}' \rangle$) reside in layer 1 (2). However,  due to the permanent dipole moment of interlayer excitons, IX-IX, IX'-IX' and IX-IX' interactions  are much larger than other interaction terms. Hence,  we write the interaction Hamiltonian for the direct Coulomb interactions as
\begin{align}
&H_{int} = \frac{1}{2A}\sum_{\substack{\bk\bk'\bq \\ t\tilde{t}}}g^{dir}_{t\tilde{t}}(\bq) x^\dag_{t,\bk+\bq} x^\dag_{\tilde{t},\bk'-\bq} x_{\tilde{t},\bk'}x_{t,\bk}  
\end{align}
where $A$ is the system area, $t,\tilde{t} \in \{$IX,IX'$\}$, $x_{t,\bk}$ annihilates an interlayer exciton of momentum $\bk$ and type $t$, $g^{dir}_{tt}(\bq)$ is the interaction vertex and the momenta sums are \textit{not} limited to mBZ. We transfer to the \moire exciton basis:
\begin{align}
\label{Hint}
&H_{int} =\sum_{\substack{\bk\bk'\bq \in \textrm{mBZ} \\ ijkl}} \frac{g^{dir}_{ijkl}(\bk,\bk',\bq)}{2A} \gamma^\dag_{\bk+\bq i}\gamma^\dag_{\bk'-\bq j}\gamma^{}_{\bk' k}\gamma^{}_{\bk l}, 
\end{align}
with 
\begin{align}
&g^{dir}_{ijkl}(\bk,\bk',\bq) = \sum_{t,\tilde{t}}\sum_{\alpha\beta\gamma} g^{dir}_{t\tilde{t}}(\bq + \textbf{G}_\gamma) \langle  u_{i,\bk+\bq} | t,\alpha +\gamma \rangle \times  \nonumber \\ & \langle u_{j,\bk'-\bq} |  \tilde{t},\beta -\gamma \rangle \langle  \tilde{t},\beta | u_{k,\bk'} \rangle \langle  t,\alpha | u_{l,\bk} \rangle.
\end{align}
Here $| u_{i,\bk} \rangle$ is the periodic part of the \moire Bloch function for the $i$th band of momentum $\bk$, and matrix elements $\langle t, \alpha |  u_{i,\bk} \rangle$ represent its components related to the exciton of type $t$ at momentum $\bk + \textbf{G}_\alpha$ with $\textbf{G}_\alpha \equiv q_\alpha \bb^m_1 + p_\alpha \bb^m_2$ ($q_\alpha$ and $p_\alpha$ are integers).

The Coulomb interaction vertex $g^{dir}_{t\tilde{t}}(\bq)$ can be straightforwardly computed by deploying the excitonic wavefunctions $\phi(\bk)$~\cite{SM}. For example, the interaction vertex between two IX-excitons is
%\begin{align}
%\label{g_dir}
%&g^{dir}_{\text{IX,IX}}(\bq) \approx \frac{e^2}{2 q} \Bigg\{ \frac{f(x^{\text{IX}}_{h}\bq)^2 + f(x^{\text{IX}}_{e}\bq)^2}{\epsilon_{\textrm{intra}}(q)} -\frac{2 f(x_{e}^{\text{IX}}\bq) f(x_{h}^{\text{IX}}\bq)}{\epsilon_{\textrm{inter}}(q)} \Bigg\}.
%\end{align}
\begin{align}
\label{g_dir}
g^{dir}_{\text{IX,IX}}(\bq) \approx \frac{e^2}{2 q} &\Bigg\{ \frac{f(x^{\text{IX}}_{h}\bq)^2}{\epsilon_{\textrm{intra},2}(q)} + \frac{f(x^{\text{IX}}_{e}\bq)^2}{\epsilon_{\textrm{intra},1}(q)} \nonumber \\
&-\frac{2 f(x_{e}^{\text{IX}}\bq) f(x_{h}^{\text{IX}}\bq)}{\epsilon_{\textrm{inter}}(q)} \Bigg\}.
\end{align}
Here $e$ is the elementary charge, $f(\bk) = \sum_{\tilde{\bq}} \phi^*_{IX} (\tilde{\bq}) \phi_{IX}(\tilde{\bq} + \bk)$ and $x_e^{\text{IX}}$ ($x_h^{\text{IX}}$) is the relative electron (hole) mass of the IX-exciton such that $x_e^{\text{IX}} +x_h^{\text{IX}} =1$~\cite{SM}. The terms inside the wave brackets arise due to electron-electron, hole-hole and electron-hole Coulomb interactions, respectively. We have approximated the excitons to be tightly localized in the momentum space around the $K$-point~\cite{SM}. Furthermore, we take into account the two-layer geometry via the momentum-dependent intra- and interlayer Keldysh-like dielectric functions, $\epsilon_{\text{intra}}(q)$ and $\epsilon_{\text{inter},l}(q)$~\cite{Danovich2018}, derived in SM~\cite{SM}.

We rewrite Eq.~\eqref{Hint} for the tight-binding model~\eqref{mainHam} by discarding all but the lowest \moire band and using the Wannier function expansion, i.e. $\gamma_{\bk 1} =\frac{1}{\sqrt{N}} \sum_i e^{i\bk \cdot \brr_i} x_{i}$~\cite{SM}, to obtain
\begin{align}
H_{int} &\approx \frac{1}{2A}\sum_{\bk\bk'\bq \in \text{mBZ}} g^{dir}_{1111}(\bk,\bk',\bq) \gamma^\dag_{\bk+\bq 1}\gamma^\dag_{\bk'-\bq 1}\gamma_{\bk' 1}\gamma_{\bk 1} \nonumber \\
& = \sum_{a,b,c,d} g_{abcd} x^\dag_{a} x^\dag_{b} x_{c} x_{d} 
\end{align}
with 
\begin{align}
\label{int_TB}
& g_{abcd} = \sum_{\bk\bk'\bq} \frac{g^{dir}_{1111}(\bk,\bk',\bq)}{2A N^2} \frac{e^{i\bk'\cdot \brr_c + i\bk \cdot \brr_d}}{e^{i(\bk+\bq)\cdot \brr_a +i(\bk'-\bq)\cdot \brr_b}}
\end{align}
%e^{-i(\bk+\bq)\cdot \brr_a   -i(\bk'-\bq)\cdot \brr_b + i\bk'\cdot \brr_c + i\bk \cdot \brr_d}.  
Equation~\eqref{int_TB} gives rise to different  scattering processes such as direct and exchange interactions ($g_{abba}$, $g_{abab}$), interaction-assisted hopping $g_{aaab}$ and pair hopping $g_{aabb}$. The importance of such terms was highlighted in Ref.~\cite{Morales-Duran2021} for the case of \moire electrons. Here, however, the direct interaction is the dominant one and we discard non-direct terms to obtain Eq.~\eqref{mainHam}.

\textit{Supersolidity of \moire excitons---} As $U_0$ in~\eqref{mainHam} is much larger than other energy scales, it is presumable that the ground state is a Mott insulator when the exciton density $n$, i.e. the number of exctions per lattice site, is $n=1$. However, for \textit{smaller densities}, the ground state can be very different. Namely, $U_0$ is so large that for  $n <1$, one can employ the hard-core constraint (HCC), i.e. to limit the occupation number of each lattice site to be less than $2$. HCC is accurate when $n < 1/A_{\text{uc}}$, where $A_{\text{uc}}$ is the area of the \moire unit cell. For example, with twist angle $\theta = 2^\circ$,  $1/A^m_{uc} = 3.1\times 10^{-12}$ cm$^{-2}$ for a MoSe$_2$/WS$_2$ structure. This density is to be contrasted with experimentally measured critical density $n_c$ above which interlayer \moire excitons dissociate to free electron-hole plasma~\cite{Wang2019b,Wang2021b}. In case of MoSe$_2$/WSe$_2$, $n_c$ was measured and theoretically computed to be roughly  $n_c \sim 1.6-3\times 10^{-12}$  cm$^{-2}$ ~\cite{Wang2019b}.  We therefore restrict our analysis to $n<\frac{1}{2}$ and employ HCC, as justified by the experiments.

To study competition between the hopping and non-local interaction terms, we treat \moire excitons as ideal bosons and  employ cluster mean-field (CMF) theory~\cite{Luhmann2013,Yamamoto2012,Yamamoto2012b,Malakar2020,Hassan2007}. As we are using sufficiently small exciton densities, excitons follow to a good approximation bosonic commutation relations and therefore our bosonic model is well justified~\cite{SM}. CMF theory has been used in earlier works to investigate non-locally interacting hard-core bosons in square and triangular lattices, revealing that such systems can feature Mott states, superfluid states and supersolid phases~\cite{Yamamoto2012,Yamamoto2012b,Malakar2020}. Moreover, CMF theory has been shown to agree well with Monte Carlo calculations~\cite{Yamamoto2012,Wessel2005}. Previous studies have considered only real-valued hopping parameters, whereas here the \moire potential can render $t_{\mathrm{NN}}$ complex-valued in case of $H_H$. One therefore cannot apply directly the results of previous studies~\cite{Yamamoto2012} here.

In CMF method, a cluster of sufficiently many lattice sites is solved exactly, and the coupling between the cluster and sites outside the cluster is treated in the mean-field level. Specifically, the cluster Hamiltonian reads
%\begin{align}
%\label{clusterH}
%H_C &= \sum_{i_c,j_c}( t_{i_c,j_c} -\mu \delta_{i_c,j_c} )x^\dag_{i_c} x_{j_c} + \sum_{\langle i_c, j_c \rangle} U_{\mathrm{NN}} x^\dag_{i_c} x^\dag_{j_c} x_{j_c}  x_{i_c} \nonumber \\
%& + \sum_{i,j_c} \Big(  t_{i,j_c} \psi_i^* x_{j_c} + h.c\Big)  + \sum_{\langle i_c,j \rangle}2U_{\mathrm{NN}} n_j x^\dag_{i_c} x_{i_c}.
%\end{align}
\begin{align}
\label{clusterH}
H_C &= \sum_{i_c,j_c}( t_{i_c j_c} -\mu \delta_{i_c,j_c} )x^\dag_{i_c} x_{j_c} + \sum_{i_c, j_c } U_{i_c j_c} x^\dag_{i_c} x^\dag_{j_c} x_{j_c}  x_{i_c} \nonumber \\
& + \sum_{i,j_c} \Big(  t_{i j_c} \psi_i^* x_{j_c} + h.c\Big)  + \sum_{i_c,j }2U_{i_c j} n_j x^\dag_{i_c} x_{i_c}.
\end{align}
Here $i_c$ ($i$) refers to the sites within (outside) the cluster and we have introduced the chemical potential $\mu$. The mean fields, namely the superfluid order parameter $\psi_i$ and exciton density $n_i$, are solved self-consistently. This is done by solving sufficiently many cluster problems, centered at different lattice sites. If site $i$ belongs to $M_i$ different clusters (as clusters can overlap), an average over these clusters is taken, i.e. $\psi_i = \frac{1}{M_i}\sum_C \langle x_{i} \rangle$, where $C$ is the cluster index (details are provided in SM~\cite{SM}). The expectation values $\langle x_{i_c} \rangle$ in cluster $C$ are computed by exactly diagonalizing the cluster Hamiltonian $H_C$ and taking the thermal average, i.e. $\langle x_{i_c} \rangle = \frac{1}{Z}\Tr{ e^{-\beta H_C} x_{i_c}}$, where $\beta = k_B T$ with $k_B$ and $T$ being the Boltzmann constant and temperature, and $Z_C = \Tr{e^{-\beta H_C}}$ is the partition function for cluster $C$. Obtained mean fields $\psi_i$ and $n_i$ are inserted back to the cluster Hamiltonian~\eqref{clusterH} and the iterative procedure is continued till $\psi_i$ and $n_i$ converge for all $i$. We use several different initial ansatzes for $\psi_i$ and $n_i$ and select the result with the lowest free energy $\Omega = -k_B T \log Z$, where $Z$ is the total partition function of the system. %In SM~\cite{SM}, we furthermore show that more complicated CMF calculations without the assumption about the two-sublattice division lead to the similar results.

To exactly diagonalize Eq.~\eqref{clusterH}, we consider the Hilbert subspace spanned by the Fock states which have, at maximum, one particle per each site. Moreover, we do not fix the particle number as the average density is controlled by $\mu$ in Eq.~\eqref{clusterH}.  This ensures that we can access the superfluid order parameter that breaks the $U(1)$-gauge symmetry. 

%%%%%%%%%%%%%%%%%%%%%%%%%%%%%%%%%%%%%%%%%%%%%%%%%%%%%%%%%%%%%%%%%%%%%%%%%%%

\begin{figure}
  \centering
    \includegraphics[width=1.0\columnwidth]{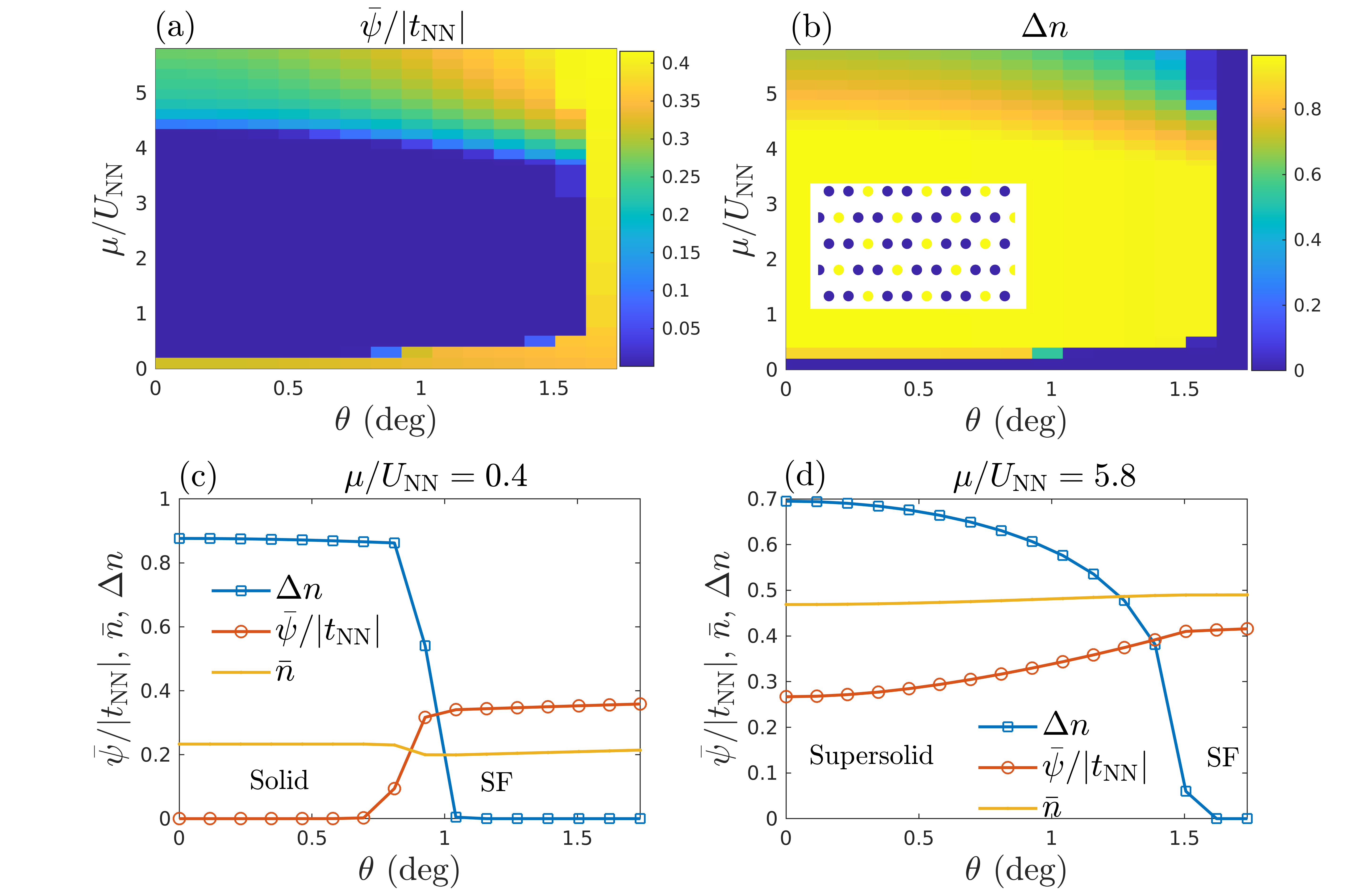}
    \caption{(a)-(b) CMF results for $\bar{\psi}$ and $\Delta n$, respectively, as a function of $\theta$ and $\mu$ at $T=0$ by using $H_H$. Inset of panel (b) shows the spatial profile of $n_i$ of the solid phase (yellow color denotes $n_i=1$ and blue $n_i =0$) with $\bar n = 1/3$. (c)-(d) $\bar{\psi}$, $\Delta n$ and $\bar{n}$ as a function of $\theta$ for $\mu/U_{\mathrm{NN}} = 0.4$ and  $\mu/U_{\mathrm{NN}} = 5.8$, respectively, at $T=0$.}
   \label{Fig:R1}
\end{figure}

%%%%%%%%%%%%%%%%%%%%%%%%%%%%%%%%%%%%%%%%%%%%%%%%%%%%%%%%%%%%%%%%%%%%%%%%%%%

 We present in Figs.~\ref{Fig:R1}(a)-(b) our CMF results for MoSe$_2$/WS$_2$ as a function of $\mu$ and $\theta$, obtained with 10-site clusters (see SM~\cite{SM}) by using $t_{\text{NN}}$, $U_0$ and $U_{ij}$ of $H_H$.  To study possible broken spatial symmetries, we define the staggered density as $\Delta n \equiv \frac{\max n_i - \min n_i}{\max n_i}$. We show both the average superfluid order parameter $\bar{\psi}$ [Fig.~\ref{Fig:R1}(a)] and $\Delta n$ [Fig.~\ref{Fig:R1}(b)]. For clarity, $\bar{\psi}$, $\Delta n$ and average density $\bar n$ are also depicted in Figs.~\ref{Fig:R1}(c)-(d) as a function of $\theta$ for $\mu/U_{\mathrm{NN}} = 0.4$ and $\mu/U_{\mathrm{NN}} = 5.8$, respectively.  
 
 From Fig.~\ref{Fig:R1} we see that by tuning $\theta$ and $\mu$ (i.e. $n$), one can reach different many-body phases: spatially homogeneous superfluid (SF) phase (characterized by $\bar{\psi} \neq 0$, $ \Delta n =0$), solid phase with broken translational symmetry ($\bar{\psi} = 0$, $\Delta n \neq 0$) and, importantly, supersolid ($\bar{\psi} \neq 0$, $\Delta n \neq 0$). The solid phase has the average density of $\bar n = 1/3$ and its spatial density profile, depicted in the inset of Fig.~\ref{Fig:R1}(b), is characterized by vanishing density within two thirds of the sites. The supersolid phase has a similar staggered density pattern, with the exception of having finite density in all the sites so that $\bar{n} > 1/3$.

\begin{figure}
  \centering
    \includegraphics[width=1.0\columnwidth]{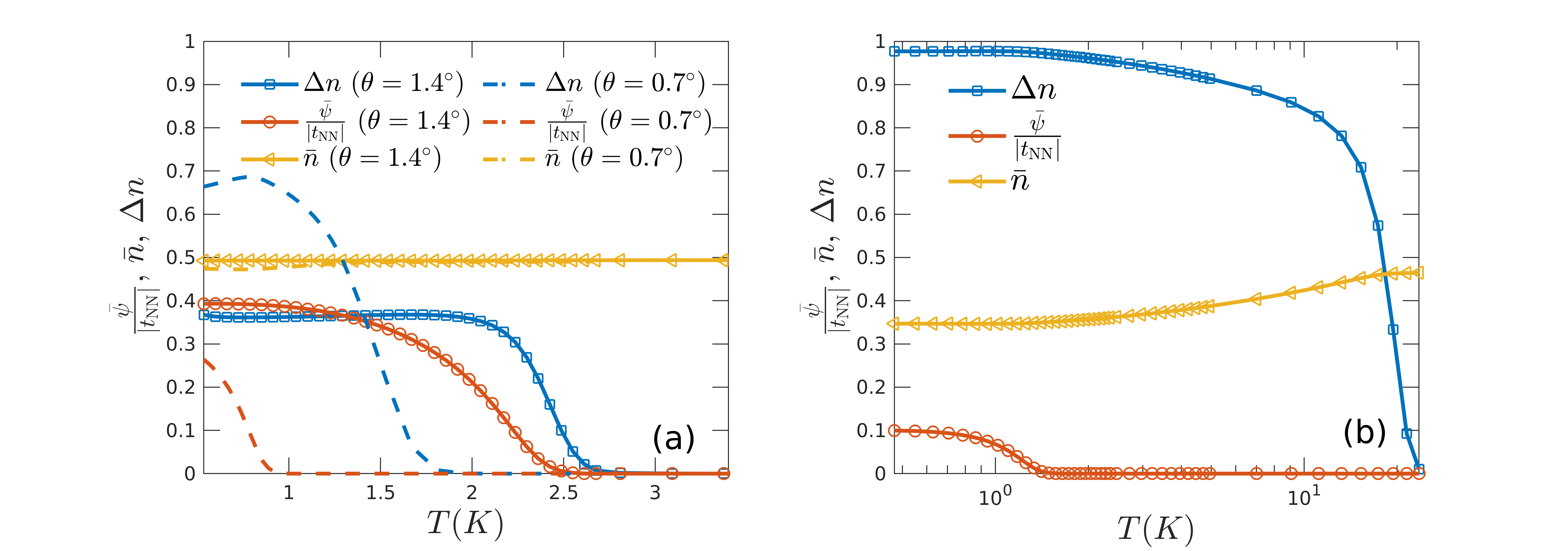}
    \caption{(a) Calculated $\Delta n$, $\bar{\psi}$ and $\bar{n}$ as a function of $T$ at $\mu/U_{\mathrm{NN}} = 5.88$, $\theta = 1.4^\circ$ (symbols) and at $\mu/U_{\mathrm{NN}} = 5.8$, $\theta = 0.7^\circ$ (dashed lines) by using $H_H$. (b) Corresponding results with $H_E$ for $\mu/U_{\mathrm{NN}} = 5.7$ and $\theta = 3^\circ$.}
   \label{Fig:T} 
\end{figure}

To study how finite temperature affects the supersolid phase, we plot in Fig.~\ref{Fig:T}(a) $\bar{\psi}$ and $\Delta n$ as a function of $T$ for $\mu/U_{\mathrm{NN}} = 5.88$ at $\theta = 1.4^\circ$ (symbols) and at  $\mu/U_{\mathrm{NN}} = 5.8$ with $\theta = 0.7^\circ$ (dashed lines). At $\theta = 1.4^\circ$, the superfluid component of the supersolid vanishes around $T\sim 2.4$ K, whereas the staggered density pattern survives to slightly higher temperatures. Similar trend can be seen more clearly in case of $\theta = 0.7^\circ$ for which the staggered solid phase vanishes at considerably higher temperatures compared to the superfluid order. This is understandable as the superfluidity emerges due to  $U(1)$ symmetry breaking and is thus more susceptible to thermal phase fluctuations. Notably, the superfluid critical temperatures $T_c$ obtained here are experimentally accessible ~\cite{Seyler2019,Wang2022}. One should note, though, that CMF accounts for exactly local and short-ranged quantum fluctuations but treats long-range flutuations in the mean-field level. Thus, CMF most likely overestimates $T_c$. Our prediction, however, should be better than that given by a simple Gutzwiller mean-field theory. To improve the prediction for $T_c$, one should perform a fluctuation analysis for the complex phase of $\psi_i$ to access the BKT-transition temperature. With CMF, this could be done as in Ref.~\cite{Malakar2020}, where fluctuations of the density matrix were studied, or computing the superfluid density by extending the quantum Gutzwiller theory~\cite{Caleffi2020} for our cluster approach. We leave this aspect to future studies.

For completeness, we performed CMF computations for MoSe$_2$/WSe$_2$ by using $t_{\text{NN}}$, $U_0$ and $U_{ij}$, obtained from $H_E$.  With experimentally feasible parameters~\cite{Tran2019}, twist angles of $\theta \sim 3^\circ$ yield supersolidity (see SM~\cite{SM}). In Fig.~\ref{Fig:T}(b) we show $\Delta n$, $\bar{\psi}$ and $\bar{n}$ as a function of $T$ at $\mu/U_{\mathrm{NN}} = 5.7$ and $\theta = 3^\circ$. We see that $T_c \sim 1$ K.  Our prediction for excitonic supersolidity is thus not model-dependent but an intrinsic property of \moire excitons that feature a finite interlayer exciton component.

\textit{Discussion---} We have demonstrated that \moire excitons, in addition to previously predicted Mott and conventional superfluid states~\cite{Gotting2022,Lagoin2021}, can also host superfluid and insulating states that break the periodicity of of the original triangular \moire lattice. By tuning the density of excitons and the twist angle, one can reach these many-body phases within a reasonable parameter regime and at experimentally accessible temperatures. We employed two different continuum models to build the tight-binding models and interactions for \moire excitons. The common feature of these models is the presence of interlayer excitons. Non-trivial states of broken translational invariance emerge then from strong non-local Coulomb interactions, which cannot be ignored in experimentally feasible density regime. %The presence of the interlayer exciton component is therefore crucial to achieve supersolidity of \moire excitons. 

We used here equilibrium CMF theory, whereas experimental systems are inherently in non-equilibrium due to optical pumping and decay processes of excitons.  However, interlayer excitons can exhibit relatively long lifetimes in \moire systems; in Ref.~\cite{Choi2021} the lifetime of interlayer excitons was measured to be around 1-10 nm for small twist angles, giving the decay rate of $\gamma \sim 10^{-4} – 10^{-3}$ meV. 
This is still much smaller than other energy scales. For example, $|t_{\text{NN}}| \sim 0.1$ meV at $\theta \sim 0.7^\circ$ used in Fig.~\ref{Fig:T}(a). Thus, we expect that including a decay term in Eq.~\eqref{mainHam} does not change our conclusions qualitatively. Moreover, recently realized dual-\moire systems~\cite{Yihang2022,YaHui2022}, where the system geometry suppresses the recombination of  electron-hole pairs, can further enhance the lifetime of interlayer excitons and can thus be a strong candidate for realizing supersolid phases similar to the ones predicted here.  Non-equilibrium dynamics of excitons provides a rich playground to reveal new properties of \moire systems~\cite{Camacho-Guardian2021moire,Choi2021} and remains an important topic for future studies.

Our work considers excitons in the $K$-valley only. On the other hand, the hybridized exciton model $H_H$ allows us to simultaneously consider the $K'$-valley excitons. However, the intralayer excitons of different valleys are coupled via the intervalley exchange interaction~\cite{Yu2014b,Yu2014,Wu2015b}. Consequently, the lowest \moire exciton bands of two valleys hybridize and a two-band tight-binding model is required~\cite{Marzari1997} to faithfully study such intervalley \moire excitons. Intervalley \moire physics can be very rich, e.g. possibly leading to topological band structures~\cite{Wu2017} and excitonic phases of broken crystal symmetries~\cite{Remez2021}. 

\Moire excitons can also easily couple with light, forming \moire exciton polaritons~\cite{Zhang2021} and therefore allowing the creation of strong non-linearities and many-body phases of light~\cite{Camacho-Guardian2021moire,Camacho-Guardian2022moire}. Furthermore, electron-exciton interactions~\cite{Marcellina2021,Liu2021b,Tang2021,Wang2021,Shimazaki2020,Tang2020,Campbell2020,Liu2021b,Jin2021,Miao2021,Wang2022} provide a promising platform for studying strongly correlated Bose-Fermi mixtures. Our work, describing the possibility to access supersolid phases, manifests vast opportunities of \moire excitons and highlights their significant role on studying strongly correalated bosonic many-body phases.

%\begin{acknowledgments} 
\textit{Acknowledgements}--- The author is  grateful to Georg M. Bruun and Arturo Camacho-Guardian for very useful discussions. The author acknowledges financial support from the Jenny and Antti Wihuri Foundation.
%\end{acknowledgments}

%-lifetime of intralayer excitons -> polaritons, hybrid excitons

%\begin{acknowledgments} 
%\textit{Acknowledgements}---  A.J. \ acknowledges financial support from the Jenny and Antti Wihuri Foundation. 
%\end{acknowledgments}

\vspace{2cm}


\begin{thebibliography}{80}%
\makeatletter
\providecommand \@ifxundefined [1]{%
 \@ifx{#1\undefined}
}%
\providecommand \@ifnum [1]{%
 \ifnum #1\expandafter \@firstoftwo
 \else \expandafter \@secondoftwo
 \fi
}%
\providecommand \@ifx [1]{%
 \ifx #1\expandafter \@firstoftwo
 \else \expandafter \@secondoftwo
 \fi
}%
\providecommand \natexlab [1]{#1}%
\providecommand \enquote  [1]{``#1''}%
\providecommand \bibnamefont  [1]{#1}%
\providecommand \bibfnamefont [1]{#1}%
\providecommand \citenamefont [1]{#1}%
\providecommand \href@noop [0]{\@secondoftwo}%
\providecommand \href [0]{\begingroup \@sanitize@url \@href}%
\providecommand \@href[1]{\@@startlink{#1}\@@href}%
\providecommand \@@href[1]{\endgroup#1\@@endlink}%
\providecommand \@sanitize@url [0]{\catcode `\\12\catcode `\$12\catcode
  `\&12\catcode `\#12\catcode `\^12\catcode `\_12\catcode `\%12\relax}%
\providecommand \@@startlink[1]{}%
\providecommand \@@endlink[0]{}%
\providecommand \url  [0]{\begingroup\@sanitize@url \@url }%
\providecommand \@url [1]{\endgroup\@href {#1}{\urlprefix }}%
\providecommand \urlprefix  [0]{URL }%
\providecommand \Eprint [0]{\href }%
\providecommand \doibase [0]{http://dx.doi.org/}%
\providecommand \selectlanguage [0]{\@gobble}%
\providecommand \bibinfo  [0]{\@secondoftwo}%
\providecommand \bibfield  [0]{\@secondoftwo}%
\providecommand \translation [1]{[#1]}%
\providecommand \BibitemOpen [0]{}%
\providecommand \bibitemStop [0]{}%
\providecommand \bibitemNoStop [0]{.\EOS\space}%
\providecommand \EOS [0]{\spacefactor3000\relax}%
\providecommand \BibitemShut  [1]{\csname bibitem#1\endcsname}%
\let\auto@bib@innerbib\@empty
%</preamble>
\bibitem [{\citenamefont {Balents}\ \emph {et~al.}(2020)\citenamefont
  {Balents}, \citenamefont {Dean}, \citenamefont {Efetov},\ and\ \citenamefont
  {Young}}]{Balents2020}%
  \BibitemOpen
  \bibfield  {author} {\bibinfo {author} {\bibfnamefont {L.}~\bibnamefont
  {Balents}}, \bibinfo {author} {\bibfnamefont {C.~R.}\ \bibnamefont {Dean}},
  \bibinfo {author} {\bibfnamefont {D.~K.}\ \bibnamefont {Efetov}}, \ and\
  \bibinfo {author} {\bibfnamefont {A.~F.}\ \bibnamefont {Young}},\ }\href
  {\doibase 10.1038/s41567-020-0906-9} {\bibfield  {journal} {\bibinfo
  {journal} {Nature Physics}\ }\textbf {\bibinfo {volume} {16}},\ \bibinfo
  {pages} {725} (\bibinfo {year} {2020})}\BibitemShut {NoStop}%
\bibitem [{\citenamefont {Kennes}\ \emph {et~al.}(2021)\citenamefont {Kennes},
  \citenamefont {Claassen}, \citenamefont {Xian}, \citenamefont {Georges},
  \citenamefont {Millis}, \citenamefont {Hone}, \citenamefont {Dean},
  \citenamefont {Basov}, \citenamefont {Pasupathy},\ and\ \citenamefont
  {Rubio}}]{Kennes2021}%
  \BibitemOpen
  \bibfield  {author} {\bibinfo {author} {\bibfnamefont {D.~M.}\ \bibnamefont
  {Kennes}}, \bibinfo {author} {\bibfnamefont {M.}~\bibnamefont {Claassen}},
  \bibinfo {author} {\bibfnamefont {L.}~\bibnamefont {Xian}}, \bibinfo {author}
  {\bibfnamefont {A.}~\bibnamefont {Georges}}, \bibinfo {author} {\bibfnamefont
  {A.~J.}\ \bibnamefont {Millis}}, \bibinfo {author} {\bibfnamefont
  {J.}~\bibnamefont {Hone}}, \bibinfo {author} {\bibfnamefont {C.~R.}\
  \bibnamefont {Dean}}, \bibinfo {author} {\bibfnamefont {D.~N.}\ \bibnamefont
  {Basov}}, \bibinfo {author} {\bibfnamefont {A.~N.}\ \bibnamefont
  {Pasupathy}}, \ and\ \bibinfo {author} {\bibfnamefont {A.}~\bibnamefont
  {Rubio}},\ }\href {\doibase 10.1038/s41567-020-01154-3} {\bibfield  {journal}
  {\bibinfo  {journal} {Nature Physics}\ }\textbf {\bibinfo {volume} {17}},\
  \bibinfo {pages} {155} (\bibinfo {year} {2021})}\BibitemShut {NoStop}%
\bibitem [{\citenamefont {Andrei}\ \emph {et~al.}(2021)\citenamefont {Andrei},
  \citenamefont {Efetov}, \citenamefont {Jarillo-Herrero}, \citenamefont
  {MacDonald}, \citenamefont {Mak}, \citenamefont {Senthil}, \citenamefont
  {Tutuc}, \citenamefont {Yazdani},\ and\ \citenamefont {Young}}]{Andrei2021}%
  \BibitemOpen
  \bibfield  {author} {\bibinfo {author} {\bibfnamefont {E.~Y.}\ \bibnamefont
  {Andrei}}, \bibinfo {author} {\bibfnamefont {D.~K.}\ \bibnamefont {Efetov}},
  \bibinfo {author} {\bibfnamefont {P.}~\bibnamefont {Jarillo-Herrero}},
  \bibinfo {author} {\bibfnamefont {A.~H.}\ \bibnamefont {MacDonald}}, \bibinfo
  {author} {\bibfnamefont {K.~F.}\ \bibnamefont {Mak}}, \bibinfo {author}
  {\bibfnamefont {T.}~\bibnamefont {Senthil}}, \bibinfo {author} {\bibfnamefont
  {E.}~\bibnamefont {Tutuc}}, \bibinfo {author} {\bibfnamefont
  {A.}~\bibnamefont {Yazdani}}, \ and\ \bibinfo {author} {\bibfnamefont
  {A.~F.}\ \bibnamefont {Young}},\ }\href {\doibase 10.1038/s41578-021-00284-1}
  {\bibfield  {journal} {\bibinfo  {journal} {Nature Reviews Materials}\
  }\textbf {\bibinfo {volume} {6}},\ \bibinfo {pages} {201} (\bibinfo {year}
  {2021})}\BibitemShut {NoStop}%
\bibitem [{\citenamefont {Cao}\ \emph {et~al.}(2018)\citenamefont {Cao},
  \citenamefont {Fatemi}, \citenamefont {Fang}, \citenamefont {Watanabe},
  \citenamefont {Taniguchi}, \citenamefont {Kaxiras},\ and\ \citenamefont
  {Jarillo-Herrero}}]{Cao2018}%
  \BibitemOpen
  \bibfield  {author} {\bibinfo {author} {\bibfnamefont {Y.}~\bibnamefont
  {Cao}}, \bibinfo {author} {\bibfnamefont {V.}~\bibnamefont {Fatemi}},
  \bibinfo {author} {\bibfnamefont {S.}~\bibnamefont {Fang}}, \bibinfo {author}
  {\bibfnamefont {K.}~\bibnamefont {Watanabe}}, \bibinfo {author}
  {\bibfnamefont {T.}~\bibnamefont {Taniguchi}}, \bibinfo {author}
  {\bibfnamefont {E.}~\bibnamefont {Kaxiras}}, \ and\ \bibinfo {author}
  {\bibfnamefont {P.}~\bibnamefont {Jarillo-Herrero}},\ }\href {\doibase
  10.1038/nature26160} {\bibfield  {journal} {\bibinfo  {journal} {Nature}\
  }\textbf {\bibinfo {volume} {556}},\ \bibinfo {pages} {43} (\bibinfo {year}
  {2018})}\BibitemShut {NoStop}%
\bibitem [{\citenamefont {Yankowitz}\ \emph {et~al.}(2019)\citenamefont
  {Yankowitz}, \citenamefont {Chen}, \citenamefont {Polshyn}, \citenamefont
  {Zhang}, \citenamefont {Watanabe}, \citenamefont {Taniguchi}, \citenamefont
  {Graf}, \citenamefont {Young},\ and\ \citenamefont {Dean}}]{Yankowitz2019}%
  \BibitemOpen
  \bibfield  {author} {\bibinfo {author} {\bibfnamefont {M.}~\bibnamefont
  {Yankowitz}}, \bibinfo {author} {\bibfnamefont {S.}~\bibnamefont {Chen}},
  \bibinfo {author} {\bibfnamefont {H.}~\bibnamefont {Polshyn}}, \bibinfo
  {author} {\bibfnamefont {Y.}~\bibnamefont {Zhang}}, \bibinfo {author}
  {\bibfnamefont {K.}~\bibnamefont {Watanabe}}, \bibinfo {author}
  {\bibfnamefont {T.}~\bibnamefont {Taniguchi}}, \bibinfo {author}
  {\bibfnamefont {D.}~\bibnamefont {Graf}}, \bibinfo {author} {\bibfnamefont
  {A.~F.}\ \bibnamefont {Young}}, \ and\ \bibinfo {author} {\bibfnamefont
  {C.~R.}\ \bibnamefont {Dean}},\ }\href {\doibase 10.1126/science.aav1910}
  {\bibfield  {journal} {\bibinfo  {journal} {Science}\ }\textbf {\bibinfo
  {volume} {363}},\ \bibinfo {pages} {1059} (\bibinfo {year} {2019})},\ \Eprint
  {http://arxiv.org/abs/https://www.science.org/doi/pdf/10.1126/science.aav1910}
  {https://www.science.org/doi/pdf/10.1126/science.aav1910} \BibitemShut
  {NoStop}%
\bibitem [{\citenamefont {Chen}\ \emph {et~al.}(2019)\citenamefont {Chen},
  \citenamefont {Sharpe}, \citenamefont {Gallagher}, \citenamefont {Rosen},
  \citenamefont {Fox}, \citenamefont {Jiang}, \citenamefont {Lyu},
  \citenamefont {Li}, \citenamefont {Watanabe}, \citenamefont {Taniguchi},
  \citenamefont {Jung}, \citenamefont {Shi}, \citenamefont {Goldhaber-Gordon},
  \citenamefont {Zhang},\ and\ \citenamefont {Wang}}]{Chen2019}%
  \BibitemOpen
  \bibfield  {author} {\bibinfo {author} {\bibfnamefont {G.}~\bibnamefont
  {Chen}}, \bibinfo {author} {\bibfnamefont {A.~L.}\ \bibnamefont {Sharpe}},
  \bibinfo {author} {\bibfnamefont {P.}~\bibnamefont {Gallagher}}, \bibinfo
  {author} {\bibfnamefont {I.~T.}\ \bibnamefont {Rosen}}, \bibinfo {author}
  {\bibfnamefont {E.~J.}\ \bibnamefont {Fox}}, \bibinfo {author} {\bibfnamefont
  {L.}~\bibnamefont {Jiang}}, \bibinfo {author} {\bibfnamefont
  {B.}~\bibnamefont {Lyu}}, \bibinfo {author} {\bibfnamefont {H.}~\bibnamefont
  {Li}}, \bibinfo {author} {\bibfnamefont {K.}~\bibnamefont {Watanabe}},
  \bibinfo {author} {\bibfnamefont {T.}~\bibnamefont {Taniguchi}}, \bibinfo
  {author} {\bibfnamefont {J.}~\bibnamefont {Jung}}, \bibinfo {author}
  {\bibfnamefont {Z.}~\bibnamefont {Shi}}, \bibinfo {author} {\bibfnamefont
  {D.}~\bibnamefont {Goldhaber-Gordon}}, \bibinfo {author} {\bibfnamefont
  {Y.}~\bibnamefont {Zhang}}, \ and\ \bibinfo {author} {\bibfnamefont
  {F.}~\bibnamefont {Wang}},\ }\href {\doibase 10.1038/s41586-019-1393-y}
  {\bibfield  {journal} {\bibinfo  {journal} {Nature}\ }\textbf {\bibinfo
  {volume} {572}},\ \bibinfo {pages} {215} (\bibinfo {year}
  {2019})}\BibitemShut {NoStop}%
\bibitem [{\citenamefont {Wang}\ \emph {et~al.}(2020)\citenamefont {Wang},
  \citenamefont {Shih}, \citenamefont {Ghiotto}, \citenamefont {Xian},
  \citenamefont {Rhodes}, \citenamefont {Tan}, \citenamefont {Claassen},
  \citenamefont {Kennes}, \citenamefont {Bai}, \citenamefont {Kim},
  \citenamefont {Watanabe}, \citenamefont {Taniguchi}, \citenamefont {Zhu},
  \citenamefont {Hone}, \citenamefont {Rubio}, \citenamefont {Pasupathy},\ and\
  \citenamefont {Dean}}]{Wang2020}%
  \BibitemOpen
  \bibfield  {author} {\bibinfo {author} {\bibfnamefont {L.}~\bibnamefont
  {Wang}}, \bibinfo {author} {\bibfnamefont {E.-M.}\ \bibnamefont {Shih}},
  \bibinfo {author} {\bibfnamefont {A.}~\bibnamefont {Ghiotto}}, \bibinfo
  {author} {\bibfnamefont {L.}~\bibnamefont {Xian}}, \bibinfo {author}
  {\bibfnamefont {D.~A.}\ \bibnamefont {Rhodes}}, \bibinfo {author}
  {\bibfnamefont {C.}~\bibnamefont {Tan}}, \bibinfo {author} {\bibfnamefont
  {M.}~\bibnamefont {Claassen}}, \bibinfo {author} {\bibfnamefont {D.~M.}\
  \bibnamefont {Kennes}}, \bibinfo {author} {\bibfnamefont {Y.}~\bibnamefont
  {Bai}}, \bibinfo {author} {\bibfnamefont {B.}~\bibnamefont {Kim}}, \bibinfo
  {author} {\bibfnamefont {K.}~\bibnamefont {Watanabe}}, \bibinfo {author}
  {\bibfnamefont {T.}~\bibnamefont {Taniguchi}}, \bibinfo {author}
  {\bibfnamefont {X.}~\bibnamefont {Zhu}}, \bibinfo {author} {\bibfnamefont
  {J.}~\bibnamefont {Hone}}, \bibinfo {author} {\bibfnamefont {A.}~\bibnamefont
  {Rubio}}, \bibinfo {author} {\bibfnamefont {A.~N.}\ \bibnamefont
  {Pasupathy}}, \ and\ \bibinfo {author} {\bibfnamefont {C.~R.}\ \bibnamefont
  {Dean}},\ }\href {\doibase 10.1038/s41563-020-0708-6} {\bibfield  {journal}
  {\bibinfo  {journal} {Nature Materials}\ }\textbf {\bibinfo {volume} {19}},\
  \bibinfo {pages} {861} (\bibinfo {year} {2020})}\BibitemShut {NoStop}%
\bibitem [{\citenamefont {Huang}\ \emph {et~al.}(2021)\citenamefont {Huang},
  \citenamefont {Wang}, \citenamefont {Miao}, \citenamefont {Wang},
  \citenamefont {Li}, \citenamefont {Lian}, \citenamefont {Taniguchi},
  \citenamefont {Watanabe}, \citenamefont {Okamoto}, \citenamefont {Xiao},
  \citenamefont {Shi},\ and\ \citenamefont {Cui}}]{Huang2021}%
  \BibitemOpen
  \bibfield  {author} {\bibinfo {author} {\bibfnamefont {X.}~\bibnamefont
  {Huang}}, \bibinfo {author} {\bibfnamefont {T.}~\bibnamefont {Wang}},
  \bibinfo {author} {\bibfnamefont {S.}~\bibnamefont {Miao}}, \bibinfo {author}
  {\bibfnamefont {C.}~\bibnamefont {Wang}}, \bibinfo {author} {\bibfnamefont
  {Z.}~\bibnamefont {Li}}, \bibinfo {author} {\bibfnamefont {Z.}~\bibnamefont
  {Lian}}, \bibinfo {author} {\bibfnamefont {T.}~\bibnamefont {Taniguchi}},
  \bibinfo {author} {\bibfnamefont {K.}~\bibnamefont {Watanabe}}, \bibinfo
  {author} {\bibfnamefont {S.}~\bibnamefont {Okamoto}}, \bibinfo {author}
  {\bibfnamefont {D.}~\bibnamefont {Xiao}}, \bibinfo {author} {\bibfnamefont
  {S.-F.}\ \bibnamefont {Shi}}, \ and\ \bibinfo {author} {\bibfnamefont
  {Y.-T.}\ \bibnamefont {Cui}},\ }\href {\doibase 10.1038/s41567-021-01171-w}
  {\bibfield  {journal} {\bibinfo  {journal} {Nature Physics}\ }\textbf
  {\bibinfo {volume} {17}},\ \bibinfo {pages} {715} (\bibinfo {year}
  {2021})}\BibitemShut {NoStop}%
\bibitem [{\citenamefont {Shimazaki}\ \emph {et~al.}(2020)\citenamefont
  {Shimazaki}, \citenamefont {Schwartz}, \citenamefont {Watanabe},
  \citenamefont {Taniguchi}, \citenamefont {Kroner},\ and\ \citenamefont
  {Imamo{\u{g}}lu}}]{Shimazaki2020}%
  \BibitemOpen
  \bibfield  {author} {\bibinfo {author} {\bibfnamefont {Y.}~\bibnamefont
  {Shimazaki}}, \bibinfo {author} {\bibfnamefont {I.}~\bibnamefont {Schwartz}},
  \bibinfo {author} {\bibfnamefont {K.}~\bibnamefont {Watanabe}}, \bibinfo
  {author} {\bibfnamefont {T.}~\bibnamefont {Taniguchi}}, \bibinfo {author}
  {\bibfnamefont {M.}~\bibnamefont {Kroner}}, \ and\ \bibinfo {author}
  {\bibfnamefont {A.}~\bibnamefont {Imamo{\u{g}}lu}},\ }\href {\doibase
  10.1038/s41586-020-2191-2} {\bibfield  {journal} {\bibinfo  {journal}
  {Nature}\ }\textbf {\bibinfo {volume} {580}},\ \bibinfo {pages} {472}
  (\bibinfo {year} {2020})}\BibitemShut {NoStop}%
\bibitem [{\citenamefont {Xu}\ \emph {et~al.}(2020)\citenamefont {Xu},
  \citenamefont {Liu}, \citenamefont {Rhodes}, \citenamefont {Watanabe},
  \citenamefont {Taniguchi}, \citenamefont {Hone}, \citenamefont {Elser},
  \citenamefont {Mak},\ and\ \citenamefont {Shan}}]{Xu2020}%
  \BibitemOpen
  \bibfield  {author} {\bibinfo {author} {\bibfnamefont {Y.}~\bibnamefont
  {Xu}}, \bibinfo {author} {\bibfnamefont {S.}~\bibnamefont {Liu}}, \bibinfo
  {author} {\bibfnamefont {D.~A.}\ \bibnamefont {Rhodes}}, \bibinfo {author}
  {\bibfnamefont {K.}~\bibnamefont {Watanabe}}, \bibinfo {author}
  {\bibfnamefont {T.}~\bibnamefont {Taniguchi}}, \bibinfo {author}
  {\bibfnamefont {J.}~\bibnamefont {Hone}}, \bibinfo {author} {\bibfnamefont
  {V.}~\bibnamefont {Elser}}, \bibinfo {author} {\bibfnamefont {K.~F.}\
  \bibnamefont {Mak}}, \ and\ \bibinfo {author} {\bibfnamefont
  {J.}~\bibnamefont {Shan}},\ }\href {\doibase 10.1038/s41586-020-2868-6}
  {\bibfield  {journal} {\bibinfo  {journal} {Nature}\ }\textbf {\bibinfo
  {volume} {587}},\ \bibinfo {pages} {214} (\bibinfo {year}
  {2020})}\BibitemShut {NoStop}%
\bibitem [{\citenamefont {Tang}\ \emph {et~al.}(2020)\citenamefont {Tang},
  \citenamefont {Li}, \citenamefont {Li}, \citenamefont {Xu}, \citenamefont
  {Liu}, \citenamefont {Barmak}, \citenamefont {Watanabe}, \citenamefont
  {Taniguchi}, \citenamefont {MacDonald}, \citenamefont {Shan},\ and\
  \citenamefont {Mak}}]{Tang2020}%
  \BibitemOpen
  \bibfield  {author} {\bibinfo {author} {\bibfnamefont {Y.}~\bibnamefont
  {Tang}}, \bibinfo {author} {\bibfnamefont {L.}~\bibnamefont {Li}}, \bibinfo
  {author} {\bibfnamefont {T.}~\bibnamefont {Li}}, \bibinfo {author}
  {\bibfnamefont {Y.}~\bibnamefont {Xu}}, \bibinfo {author} {\bibfnamefont
  {S.}~\bibnamefont {Liu}}, \bibinfo {author} {\bibfnamefont {K.}~\bibnamefont
  {Barmak}}, \bibinfo {author} {\bibfnamefont {K.}~\bibnamefont {Watanabe}},
  \bibinfo {author} {\bibfnamefont {T.}~\bibnamefont {Taniguchi}}, \bibinfo
  {author} {\bibfnamefont {A.~H.}\ \bibnamefont {MacDonald}}, \bibinfo {author}
  {\bibfnamefont {J.}~\bibnamefont {Shan}}, \ and\ \bibinfo {author}
  {\bibfnamefont {K.~F.}\ \bibnamefont {Mak}},\ }\href {\doibase
  10.1038/s41586-020-2085-3} {\bibfield  {journal} {\bibinfo  {journal}
  {Nature}\ }\textbf {\bibinfo {volume} {579}},\ \bibinfo {pages} {353}
  (\bibinfo {year} {2020})}\BibitemShut {NoStop}%
\bibitem [{\citenamefont {Campbell}\ \emph {et~al.}(2022)\citenamefont
  {Campbell}, \citenamefont {Brotons-Gisbert}, \citenamefont {Baek},
  \citenamefont {Vitale}, \citenamefont {Taniguchi}, \citenamefont {Watanabe},
  \citenamefont {Lischner},\ and\ \citenamefont {Gerardot}}]{Campbell2020}%
  \BibitemOpen
  \bibfield  {author} {\bibinfo {author} {\bibfnamefont {A.~J.}\ \bibnamefont
  {Campbell}}, \bibinfo {author} {\bibfnamefont {M.}~\bibnamefont
  {Brotons-Gisbert}}, \bibinfo {author} {\bibfnamefont {H.}~\bibnamefont
  {Baek}}, \bibinfo {author} {\bibfnamefont {V.}~\bibnamefont {Vitale}},
  \bibinfo {author} {\bibfnamefont {T.}~\bibnamefont {Taniguchi}}, \bibinfo
  {author} {\bibfnamefont {K.}~\bibnamefont {Watanabe}}, \bibinfo {author}
  {\bibfnamefont {J.}~\bibnamefont {Lischner}}, \ and\ \bibinfo {author}
  {\bibfnamefont {B.~D.}\ \bibnamefont {Gerardot}},\ }\href {\doibase
  10.48550/ARXIV.2202.08879} {\enquote {\bibinfo {title} {Strongly correlated
  electronic states in a fermi sea spatially pinned by a mose$_2$/wse$_2$
  moiré superlattice},}\ } (\bibinfo {year} {2022})\BibitemShut {NoStop}%
\bibitem [{\citenamefont {Liu}\ \emph {et~al.}(2021{\natexlab{a}})\citenamefont
  {Liu}, \citenamefont {Taniguchi}, \citenamefont {Watanabe}, \citenamefont
  {Gabor}, \citenamefont {Cui},\ and\ \citenamefont {Lui}}]{Liu2021}%
  \BibitemOpen
  \bibfield  {author} {\bibinfo {author} {\bibfnamefont {E.}~\bibnamefont
  {Liu}}, \bibinfo {author} {\bibfnamefont {T.}~\bibnamefont {Taniguchi}},
  \bibinfo {author} {\bibfnamefont {K.}~\bibnamefont {Watanabe}}, \bibinfo
  {author} {\bibfnamefont {N.~M.}\ \bibnamefont {Gabor}}, \bibinfo {author}
  {\bibfnamefont {Y.-T.}\ \bibnamefont {Cui}}, \ and\ \bibinfo {author}
  {\bibfnamefont {C.~H.}\ \bibnamefont {Lui}},\ }\href {\doibase
  10.1103/PhysRevLett.127.037402} {\bibfield  {journal} {\bibinfo  {journal}
  {Phys. Rev. Lett.}\ }\textbf {\bibinfo {volume} {127}},\ \bibinfo {pages}
  {037402} (\bibinfo {year} {2021}{\natexlab{a}})}\BibitemShut {NoStop}%
\bibitem [{\citenamefont {Jin}\ \emph {et~al.}(2021)\citenamefont {Jin},
  \citenamefont {Tao}, \citenamefont {Li}, \citenamefont {Xu}, \citenamefont
  {Tang}, \citenamefont {Zhu}, \citenamefont {Liu}, \citenamefont {Watanabe},
  \citenamefont {Taniguchi}, \citenamefont {Hone}, \citenamefont {Fu},
  \citenamefont {Shan},\ and\ \citenamefont {Mak}}]{Jin2021}%
  \BibitemOpen
  \bibfield  {author} {\bibinfo {author} {\bibfnamefont {C.}~\bibnamefont
  {Jin}}, \bibinfo {author} {\bibfnamefont {Z.}~\bibnamefont {Tao}}, \bibinfo
  {author} {\bibfnamefont {T.}~\bibnamefont {Li}}, \bibinfo {author}
  {\bibfnamefont {Y.}~\bibnamefont {Xu}}, \bibinfo {author} {\bibfnamefont
  {Y.}~\bibnamefont {Tang}}, \bibinfo {author} {\bibfnamefont {J.}~\bibnamefont
  {Zhu}}, \bibinfo {author} {\bibfnamefont {S.}~\bibnamefont {Liu}}, \bibinfo
  {author} {\bibfnamefont {K.}~\bibnamefont {Watanabe}}, \bibinfo {author}
  {\bibfnamefont {T.}~\bibnamefont {Taniguchi}}, \bibinfo {author}
  {\bibfnamefont {J.~C.}\ \bibnamefont {Hone}}, \bibinfo {author}
  {\bibfnamefont {L.}~\bibnamefont {Fu}}, \bibinfo {author} {\bibfnamefont
  {J.}~\bibnamefont {Shan}}, \ and\ \bibinfo {author} {\bibfnamefont {K.~F.}\
  \bibnamefont {Mak}},\ }\href {\doibase 10.1038/s41563-021-00959-8} {\bibfield
   {journal} {\bibinfo  {journal} {Nature Materials}\ }\textbf {\bibinfo
  {volume} {20}},\ \bibinfo {pages} {940} (\bibinfo {year} {2021})}\BibitemShut
  {NoStop}%
\bibitem [{\citenamefont {Miao}\ \emph {et~al.}(2021)\citenamefont {Miao},
  \citenamefont {Wang}, \citenamefont {Huang}, \citenamefont {Chen},
  \citenamefont {Lian}, \citenamefont {Wang}, \citenamefont {Blei},
  \citenamefont {Taniguchi}, \citenamefont {Watanabe}, \citenamefont {Tongay},
  \citenamefont {Wang}, \citenamefont {Xiao}, \citenamefont {Cui},\ and\
  \citenamefont {Shi}}]{Miao2021}%
  \BibitemOpen
  \bibfield  {author} {\bibinfo {author} {\bibfnamefont {S.}~\bibnamefont
  {Miao}}, \bibinfo {author} {\bibfnamefont {T.}~\bibnamefont {Wang}}, \bibinfo
  {author} {\bibfnamefont {X.}~\bibnamefont {Huang}}, \bibinfo {author}
  {\bibfnamefont {D.}~\bibnamefont {Chen}}, \bibinfo {author} {\bibfnamefont
  {Z.}~\bibnamefont {Lian}}, \bibinfo {author} {\bibfnamefont {C.}~\bibnamefont
  {Wang}}, \bibinfo {author} {\bibfnamefont {M.}~\bibnamefont {Blei}}, \bibinfo
  {author} {\bibfnamefont {T.}~\bibnamefont {Taniguchi}}, \bibinfo {author}
  {\bibfnamefont {K.}~\bibnamefont {Watanabe}}, \bibinfo {author}
  {\bibfnamefont {S.}~\bibnamefont {Tongay}}, \bibinfo {author} {\bibfnamefont
  {Z.}~\bibnamefont {Wang}}, \bibinfo {author} {\bibfnamefont {D.}~\bibnamefont
  {Xiao}}, \bibinfo {author} {\bibfnamefont {Y.-T.}\ \bibnamefont {Cui}}, \
  and\ \bibinfo {author} {\bibfnamefont {S.-F.}\ \bibnamefont {Shi}},\ }\href
  {\doibase 10.1038/s41467-021-23732-6} {\bibfield  {journal} {\bibinfo
  {journal} {Nature Communications}\ }\textbf {\bibinfo {volume} {12}},\
  \bibinfo {pages} {3608} (\bibinfo {year} {2021})}\BibitemShut {NoStop}%
\bibitem [{\citenamefont {Tran}\ \emph {et~al.}(2019)\citenamefont {Tran},
  \citenamefont {Moody}, \citenamefont {Wu}, \citenamefont {Lu}, \citenamefont
  {Choi}, \citenamefont {Kim}, \citenamefont {Rai}, \citenamefont {Sanchez},
  \citenamefont {Quan}, \citenamefont {Singh}, \citenamefont {Embley},
  \citenamefont {Zepeda}, \citenamefont {Campbell}, \citenamefont {Autry},
  \citenamefont {Taniguchi}, \citenamefont {Watanabe}, \citenamefont {Lu},
  \citenamefont {Banerjee}, \citenamefont {Silverman}, \citenamefont {Kim},
  \citenamefont {Tutuc}, \citenamefont {Yang}, \citenamefont {MacDonald},\ and\
  \citenamefont {Li}}]{Tran2019}%
  \BibitemOpen
  \bibfield  {author} {\bibinfo {author} {\bibfnamefont {K.}~\bibnamefont
  {Tran}}, \bibinfo {author} {\bibfnamefont {G.}~\bibnamefont {Moody}},
  \bibinfo {author} {\bibfnamefont {F.}~\bibnamefont {Wu}}, \bibinfo {author}
  {\bibfnamefont {X.}~\bibnamefont {Lu}}, \bibinfo {author} {\bibfnamefont
  {J.}~\bibnamefont {Choi}}, \bibinfo {author} {\bibfnamefont {K.}~\bibnamefont
  {Kim}}, \bibinfo {author} {\bibfnamefont {A.}~\bibnamefont {Rai}}, \bibinfo
  {author} {\bibfnamefont {D.~A.}\ \bibnamefont {Sanchez}}, \bibinfo {author}
  {\bibfnamefont {J.}~\bibnamefont {Quan}}, \bibinfo {author} {\bibfnamefont
  {A.}~\bibnamefont {Singh}}, \bibinfo {author} {\bibfnamefont
  {J.}~\bibnamefont {Embley}}, \bibinfo {author} {\bibfnamefont
  {A.}~\bibnamefont {Zepeda}}, \bibinfo {author} {\bibfnamefont
  {M.}~\bibnamefont {Campbell}}, \bibinfo {author} {\bibfnamefont
  {T.}~\bibnamefont {Autry}}, \bibinfo {author} {\bibfnamefont
  {T.}~\bibnamefont {Taniguchi}}, \bibinfo {author} {\bibfnamefont
  {K.}~\bibnamefont {Watanabe}}, \bibinfo {author} {\bibfnamefont
  {N.}~\bibnamefont {Lu}}, \bibinfo {author} {\bibfnamefont {S.~K.}\
  \bibnamefont {Banerjee}}, \bibinfo {author} {\bibfnamefont {K.~L.}\
  \bibnamefont {Silverman}}, \bibinfo {author} {\bibfnamefont {S.}~\bibnamefont
  {Kim}}, \bibinfo {author} {\bibfnamefont {E.}~\bibnamefont {Tutuc}}, \bibinfo
  {author} {\bibfnamefont {L.}~\bibnamefont {Yang}}, \bibinfo {author}
  {\bibfnamefont {A.~H.}\ \bibnamefont {MacDonald}}, \ and\ \bibinfo {author}
  {\bibfnamefont {X.}~\bibnamefont {Li}},\ }\href {\doibase
  10.1038/s41586-019-0975-z} {\bibfield  {journal} {\bibinfo  {journal}
  {Nature}\ }\textbf {\bibinfo {volume} {567}},\ \bibinfo {pages} {71}
  (\bibinfo {year} {2019})}\BibitemShut {NoStop}%
\bibitem [{\citenamefont {Alexeev}\ \emph {et~al.}(2019)\citenamefont
  {Alexeev}, \citenamefont {Ruiz-Tijerina}, \citenamefont {Danovich},
  \citenamefont {Hamer}, \citenamefont {Terry}, \citenamefont {Nayak},
  \citenamefont {Ahn}, \citenamefont {Pak}, \citenamefont {Lee}, \citenamefont
  {Sohn}, \citenamefont {Molas}, \citenamefont {Koperski}, \citenamefont
  {Watanabe}, \citenamefont {Taniguchi}, \citenamefont {Novoselov},
  \citenamefont {Gorbachev}, \citenamefont {Shin}, \citenamefont {Fal'ko},\
  and\ \citenamefont {Tartakovskii}}]{Alexeev2019}%
  \BibitemOpen
  \bibfield  {author} {\bibinfo {author} {\bibfnamefont {E.~M.}\ \bibnamefont
  {Alexeev}}, \bibinfo {author} {\bibfnamefont {D.~A.}\ \bibnamefont
  {Ruiz-Tijerina}}, \bibinfo {author} {\bibfnamefont {M.}~\bibnamefont
  {Danovich}}, \bibinfo {author} {\bibfnamefont {M.~J.}\ \bibnamefont {Hamer}},
  \bibinfo {author} {\bibfnamefont {D.~J.}\ \bibnamefont {Terry}}, \bibinfo
  {author} {\bibfnamefont {P.~K.}\ \bibnamefont {Nayak}}, \bibinfo {author}
  {\bibfnamefont {S.}~\bibnamefont {Ahn}}, \bibinfo {author} {\bibfnamefont
  {S.}~\bibnamefont {Pak}}, \bibinfo {author} {\bibfnamefont {J.}~\bibnamefont
  {Lee}}, \bibinfo {author} {\bibfnamefont {J.~I.}\ \bibnamefont {Sohn}},
  \bibinfo {author} {\bibfnamefont {M.~R.}\ \bibnamefont {Molas}}, \bibinfo
  {author} {\bibfnamefont {M.}~\bibnamefont {Koperski}}, \bibinfo {author}
  {\bibfnamefont {K.}~\bibnamefont {Watanabe}}, \bibinfo {author}
  {\bibfnamefont {T.}~\bibnamefont {Taniguchi}}, \bibinfo {author}
  {\bibfnamefont {K.~S.}\ \bibnamefont {Novoselov}}, \bibinfo {author}
  {\bibfnamefont {R.~V.}\ \bibnamefont {Gorbachev}}, \bibinfo {author}
  {\bibfnamefont {H.~S.}\ \bibnamefont {Shin}}, \bibinfo {author}
  {\bibfnamefont {V.~I.}\ \bibnamefont {Fal'ko}}, \ and\ \bibinfo {author}
  {\bibfnamefont {A.~I.}\ \bibnamefont {Tartakovskii}},\ }\href {\doibase
  10.1038/s41586-019-0986-9} {\bibfield  {journal} {\bibinfo  {journal}
  {Nature}\ }\textbf {\bibinfo {volume} {567}},\ \bibinfo {pages} {81}
  (\bibinfo {year} {2019})}\BibitemShut {NoStop}%
\bibitem [{\citenamefont {Ruiz-Tijerina}\ and\ \citenamefont
  {Fal'ko}(2019)}]{Ruiz-Tijerina2019}%
  \BibitemOpen
  \bibfield  {author} {\bibinfo {author} {\bibfnamefont {D.~A.}\ \bibnamefont
  {Ruiz-Tijerina}}\ and\ \bibinfo {author} {\bibfnamefont {V.~I.}\ \bibnamefont
  {Fal'ko}},\ }\href {\doibase 10.1103/PhysRevB.99.125424} {\bibfield
  {journal} {\bibinfo  {journal} {Phys. Rev. B}\ }\textbf {\bibinfo {volume}
  {99}},\ \bibinfo {pages} {125424} (\bibinfo {year} {2019})}\BibitemShut
  {NoStop}%
\bibitem [{\citenamefont {Yu}\ \emph {et~al.}(2017)\citenamefont {Yu},
  \citenamefont {Liu}, \citenamefont {Tang}, \citenamefont {Xu},\ and\
  \citenamefont {Yao}}]{Yu2017}%
  \BibitemOpen
  \bibfield  {author} {\bibinfo {author} {\bibfnamefont {H.}~\bibnamefont
  {Yu}}, \bibinfo {author} {\bibfnamefont {G.-B.}\ \bibnamefont {Liu}},
  \bibinfo {author} {\bibfnamefont {J.}~\bibnamefont {Tang}}, \bibinfo {author}
  {\bibfnamefont {X.}~\bibnamefont {Xu}}, \ and\ \bibinfo {author}
  {\bibfnamefont {W.}~\bibnamefont {Yao}},\ }\href {\doibase
  10.1126/sciadv.1701696} {\bibfield  {journal} {\bibinfo  {journal} {Science
  Advances}\ }\textbf {\bibinfo {volume} {3}},\ \bibinfo {pages} {e1701696}
  (\bibinfo {year} {2017})},\ \Eprint
  {http://arxiv.org/abs/https://www.science.org/doi/pdf/10.1126/sciadv.1701696}
  {https://www.science.org/doi/pdf/10.1126/sciadv.1701696} \BibitemShut
  {NoStop}%
\bibitem [{\citenamefont {Seyler}\ \emph {et~al.}(2019)\citenamefont {Seyler},
  \citenamefont {Rivera}, \citenamefont {Yu}, \citenamefont {Wilson},
  \citenamefont {Ray}, \citenamefont {Mandrus}, \citenamefont {Yan},
  \citenamefont {Yao},\ and\ \citenamefont {Xu}}]{Seyler2019}%
  \BibitemOpen
  \bibfield  {author} {\bibinfo {author} {\bibfnamefont {K.~L.}\ \bibnamefont
  {Seyler}}, \bibinfo {author} {\bibfnamefont {P.}~\bibnamefont {Rivera}},
  \bibinfo {author} {\bibfnamefont {H.}~\bibnamefont {Yu}}, \bibinfo {author}
  {\bibfnamefont {N.~P.}\ \bibnamefont {Wilson}}, \bibinfo {author}
  {\bibfnamefont {E.~L.}\ \bibnamefont {Ray}}, \bibinfo {author} {\bibfnamefont
  {D.~G.}\ \bibnamefont {Mandrus}}, \bibinfo {author} {\bibfnamefont
  {J.}~\bibnamefont {Yan}}, \bibinfo {author} {\bibfnamefont {W.}~\bibnamefont
  {Yao}}, \ and\ \bibinfo {author} {\bibfnamefont {X.}~\bibnamefont {Xu}},\
  }\href {\doibase 10.1038/s41586-019-0957-1} {\bibfield  {journal} {\bibinfo
  {journal} {Nature}\ }\textbf {\bibinfo {volume} {567}},\ \bibinfo {pages}
  {66} (\bibinfo {year} {2019})}\BibitemShut {NoStop}%
\bibitem [{\citenamefont {Jin}\ \emph {et~al.}(2019)\citenamefont {Jin},
  \citenamefont {Regan}, \citenamefont {Yan}, \citenamefont {Iqbal
  Bakti~Utama}, \citenamefont {Wang}, \citenamefont {Zhao}, \citenamefont
  {Qin}, \citenamefont {Yang}, \citenamefont {Zheng}, \citenamefont {Shi},
  \citenamefont {Watanabe}, \citenamefont {Taniguchi}, \citenamefont {Tongay},
  \citenamefont {Zettl},\ and\ \citenamefont {Wang}}]{Jin2019}%
  \BibitemOpen
  \bibfield  {author} {\bibinfo {author} {\bibfnamefont {C.}~\bibnamefont
  {Jin}}, \bibinfo {author} {\bibfnamefont {E.~C.}\ \bibnamefont {Regan}},
  \bibinfo {author} {\bibfnamefont {A.}~\bibnamefont {Yan}}, \bibinfo {author}
  {\bibfnamefont {M.}~\bibnamefont {Iqbal Bakti~Utama}}, \bibinfo {author}
  {\bibfnamefont {D.}~\bibnamefont {Wang}}, \bibinfo {author} {\bibfnamefont
  {S.}~\bibnamefont {Zhao}}, \bibinfo {author} {\bibfnamefont {Y.}~\bibnamefont
  {Qin}}, \bibinfo {author} {\bibfnamefont {S.}~\bibnamefont {Yang}}, \bibinfo
  {author} {\bibfnamefont {Z.}~\bibnamefont {Zheng}}, \bibinfo {author}
  {\bibfnamefont {S.}~\bibnamefont {Shi}}, \bibinfo {author} {\bibfnamefont
  {K.}~\bibnamefont {Watanabe}}, \bibinfo {author} {\bibfnamefont
  {T.}~\bibnamefont {Taniguchi}}, \bibinfo {author} {\bibfnamefont
  {S.}~\bibnamefont {Tongay}}, \bibinfo {author} {\bibfnamefont
  {A.}~\bibnamefont {Zettl}}, \ and\ \bibinfo {author} {\bibfnamefont
  {F.}~\bibnamefont {Wang}},\ }\href {\doibase 10.1038/s41586-019-0976-y}
  {\bibfield  {journal} {\bibinfo  {journal} {Nature}\ }\textbf {\bibinfo
  {volume} {567}},\ \bibinfo {pages} {76} (\bibinfo {year} {2019})}\BibitemShut
  {NoStop}%
\bibitem [{\citenamefont {Jiang}\ \emph {et~al.}(2021)\citenamefont {Jiang},
  \citenamefont {Chen}, \citenamefont {Zheng}, \citenamefont {Zheng},\ and\
  \citenamefont {Pan}}]{Jiang2021}%
  \BibitemOpen
  \bibfield  {author} {\bibinfo {author} {\bibfnamefont {Y.}~\bibnamefont
  {Jiang}}, \bibinfo {author} {\bibfnamefont {S.}~\bibnamefont {Chen}},
  \bibinfo {author} {\bibfnamefont {W.}~\bibnamefont {Zheng}}, \bibinfo
  {author} {\bibfnamefont {B.}~\bibnamefont {Zheng}}, \ and\ \bibinfo {author}
  {\bibfnamefont {A.}~\bibnamefont {Pan}},\ }\href {\doibase
  10.1038/s41377-021-00500-1} {\bibfield  {journal} {\bibinfo  {journal}
  {Light: Science {\&} Applications}\ }\textbf {\bibinfo {volume} {10}},\
  \bibinfo {pages} {72} (\bibinfo {year} {2021})}\BibitemShut {NoStop}%
\bibitem [{\citenamefont {Montblanch}\ \emph {et~al.}(2021)\citenamefont
  {Montblanch}, \citenamefont {Kara}, \citenamefont {Paradisanos},
  \citenamefont {Purser}, \citenamefont {Feuer}, \citenamefont {Alexeev},
  \citenamefont {Stefan}, \citenamefont {Qin}, \citenamefont {Blei},
  \citenamefont {Wang}, \citenamefont {Cadore}, \citenamefont {Latawiec},
  \citenamefont {Lon{\v{c}}ar}, \citenamefont {Tongay}, \citenamefont
  {Ferrari},\ and\ \citenamefont {Atat{\"u}re}}]{Montblanch2021}%
  \BibitemOpen
  \bibfield  {author} {\bibinfo {author} {\bibfnamefont {A.~R.-P.}\
  \bibnamefont {Montblanch}}, \bibinfo {author} {\bibfnamefont {D.~M.}\
  \bibnamefont {Kara}}, \bibinfo {author} {\bibfnamefont {I.}~\bibnamefont
  {Paradisanos}}, \bibinfo {author} {\bibfnamefont {C.~M.}\ \bibnamefont
  {Purser}}, \bibinfo {author} {\bibfnamefont {M.~S.~G.}\ \bibnamefont
  {Feuer}}, \bibinfo {author} {\bibfnamefont {E.~M.}\ \bibnamefont {Alexeev}},
  \bibinfo {author} {\bibfnamefont {L.}~\bibnamefont {Stefan}}, \bibinfo
  {author} {\bibfnamefont {Y.}~\bibnamefont {Qin}}, \bibinfo {author}
  {\bibfnamefont {M.}~\bibnamefont {Blei}}, \bibinfo {author} {\bibfnamefont
  {G.}~\bibnamefont {Wang}}, \bibinfo {author} {\bibfnamefont {A.~R.}\
  \bibnamefont {Cadore}}, \bibinfo {author} {\bibfnamefont {P.}~\bibnamefont
  {Latawiec}}, \bibinfo {author} {\bibfnamefont {M.}~\bibnamefont
  {Lon{\v{c}}ar}}, \bibinfo {author} {\bibfnamefont {S.}~\bibnamefont
  {Tongay}}, \bibinfo {author} {\bibfnamefont {A.~C.}\ \bibnamefont {Ferrari}},
  \ and\ \bibinfo {author} {\bibfnamefont {M.}~\bibnamefont {Atat{\"u}re}},\
  }\href {\doibase 10.1038/s42005-021-00625-0} {\bibfield  {journal} {\bibinfo
  {journal} {Communications Physics}\ }\textbf {\bibinfo {volume} {4}},\
  \bibinfo {pages} {119} (\bibinfo {year} {2021})}\BibitemShut {NoStop}%
\bibitem [{\citenamefont {Huang}\ \emph {et~al.}(2022)\citenamefont {Huang},
  \citenamefont {Choi}, \citenamefont {Shih},\ and\ \citenamefont
  {Li}}]{Huang2022}%
  \BibitemOpen
  \bibfield  {author} {\bibinfo {author} {\bibfnamefont {D.}~\bibnamefont
  {Huang}}, \bibinfo {author} {\bibfnamefont {J.}~\bibnamefont {Choi}},
  \bibinfo {author} {\bibfnamefont {C.-K.}\ \bibnamefont {Shih}}, \ and\
  \bibinfo {author} {\bibfnamefont {X.}~\bibnamefont {Li}},\ }\href {\doibase
  10.1038/s41565-021-01068-y} {\bibfield  {journal} {\bibinfo  {journal}
  {Nature Nanotechnology}\ }\textbf {\bibinfo {volume} {17}},\ \bibinfo {pages}
  {227} (\bibinfo {year} {2022})}\BibitemShut {NoStop}%
\bibitem [{\citenamefont {Wang}\ \emph {et~al.}(2022)\citenamefont {Wang},
  \citenamefont {Xiao}, \citenamefont {Park}, \citenamefont {Zhu},
  \citenamefont {Wang}, \citenamefont {Taniguchi}, \citenamefont {Watanabe},
  \citenamefont {Yan}, \citenamefont {Xiao}, \citenamefont {Gamelin},
  \citenamefont {Yao},\ and\ \citenamefont {Xu}}]{Wang2022}%
  \BibitemOpen
  \bibfield  {author} {\bibinfo {author} {\bibfnamefont {X.}~\bibnamefont
  {Wang}}, \bibinfo {author} {\bibfnamefont {C.}~\bibnamefont {Xiao}}, \bibinfo
  {author} {\bibfnamefont {H.}~\bibnamefont {Park}}, \bibinfo {author}
  {\bibfnamefont {J.}~\bibnamefont {Zhu}}, \bibinfo {author} {\bibfnamefont
  {C.}~\bibnamefont {Wang}}, \bibinfo {author} {\bibfnamefont {T.}~\bibnamefont
  {Taniguchi}}, \bibinfo {author} {\bibfnamefont {K.}~\bibnamefont {Watanabe}},
  \bibinfo {author} {\bibfnamefont {J.}~\bibnamefont {Yan}}, \bibinfo {author}
  {\bibfnamefont {D.}~\bibnamefont {Xiao}}, \bibinfo {author} {\bibfnamefont
  {D.~R.}\ \bibnamefont {Gamelin}}, \bibinfo {author} {\bibfnamefont
  {W.}~\bibnamefont {Yao}}, \ and\ \bibinfo {author} {\bibfnamefont
  {X.}~\bibnamefont {Xu}},\ }\href {\doibase 10.1038/s41586-022-04472-z}
  {\bibfield  {journal} {\bibinfo  {journal} {Nature}\ }\textbf {\bibinfo
  {volume} {604}},\ \bibinfo {pages} {468} (\bibinfo {year}
  {2022})}\BibitemShut {NoStop}%
\bibitem [{\citenamefont {Liu}\ \emph {et~al.}(2021{\natexlab{b}})\citenamefont
  {Liu}, \citenamefont {Barr{\'e}}, \citenamefont {van Baren}, \citenamefont
  {Wilson}, \citenamefont {Taniguchi}, \citenamefont {Watanabe}, \citenamefont
  {Cui}, \citenamefont {Gabor}, \citenamefont {Heinz}, \citenamefont {Chang},\
  and\ \citenamefont {Lui}}]{Liu2021b}%
  \BibitemOpen
  \bibfield  {author} {\bibinfo {author} {\bibfnamefont {E.}~\bibnamefont
  {Liu}}, \bibinfo {author} {\bibfnamefont {E.}~\bibnamefont {Barr{\'e}}},
  \bibinfo {author} {\bibfnamefont {J.}~\bibnamefont {van Baren}}, \bibinfo
  {author} {\bibfnamefont {M.}~\bibnamefont {Wilson}}, \bibinfo {author}
  {\bibfnamefont {T.}~\bibnamefont {Taniguchi}}, \bibinfo {author}
  {\bibfnamefont {K.}~\bibnamefont {Watanabe}}, \bibinfo {author}
  {\bibfnamefont {Y.-T.}\ \bibnamefont {Cui}}, \bibinfo {author} {\bibfnamefont
  {N.~M.}\ \bibnamefont {Gabor}}, \bibinfo {author} {\bibfnamefont {T.~F.}\
  \bibnamefont {Heinz}}, \bibinfo {author} {\bibfnamefont {Y.-C.}\ \bibnamefont
  {Chang}}, \ and\ \bibinfo {author} {\bibfnamefont {C.~H.}\ \bibnamefont
  {Lui}},\ }\href {\doibase 10.1038/s41586-021-03541-z} {\bibfield  {journal}
  {\bibinfo  {journal} {Nature}\ }\textbf {\bibinfo {volume} {594}},\ \bibinfo
  {pages} {46} (\bibinfo {year} {2021}{\natexlab{b}})}\BibitemShut {NoStop}%
\bibitem [{\citenamefont {G\"otting}\ \emph {et~al.}(2022)\citenamefont
  {G\"otting}, \citenamefont {Lohof},\ and\ \citenamefont
  {Gies}}]{Gotting2022}%
  \BibitemOpen
  \bibfield  {author} {\bibinfo {author} {\bibfnamefont {N.}~\bibnamefont
  {G\"otting}}, \bibinfo {author} {\bibfnamefont {F.}~\bibnamefont {Lohof}}, \
  and\ \bibinfo {author} {\bibfnamefont {C.}~\bibnamefont {Gies}},\ }\href
  {\doibase 10.1103/PhysRevB.105.165419} {\bibfield  {journal} {\bibinfo
  {journal} {Phys. Rev. B}\ }\textbf {\bibinfo {volume} {105}},\ \bibinfo
  {pages} {165419} (\bibinfo {year} {2022})}\BibitemShut {NoStop}%
\bibitem [{\citenamefont {Lagoin}\ and\ \citenamefont
  {Dubin}(2021)}]{Lagoin2021}%
  \BibitemOpen
  \bibfield  {author} {\bibinfo {author} {\bibfnamefont {C.}~\bibnamefont
  {Lagoin}}\ and\ \bibinfo {author} {\bibfnamefont {F.~m.~c.}\ \bibnamefont
  {Dubin}},\ }\href {\doibase 10.1103/PhysRevB.103.L041406} {\bibfield
  {journal} {\bibinfo  {journal} {Phys. Rev. B}\ }\textbf {\bibinfo {volume}
  {103}},\ \bibinfo {pages} {L041406} (\bibinfo {year} {2021})}\BibitemShut
  {NoStop}%
\bibitem [{\citenamefont {Wu}\ \emph {et~al.}(2018{\natexlab{a}})\citenamefont
  {Wu}, \citenamefont {Lovorn},\ and\ \citenamefont {MacDonald}}]{Wu2018a}%
  \BibitemOpen
  \bibfield  {author} {\bibinfo {author} {\bibfnamefont {F.}~\bibnamefont
  {Wu}}, \bibinfo {author} {\bibfnamefont {T.}~\bibnamefont {Lovorn}}, \ and\
  \bibinfo {author} {\bibfnamefont {A.~H.}\ \bibnamefont {MacDonald}},\ }\href
  {\doibase 10.1103/PhysRevB.97.035306} {\bibfield  {journal} {\bibinfo
  {journal} {Phys. Rev. B}\ }\textbf {\bibinfo {volume} {97}},\ \bibinfo
  {pages} {035306} (\bibinfo {year} {2018}{\natexlab{a}})}\BibitemShut
  {NoStop}%
\bibitem [{\citenamefont {L\"uhmann}(2013)}]{Luhmann2013}%
  \BibitemOpen
  \bibfield  {author} {\bibinfo {author} {\bibfnamefont {D.-S.}\ \bibnamefont
  {L\"uhmann}},\ }\href {\doibase 10.1103/PhysRevA.87.043619} {\bibfield
  {journal} {\bibinfo  {journal} {Phys. Rev. A}\ }\textbf {\bibinfo {volume}
  {87}},\ \bibinfo {pages} {043619} (\bibinfo {year} {2013})}\BibitemShut
  {NoStop}%
\bibitem [{\citenamefont {Yamamoto}\ \emph
  {et~al.}(2012{\natexlab{a}})\citenamefont {Yamamoto}, \citenamefont
  {Danshita},\ and\ \citenamefont {S\'a~de Melo}}]{Yamamoto2012}%
  \BibitemOpen
  \bibfield  {author} {\bibinfo {author} {\bibfnamefont {D.}~\bibnamefont
  {Yamamoto}}, \bibinfo {author} {\bibfnamefont {I.}~\bibnamefont {Danshita}},
  \ and\ \bibinfo {author} {\bibfnamefont {C.~A.~R.}\ \bibnamefont {S\'a~de
  Melo}},\ }\href {\doibase 10.1103/PhysRevA.85.021601} {\bibfield  {journal}
  {\bibinfo  {journal} {Phys. Rev. A}\ }\textbf {\bibinfo {volume} {85}},\
  \bibinfo {pages} {021601} (\bibinfo {year} {2012}{\natexlab{a}})}\BibitemShut
  {NoStop}%
\bibitem [{\citenamefont {Yamamoto}\ \emph
  {et~al.}(2012{\natexlab{b}})\citenamefont {Yamamoto}, \citenamefont
  {Masaki},\ and\ \citenamefont {Danshita}}]{Yamamoto2012b}%
  \BibitemOpen
  \bibfield  {author} {\bibinfo {author} {\bibfnamefont {D.}~\bibnamefont
  {Yamamoto}}, \bibinfo {author} {\bibfnamefont {A.}~\bibnamefont {Masaki}}, \
  and\ \bibinfo {author} {\bibfnamefont {I.}~\bibnamefont {Danshita}},\ }\href
  {\doibase 10.1103/PhysRevB.86.054516} {\bibfield  {journal} {\bibinfo
  {journal} {Phys. Rev. B}\ }\textbf {\bibinfo {volume} {86}},\ \bibinfo
  {pages} {054516} (\bibinfo {year} {2012}{\natexlab{b}})}\BibitemShut
  {NoStop}%
\bibitem [{\citenamefont {Malakar}\ \emph {et~al.}(2020)\citenamefont
  {Malakar}, \citenamefont {Ray}, \citenamefont {Sinha},\ and\ \citenamefont
  {Angom}}]{Malakar2020}%
  \BibitemOpen
  \bibfield  {author} {\bibinfo {author} {\bibfnamefont {M.}~\bibnamefont
  {Malakar}}, \bibinfo {author} {\bibfnamefont {S.}~\bibnamefont {Ray}},
  \bibinfo {author} {\bibfnamefont {S.}~\bibnamefont {Sinha}}, \ and\ \bibinfo
  {author} {\bibfnamefont {D.}~\bibnamefont {Angom}},\ }\href {\doibase
  10.1103/PhysRevB.102.184515} {\bibfield  {journal} {\bibinfo  {journal}
  {Phys. Rev. B}\ }\textbf {\bibinfo {volume} {102}},\ \bibinfo {pages}
  {184515} (\bibinfo {year} {2020})}\BibitemShut {NoStop}%
\bibitem [{\citenamefont {Hassan}\ \emph {et~al.}(2007)\citenamefont {Hassan},
  \citenamefont {de~Medici},\ and\ \citenamefont {Tremblay}}]{Hassan2007}%
  \BibitemOpen
  \bibfield  {author} {\bibinfo {author} {\bibfnamefont {S.~R.}\ \bibnamefont
  {Hassan}}, \bibinfo {author} {\bibfnamefont {L.}~\bibnamefont {de~Medici}}, \
  and\ \bibinfo {author} {\bibfnamefont {A.-M.~S.}\ \bibnamefont {Tremblay}},\
  }\href {\doibase 10.1103/PhysRevB.76.144420} {\bibfield  {journal} {\bibinfo
  {journal} {Phys. Rev. B}\ }\textbf {\bibinfo {volume} {76}},\ \bibinfo
  {pages} {144420} (\bibinfo {year} {2007})}\BibitemShut {NoStop}%
\bibitem [{\citenamefont {Wang}\ \emph {et~al.}(2018)\citenamefont {Wang},
  \citenamefont {Chernikov}, \citenamefont {Glazov}, \citenamefont {Heinz},
  \citenamefont {Marie}, \citenamefont {Amand},\ and\ \citenamefont
  {Urbaszek}}]{Wang2018}%
  \BibitemOpen
  \bibfield  {author} {\bibinfo {author} {\bibfnamefont {G.}~\bibnamefont
  {Wang}}, \bibinfo {author} {\bibfnamefont {A.}~\bibnamefont {Chernikov}},
  \bibinfo {author} {\bibfnamefont {M.~M.}\ \bibnamefont {Glazov}}, \bibinfo
  {author} {\bibfnamefont {T.~F.}\ \bibnamefont {Heinz}}, \bibinfo {author}
  {\bibfnamefont {X.}~\bibnamefont {Marie}}, \bibinfo {author} {\bibfnamefont
  {T.}~\bibnamefont {Amand}}, \ and\ \bibinfo {author} {\bibfnamefont
  {B.}~\bibnamefont {Urbaszek}},\ }\href {\doibase
  10.1103/RevModPhys.90.021001} {\bibfield  {journal} {\bibinfo  {journal}
  {Rev. Mod. Phys.}\ }\textbf {\bibinfo {volume} {90}},\ \bibinfo {pages}
  {021001} (\bibinfo {year} {2018})}\BibitemShut {NoStop}%
\bibitem [{\citenamefont {Mueller}\ and\ \citenamefont
  {Malic}(2018)}]{Mueller2018}%
  \BibitemOpen
  \bibfield  {author} {\bibinfo {author} {\bibfnamefont {T.}~\bibnamefont
  {Mueller}}\ and\ \bibinfo {author} {\bibfnamefont {E.}~\bibnamefont
  {Malic}},\ }\href {\doibase 10.1038/s41699-018-0074-2} {\bibfield  {journal}
  {\bibinfo  {journal} {npj 2D Materials and Applications}\ }\textbf {\bibinfo
  {volume} {2}},\ \bibinfo {pages} {29} (\bibinfo {year} {2018})}\BibitemShut
  {NoStop}%
\bibitem [{\citenamefont {Yu}\ \emph {et~al.}(2015{\natexlab{a}})\citenamefont
  {Yu}, \citenamefont {Cui}, \citenamefont {Xu},\ and\ \citenamefont
  {Yao}}]{Yu2015}%
  \BibitemOpen
  \bibfield  {author} {\bibinfo {author} {\bibfnamefont {H.}~\bibnamefont
  {Yu}}, \bibinfo {author} {\bibfnamefont {X.}~\bibnamefont {Cui}}, \bibinfo
  {author} {\bibfnamefont {X.}~\bibnamefont {Xu}}, \ and\ \bibinfo {author}
  {\bibfnamefont {W.}~\bibnamefont {Yao}},\ }\href {\doibase
  10.1093/nsr/nwu078} {\bibfield  {journal} {\bibinfo  {journal} {National
  Science Review}\ }\textbf {\bibinfo {volume} {2}},\ \bibinfo {pages} {57}
  (\bibinfo {year} {2015}{\natexlab{a}})}\BibitemShut {NoStop}%
\bibitem [{\citenamefont {Wang}\ \emph {et~al.}(2017)\citenamefont {Wang},
  \citenamefont {Wang}, \citenamefont {Yao}, \citenamefont {Liu},\ and\
  \citenamefont {Yu}}]{Wang2017}%
  \BibitemOpen
  \bibfield  {author} {\bibinfo {author} {\bibfnamefont {Y.}~\bibnamefont
  {Wang}}, \bibinfo {author} {\bibfnamefont {Z.}~\bibnamefont {Wang}}, \bibinfo
  {author} {\bibfnamefont {W.}~\bibnamefont {Yao}}, \bibinfo {author}
  {\bibfnamefont {G.-B.}\ \bibnamefont {Liu}}, \ and\ \bibinfo {author}
  {\bibfnamefont {H.}~\bibnamefont {Yu}},\ }\href {\doibase
  10.1103/PhysRevB.95.115429} {\bibfield  {journal} {\bibinfo  {journal} {Phys.
  Rev. B}\ }\textbf {\bibinfo {volume} {95}},\ \bibinfo {pages} {115429}
  (\bibinfo {year} {2017})}\BibitemShut {NoStop}%
\bibitem [{\citenamefont {Xiao}\ \emph {et~al.}(2012)\citenamefont {Xiao},
  \citenamefont {Liu}, \citenamefont {Feng}, \citenamefont {Xu},\ and\
  \citenamefont {Yao}}]{Xiao2012}%
  \BibitemOpen
  \bibfield  {author} {\bibinfo {author} {\bibfnamefont {D.}~\bibnamefont
  {Xiao}}, \bibinfo {author} {\bibfnamefont {G.-B.}\ \bibnamefont {Liu}},
  \bibinfo {author} {\bibfnamefont {W.}~\bibnamefont {Feng}}, \bibinfo {author}
  {\bibfnamefont {X.}~\bibnamefont {Xu}}, \ and\ \bibinfo {author}
  {\bibfnamefont {W.}~\bibnamefont {Yao}},\ }\href {\doibase
  10.1103/PhysRevLett.108.196802} {\bibfield  {journal} {\bibinfo  {journal}
  {Phys. Rev. Lett.}\ }\textbf {\bibinfo {volume} {108}},\ \bibinfo {pages}
  {196802} (\bibinfo {year} {2012})}\BibitemShut {NoStop}%
\bibitem [{\citenamefont {Cao}\ \emph {et~al.}(2012)\citenamefont {Cao},
  \citenamefont {Wang}, \citenamefont {Han}, \citenamefont {Ye}, \citenamefont
  {Zhu}, \citenamefont {Shi}, \citenamefont {Niu}, \citenamefont {Tan},
  \citenamefont {Wang}, \citenamefont {Liu},\ and\ \citenamefont
  {Feng}}]{Cao2012}%
  \BibitemOpen
  \bibfield  {author} {\bibinfo {author} {\bibfnamefont {T.}~\bibnamefont
  {Cao}}, \bibinfo {author} {\bibfnamefont {G.}~\bibnamefont {Wang}}, \bibinfo
  {author} {\bibfnamefont {W.}~\bibnamefont {Han}}, \bibinfo {author}
  {\bibfnamefont {H.}~\bibnamefont {Ye}}, \bibinfo {author} {\bibfnamefont
  {C.}~\bibnamefont {Zhu}}, \bibinfo {author} {\bibfnamefont {J.}~\bibnamefont
  {Shi}}, \bibinfo {author} {\bibfnamefont {Q.}~\bibnamefont {Niu}}, \bibinfo
  {author} {\bibfnamefont {P.}~\bibnamefont {Tan}}, \bibinfo {author}
  {\bibfnamefont {E.}~\bibnamefont {Wang}}, \bibinfo {author} {\bibfnamefont
  {B.}~\bibnamefont {Liu}}, \ and\ \bibinfo {author} {\bibfnamefont
  {J.}~\bibnamefont {Feng}},\ }\href {https://doi.org/10.1038/ncomms1882}
  {\bibfield  {journal} {\bibinfo  {journal} {Nature Communications}\ }\textbf
  {\bibinfo {volume} {3}},\ \bibinfo {pages} {887 EP } (\bibinfo {year}
  {2012})},\ \bibinfo {note} {article}\BibitemShut {NoStop}%
\bibitem [{\citenamefont {Zeng}\ \emph {et~al.}(2012)\citenamefont {Zeng},
  \citenamefont {Dai}, \citenamefont {Yao}, \citenamefont {Xiao},\ and\
  \citenamefont {Cui}}]{Zeng2012}%
  \BibitemOpen
  \bibfield  {author} {\bibinfo {author} {\bibfnamefont {H.}~\bibnamefont
  {Zeng}}, \bibinfo {author} {\bibfnamefont {J.}~\bibnamefont {Dai}}, \bibinfo
  {author} {\bibfnamefont {W.}~\bibnamefont {Yao}}, \bibinfo {author}
  {\bibfnamefont {D.}~\bibnamefont {Xiao}}, \ and\ \bibinfo {author}
  {\bibfnamefont {X.}~\bibnamefont {Cui}},\ }\href {\doibase
  10.1038/nnano.2012.95} {\bibfield  {journal} {\bibinfo  {journal} {Nature
  Nanotechnology}\ }\textbf {\bibinfo {volume} {7}},\ \bibinfo {pages} {490}
  (\bibinfo {year} {2012})}\BibitemShut {NoStop}%
\bibitem [{\citenamefont {Mak}\ \emph {et~al.}(2012)\citenamefont {Mak},
  \citenamefont {He}, \citenamefont {Shan},\ and\ \citenamefont
  {Heinz}}]{Mak2012b}%
  \BibitemOpen
  \bibfield  {author} {\bibinfo {author} {\bibfnamefont {K.~F.}\ \bibnamefont
  {Mak}}, \bibinfo {author} {\bibfnamefont {K.}~\bibnamefont {He}}, \bibinfo
  {author} {\bibfnamefont {J.}~\bibnamefont {Shan}}, \ and\ \bibinfo {author}
  {\bibfnamefont {T.~F.}\ \bibnamefont {Heinz}},\ }\href {\doibase
  10.1038/nnano.2012.96} {\bibfield  {journal} {\bibinfo  {journal} {Nature
  Nanotechnology}\ }\textbf {\bibinfo {volume} {7}},\ \bibinfo {pages} {494}
  (\bibinfo {year} {2012})}\BibitemShut {NoStop}%
\bibitem [{\citenamefont {Schaibley}\ \emph {et~al.}(2016)\citenamefont
  {Schaibley}, \citenamefont {Yu}, \citenamefont {Clark}, \citenamefont
  {Rivera}, \citenamefont {Ross}, \citenamefont {Seyler}, \citenamefont {Yao},\
  and\ \citenamefont {Xu}}]{Schaibley2016}%
  \BibitemOpen
  \bibfield  {author} {\bibinfo {author} {\bibfnamefont {J.~R.}\ \bibnamefont
  {Schaibley}}, \bibinfo {author} {\bibfnamefont {H.}~\bibnamefont {Yu}},
  \bibinfo {author} {\bibfnamefont {G.}~\bibnamefont {Clark}}, \bibinfo
  {author} {\bibfnamefont {P.}~\bibnamefont {Rivera}}, \bibinfo {author}
  {\bibfnamefont {J.~S.}\ \bibnamefont {Ross}}, \bibinfo {author}
  {\bibfnamefont {K.~L.}\ \bibnamefont {Seyler}}, \bibinfo {author}
  {\bibfnamefont {W.}~\bibnamefont {Yao}}, \ and\ \bibinfo {author}
  {\bibfnamefont {X.}~\bibnamefont {Xu}},\ }\href
  {https://doi.org/10.1038/natrevmats.2016.55} {\bibfield  {journal} {\bibinfo
  {journal} {Nature Reviews Materials}\ }\textbf {\bibinfo {volume} {1}},\
  \bibinfo {pages} {16055 EP } (\bibinfo {year} {2016})},\ \bibinfo {note}
  {review Article}\BibitemShut {NoStop}%
\bibitem [{\citenamefont {Zhang}\ \emph {et~al.}(2019)\citenamefont {Zhang},
  \citenamefont {Gogna}, \citenamefont {Burg}, \citenamefont {Horng},
  \citenamefont {Paik}, \citenamefont {Chou}, \citenamefont {Kim},
  \citenamefont {Tutuc},\ and\ \citenamefont {Deng}}]{Zhang2019}%
  \BibitemOpen
  \bibfield  {author} {\bibinfo {author} {\bibfnamefont {L.}~\bibnamefont
  {Zhang}}, \bibinfo {author} {\bibfnamefont {R.}~\bibnamefont {Gogna}},
  \bibinfo {author} {\bibfnamefont {G.~W.}\ \bibnamefont {Burg}}, \bibinfo
  {author} {\bibfnamefont {J.}~\bibnamefont {Horng}}, \bibinfo {author}
  {\bibfnamefont {E.}~\bibnamefont {Paik}}, \bibinfo {author} {\bibfnamefont
  {Y.-H.}\ \bibnamefont {Chou}}, \bibinfo {author} {\bibfnamefont
  {K.}~\bibnamefont {Kim}}, \bibinfo {author} {\bibfnamefont {E.}~\bibnamefont
  {Tutuc}}, \ and\ \bibinfo {author} {\bibfnamefont {H.}~\bibnamefont {Deng}},\
  }\href {\doibase 10.1103/PhysRevB.100.041402} {\bibfield  {journal} {\bibinfo
   {journal} {Phys. Rev. B}\ }\textbf {\bibinfo {volume} {100}},\ \bibinfo
  {pages} {041402} (\bibinfo {year} {2019})}\BibitemShut {NoStop}%
\bibitem [{SM()}]{SM}%
  \BibitemOpen
  \href@noop {} {}\bibinfo {note} {See {\it Supplemental Material} for details
  on derivation of $H_H$ and $H_E$ continuum models, details on constructing
  the tight-binding model, details on computing the interaction terms,
  derivation of the dielectric functions, details on CMF computations, the
  results of the weak-coupling Gross-Pitaevskii equation and Bogoliubov theory,
  and additional
  Refs.~\cite{Yu2015:2,Rivera2016,Leykam2018,Bistritzer2010,Angeli2021,castin:book,Moskalenko2000}}\BibitemShut
  {NoStop}%
\bibitem [{\citenamefont {Wu}\ \emph {et~al.}(2019)\citenamefont {Wu},
  \citenamefont {Lovorn}, \citenamefont {Tutuc}, \citenamefont {Martin},\ and\
  \citenamefont {MacDonald}}]{Wu2019a}%
  \BibitemOpen
  \bibfield  {author} {\bibinfo {author} {\bibfnamefont {F.}~\bibnamefont
  {Wu}}, \bibinfo {author} {\bibfnamefont {T.}~\bibnamefont {Lovorn}}, \bibinfo
  {author} {\bibfnamefont {E.}~\bibnamefont {Tutuc}}, \bibinfo {author}
  {\bibfnamefont {I.}~\bibnamefont {Martin}}, \ and\ \bibinfo {author}
  {\bibfnamefont {A.~H.}\ \bibnamefont {MacDonald}},\ }\href {\doibase
  10.1103/PhysRevLett.122.086402} {\bibfield  {journal} {\bibinfo  {journal}
  {Phys. Rev. Lett.}\ }\textbf {\bibinfo {volume} {122}},\ \bibinfo {pages}
  {086402} (\bibinfo {year} {2019})}\BibitemShut {NoStop}%
\bibitem [{\citenamefont {Wu}\ \emph {et~al.}(2018{\natexlab{b}})\citenamefont
  {Wu}, \citenamefont {Lovorn}, \citenamefont {Tutuc},\ and\ \citenamefont
  {MacDonald}}]{Wu2018}%
  \BibitemOpen
  \bibfield  {author} {\bibinfo {author} {\bibfnamefont {F.}~\bibnamefont
  {Wu}}, \bibinfo {author} {\bibfnamefont {T.}~\bibnamefont {Lovorn}}, \bibinfo
  {author} {\bibfnamefont {E.}~\bibnamefont {Tutuc}}, \ and\ \bibinfo {author}
  {\bibfnamefont {A.~H.}\ \bibnamefont {MacDonald}},\ }\href {\doibase
  10.1103/PhysRevLett.121.026402} {\bibfield  {journal} {\bibinfo  {journal}
  {Phys. Rev. Lett.}\ }\textbf {\bibinfo {volume} {121}},\ \bibinfo {pages}
  {026402} (\bibinfo {year} {2018}{\natexlab{b}})}\BibitemShut {NoStop}%
\bibitem [{\citenamefont {Pan}\ \emph {et~al.}(2020{\natexlab{a}})\citenamefont
  {Pan}, \citenamefont {Wu},\ and\ \citenamefont {Das~Sarma}}]{Pan2020}%
  \BibitemOpen
  \bibfield  {author} {\bibinfo {author} {\bibfnamefont {H.}~\bibnamefont
  {Pan}}, \bibinfo {author} {\bibfnamefont {F.}~\bibnamefont {Wu}}, \ and\
  \bibinfo {author} {\bibfnamefont {S.}~\bibnamefont {Das~Sarma}},\ }\href
  {\doibase 10.1103/PhysRevResearch.2.033087} {\bibfield  {journal} {\bibinfo
  {journal} {Phys. Rev. Research}\ }\textbf {\bibinfo {volume} {2}},\ \bibinfo
  {pages} {033087} (\bibinfo {year} {2020}{\natexlab{a}})}\BibitemShut
  {NoStop}%
\bibitem [{\citenamefont {Morales-Durán}\ \emph {et~al.}(2021)\citenamefont
  {Morales-Durán}, \citenamefont {Hu}, \citenamefont {Potasz},\ and\
  \citenamefont {MacDonald}}]{Morales-Duran2021}%
  \BibitemOpen
  \bibfield  {author} {\bibinfo {author} {\bibfnamefont {N.}~\bibnamefont
  {Morales-Durán}}, \bibinfo {author} {\bibfnamefont {N.~C.}\ \bibnamefont
  {Hu}}, \bibinfo {author} {\bibfnamefont {P.}~\bibnamefont {Potasz}}, \ and\
  \bibinfo {author} {\bibfnamefont {A.~H.}\ \bibnamefont {MacDonald}},\
  }\href@noop {} {\enquote {\bibinfo {title} {Non-local interactions in moir\'e
  hubbard systems},}\ } (\bibinfo {year} {2021}),\ \Eprint
  {http://arxiv.org/abs/2108.03313} {arXiv:2108.03313 [cond-mat.str-el]}
  \BibitemShut {NoStop}%
\bibitem [{\citenamefont {Wu}\ \emph {et~al.}(2017)\citenamefont {Wu},
  \citenamefont {Lovorn},\ and\ \citenamefont {MacDonald}}]{Wu2017}%
  \BibitemOpen
  \bibfield  {author} {\bibinfo {author} {\bibfnamefont {F.}~\bibnamefont
  {Wu}}, \bibinfo {author} {\bibfnamefont {T.}~\bibnamefont {Lovorn}}, \ and\
  \bibinfo {author} {\bibfnamefont {A.~H.}\ \bibnamefont {MacDonald}},\ }\href
  {\doibase 10.1103/PhysRevLett.118.147401} {\bibfield  {journal} {\bibinfo
  {journal} {Phys. Rev. Lett.}\ }\textbf {\bibinfo {volume} {118}},\ \bibinfo
  {pages} {147401} (\bibinfo {year} {2017})}\BibitemShut {NoStop}%
\bibitem [{\citenamefont {Naik}\ \emph {et~al.}(2022)\citenamefont {Naik},
  \citenamefont {Regan}, \citenamefont {Zhang}, \citenamefont {Chan},
  \citenamefont {Li}, \citenamefont {Wang}, \citenamefont {Yoon}, \citenamefont
  {Ong}, \citenamefont {Zhao}, \citenamefont {Zhao}, \citenamefont {Utama},
  \citenamefont {Gao}, \citenamefont {Wei}, \citenamefont {Sayyad},
  \citenamefont {Yumigeta}, \citenamefont {Watanabe}, \citenamefont
  {Taniguchi}, \citenamefont {Tongay}, \citenamefont {da~Jornada},
  \citenamefont {Wang},\ and\ \citenamefont {Louie}}]{Naik2022}%
  \BibitemOpen
  \bibfield  {author} {\bibinfo {author} {\bibfnamefont {M.~H.}\ \bibnamefont
  {Naik}}, \bibinfo {author} {\bibfnamefont {E.~C.}\ \bibnamefont {Regan}},
  \bibinfo {author} {\bibfnamefont {Z.}~\bibnamefont {Zhang}}, \bibinfo
  {author} {\bibfnamefont {Y.-h.}\ \bibnamefont {Chan}}, \bibinfo {author}
  {\bibfnamefont {Z.}~\bibnamefont {Li}}, \bibinfo {author} {\bibfnamefont
  {D.}~\bibnamefont {Wang}}, \bibinfo {author} {\bibfnamefont {Y.}~\bibnamefont
  {Yoon}}, \bibinfo {author} {\bibfnamefont {C.~S.}\ \bibnamefont {Ong}},
  \bibinfo {author} {\bibfnamefont {W.}~\bibnamefont {Zhao}}, \bibinfo {author}
  {\bibfnamefont {S.}~\bibnamefont {Zhao}}, \bibinfo {author} {\bibfnamefont
  {M.~I.~B.}\ \bibnamefont {Utama}}, \bibinfo {author} {\bibfnamefont
  {B.}~\bibnamefont {Gao}}, \bibinfo {author} {\bibfnamefont {X.}~\bibnamefont
  {Wei}}, \bibinfo {author} {\bibfnamefont {M.}~\bibnamefont {Sayyad}},
  \bibinfo {author} {\bibfnamefont {K.}~\bibnamefont {Yumigeta}}, \bibinfo
  {author} {\bibfnamefont {K.}~\bibnamefont {Watanabe}}, \bibinfo {author}
  {\bibfnamefont {T.}~\bibnamefont {Taniguchi}}, \bibinfo {author}
  {\bibfnamefont {S.}~\bibnamefont {Tongay}}, \bibinfo {author} {\bibfnamefont
  {F.~H.}\ \bibnamefont {da~Jornada}}, \bibinfo {author} {\bibfnamefont
  {F.}~\bibnamefont {Wang}}, \ and\ \bibinfo {author} {\bibfnamefont {S.~G.}\
  \bibnamefont {Louie}},\ }\href {\doibase 10.48550/ARXIV.2201.02562} {\enquote
  {\bibinfo {title} {Nature of novel moiré exciton states in wse$_2$/ws$_2$
  heterobilayers},}\ } (\bibinfo {year} {2022})\BibitemShut {NoStop}%
\bibitem [{\citenamefont {Pan}\ \emph {et~al.}(2020{\natexlab{b}})\citenamefont
  {Pan}, \citenamefont {Wu},\ and\ \citenamefont {Das~Sarma}}]{Pan2020b}%
  \BibitemOpen
  \bibfield  {author} {\bibinfo {author} {\bibfnamefont {H.}~\bibnamefont
  {Pan}}, \bibinfo {author} {\bibfnamefont {F.}~\bibnamefont {Wu}}, \ and\
  \bibinfo {author} {\bibfnamefont {S.}~\bibnamefont {Das~Sarma}},\ }\href
  {\doibase 10.1103/PhysRevB.102.201104} {\bibfield  {journal} {\bibinfo
  {journal} {Phys. Rev. B}\ }\textbf {\bibinfo {volume} {102}},\ \bibinfo
  {pages} {201104} (\bibinfo {year} {2020}{\natexlab{b}})}\BibitemShut
  {NoStop}%
\bibitem [{\citenamefont {Pan}\ and\ \citenamefont
  {Das~Sarma}(2022)}]{Pan2020c}%
  \BibitemOpen
  \bibfield  {author} {\bibinfo {author} {\bibfnamefont {H.}~\bibnamefont
  {Pan}}\ and\ \bibinfo {author} {\bibfnamefont {S.}~\bibnamefont
  {Das~Sarma}},\ }\href {\doibase 10.1103/PhysRevB.105.041109} {\bibfield
  {journal} {\bibinfo  {journal} {Phys. Rev. B}\ }\textbf {\bibinfo {volume}
  {105}},\ \bibinfo {pages} {041109} (\bibinfo {year} {2022})}\BibitemShut
  {NoStop}%
\bibitem [{\citenamefont {Pan}\ and\ \citenamefont
  {Das~Sarma}(2021)}]{Pan2021}%
  \BibitemOpen
  \bibfield  {author} {\bibinfo {author} {\bibfnamefont {H.}~\bibnamefont
  {Pan}}\ and\ \bibinfo {author} {\bibfnamefont {S.}~\bibnamefont
  {Das~Sarma}},\ }\href {\doibase 10.1103/PhysRevLett.127.096802} {\bibfield
  {journal} {\bibinfo  {journal} {Phys. Rev. Lett.}\ }\textbf {\bibinfo
  {volume} {127}},\ \bibinfo {pages} {096802} (\bibinfo {year}
  {2021})}\BibitemShut {NoStop}%
\bibitem [{\citenamefont {Danovich}\ \emph {et~al.}(2018)\citenamefont
  {Danovich}, \citenamefont {Ruiz-Tijerina}, \citenamefont {Hunt},
  \citenamefont {Szyniszewski}, \citenamefont {Drummond},\ and\ \citenamefont
  {Fal'ko}}]{Danovich2018}%
  \BibitemOpen
  \bibfield  {author} {\bibinfo {author} {\bibfnamefont {M.}~\bibnamefont
  {Danovich}}, \bibinfo {author} {\bibfnamefont {D.~A.}\ \bibnamefont
  {Ruiz-Tijerina}}, \bibinfo {author} {\bibfnamefont {R.~J.}\ \bibnamefont
  {Hunt}}, \bibinfo {author} {\bibfnamefont {M.}~\bibnamefont {Szyniszewski}},
  \bibinfo {author} {\bibfnamefont {N.~D.}\ \bibnamefont {Drummond}}, \ and\
  \bibinfo {author} {\bibfnamefont {V.~I.}\ \bibnamefont {Fal'ko}},\ }\href
  {\doibase 10.1103/PhysRevB.97.195452} {\bibfield  {journal} {\bibinfo
  {journal} {Phys. Rev. B}\ }\textbf {\bibinfo {volume} {97}},\ \bibinfo
  {pages} {195452} (\bibinfo {year} {2018})}\BibitemShut {NoStop}%
\bibitem [{\citenamefont {Wang}\ \emph {et~al.}(2019)\citenamefont {Wang},
  \citenamefont {Ardelean}, \citenamefont {Bai}, \citenamefont {Steinhoff},
  \citenamefont {Florian}, \citenamefont {Jahnke}, \citenamefont {Xu},
  \citenamefont {Kira}, \citenamefont {Hone},\ and\ \citenamefont
  {Zhu}}]{Wang2019b}%
  \BibitemOpen
  \bibfield  {author} {\bibinfo {author} {\bibfnamefont {J.}~\bibnamefont
  {Wang}}, \bibinfo {author} {\bibfnamefont {J.}~\bibnamefont {Ardelean}},
  \bibinfo {author} {\bibfnamefont {Y.}~\bibnamefont {Bai}}, \bibinfo {author}
  {\bibfnamefont {A.}~\bibnamefont {Steinhoff}}, \bibinfo {author}
  {\bibfnamefont {M.}~\bibnamefont {Florian}}, \bibinfo {author} {\bibfnamefont
  {F.}~\bibnamefont {Jahnke}}, \bibinfo {author} {\bibfnamefont
  {X.}~\bibnamefont {Xu}}, \bibinfo {author} {\bibfnamefont {M.}~\bibnamefont
  {Kira}}, \bibinfo {author} {\bibfnamefont {J.}~\bibnamefont {Hone}}, \ and\
  \bibinfo {author} {\bibfnamefont {X.-Y.}\ \bibnamefont {Zhu}},\ }\href
  {\doibase 10.1126/sciadv.aax0145} {\bibfield  {journal} {\bibinfo  {journal}
  {Science Advances}\ }\textbf {\bibinfo {volume} {5}},\ \bibinfo {pages}
  {eaax0145} (\bibinfo {year} {2019})},\ \Eprint
  {http://arxiv.org/abs/https://www.science.org/doi/pdf/10.1126/sciadv.aax0145}
  {https://www.science.org/doi/pdf/10.1126/sciadv.aax0145} \BibitemShut
  {NoStop}%
\bibitem [{\citenamefont {Wang}\ \emph
  {et~al.}(2021{\natexlab{a}})\citenamefont {Wang}, \citenamefont {Shi},
  \citenamefont {Shih}, \citenamefont {Zhou}, \citenamefont {Wu}, \citenamefont
  {Bai}, \citenamefont {Rhodes}, \citenamefont {Barmak}, \citenamefont {Hone},
  \citenamefont {Dean},\ and\ \citenamefont {Zhu}}]{Wang2021b}%
  \BibitemOpen
  \bibfield  {author} {\bibinfo {author} {\bibfnamefont {J.}~\bibnamefont
  {Wang}}, \bibinfo {author} {\bibfnamefont {Q.}~\bibnamefont {Shi}}, \bibinfo
  {author} {\bibfnamefont {E.-M.}\ \bibnamefont {Shih}}, \bibinfo {author}
  {\bibfnamefont {L.}~\bibnamefont {Zhou}}, \bibinfo {author} {\bibfnamefont
  {W.}~\bibnamefont {Wu}}, \bibinfo {author} {\bibfnamefont {Y.}~\bibnamefont
  {Bai}}, \bibinfo {author} {\bibfnamefont {D.}~\bibnamefont {Rhodes}},
  \bibinfo {author} {\bibfnamefont {K.}~\bibnamefont {Barmak}}, \bibinfo
  {author} {\bibfnamefont {J.}~\bibnamefont {Hone}}, \bibinfo {author}
  {\bibfnamefont {C.~R.}\ \bibnamefont {Dean}}, \ and\ \bibinfo {author}
  {\bibfnamefont {X.-Y.}\ \bibnamefont {Zhu}},\ }\href {\doibase
  10.1103/PhysRevLett.126.106804} {\bibfield  {journal} {\bibinfo  {journal}
  {Phys. Rev. Lett.}\ }\textbf {\bibinfo {volume} {126}},\ \bibinfo {pages}
  {106804} (\bibinfo {year} {2021}{\natexlab{a}})}\BibitemShut {NoStop}%
\bibitem [{\citenamefont {Wessel}\ and\ \citenamefont
  {Troyer}(2005)}]{Wessel2005}%
  \BibitemOpen
  \bibfield  {author} {\bibinfo {author} {\bibfnamefont {S.}~\bibnamefont
  {Wessel}}\ and\ \bibinfo {author} {\bibfnamefont {M.}~\bibnamefont
  {Troyer}},\ }\href {\doibase 10.1103/PhysRevLett.95.127205} {\bibfield
  {journal} {\bibinfo  {journal} {Phys. Rev. Lett.}\ }\textbf {\bibinfo
  {volume} {95}},\ \bibinfo {pages} {127205} (\bibinfo {year}
  {2005})}\BibitemShut {NoStop}%
\bibitem [{\citenamefont {Caleffi}\ \emph {et~al.}(2020)\citenamefont
  {Caleffi}, \citenamefont {Capone}, \citenamefont {Menotti}, \citenamefont
  {Carusotto},\ and\ \citenamefont {Recati}}]{Caleffi2020}%
  \BibitemOpen
  \bibfield  {author} {\bibinfo {author} {\bibfnamefont {F.}~\bibnamefont
  {Caleffi}}, \bibinfo {author} {\bibfnamefont {M.}~\bibnamefont {Capone}},
  \bibinfo {author} {\bibfnamefont {C.}~\bibnamefont {Menotti}}, \bibinfo
  {author} {\bibfnamefont {I.}~\bibnamefont {Carusotto}}, \ and\ \bibinfo
  {author} {\bibfnamefont {A.}~\bibnamefont {Recati}},\ }\href {\doibase
  10.1103/PhysRevResearch.2.033276} {\bibfield  {journal} {\bibinfo  {journal}
  {Phys. Rev. Research}\ }\textbf {\bibinfo {volume} {2}},\ \bibinfo {pages}
  {033276} (\bibinfo {year} {2020})}\BibitemShut {NoStop}%
\bibitem [{\citenamefont {Choi}\ \emph {et~al.}(2021)\citenamefont {Choi},
  \citenamefont {Florian}, \citenamefont {Steinhoff}, \citenamefont {Erben},
  \citenamefont {Tran}, \citenamefont {Kim}, \citenamefont {Sun}, \citenamefont
  {Quan}, \citenamefont {Claassen}, \citenamefont {Majumder}, \citenamefont
  {Hollingsworth}, \citenamefont {Taniguchi}, \citenamefont {Watanabe},
  \citenamefont {Ueno}, \citenamefont {Singh}, \citenamefont {Moody},
  \citenamefont {Jahnke},\ and\ \citenamefont {Li}}]{Choi2021}%
  \BibitemOpen
  \bibfield  {author} {\bibinfo {author} {\bibfnamefont {J.}~\bibnamefont
  {Choi}}, \bibinfo {author} {\bibfnamefont {M.}~\bibnamefont {Florian}},
  \bibinfo {author} {\bibfnamefont {A.}~\bibnamefont {Steinhoff}}, \bibinfo
  {author} {\bibfnamefont {D.}~\bibnamefont {Erben}}, \bibinfo {author}
  {\bibfnamefont {K.}~\bibnamefont {Tran}}, \bibinfo {author} {\bibfnamefont
  {D.~S.}\ \bibnamefont {Kim}}, \bibinfo {author} {\bibfnamefont
  {L.}~\bibnamefont {Sun}}, \bibinfo {author} {\bibfnamefont {J.}~\bibnamefont
  {Quan}}, \bibinfo {author} {\bibfnamefont {R.}~\bibnamefont {Claassen}},
  \bibinfo {author} {\bibfnamefont {S.}~\bibnamefont {Majumder}}, \bibinfo
  {author} {\bibfnamefont {J.~A.}\ \bibnamefont {Hollingsworth}}, \bibinfo
  {author} {\bibfnamefont {T.}~\bibnamefont {Taniguchi}}, \bibinfo {author}
  {\bibfnamefont {K.}~\bibnamefont {Watanabe}}, \bibinfo {author}
  {\bibfnamefont {K.}~\bibnamefont {Ueno}}, \bibinfo {author} {\bibfnamefont
  {A.}~\bibnamefont {Singh}}, \bibinfo {author} {\bibfnamefont
  {G.}~\bibnamefont {Moody}}, \bibinfo {author} {\bibfnamefont
  {F.}~\bibnamefont {Jahnke}}, \ and\ \bibinfo {author} {\bibfnamefont
  {X.}~\bibnamefont {Li}},\ }\href {\doibase 10.1103/PhysRevLett.126.047401}
  {\bibfield  {journal} {\bibinfo  {journal} {Phys. Rev. Lett.}\ }\textbf
  {\bibinfo {volume} {126}},\ \bibinfo {pages} {047401} (\bibinfo {year}
  {2021})}\BibitemShut {NoStop}%
\bibitem [{\citenamefont {Zeng}\ \emph {et~al.}(2022)\citenamefont {Zeng},
  \citenamefont {Xia}, \citenamefont {Dery}, \citenamefont {Watanabe},
  \citenamefont {Taniguchi}, \citenamefont {Shan},\ and\ \citenamefont
  {Mak}}]{Yihang2022}%
  \BibitemOpen
  \bibfield  {author} {\bibinfo {author} {\bibfnamefont {Y.}~\bibnamefont
  {Zeng}}, \bibinfo {author} {\bibfnamefont {Z.}~\bibnamefont {Xia}}, \bibinfo
  {author} {\bibfnamefont {R.}~\bibnamefont {Dery}}, \bibinfo {author}
  {\bibfnamefont {K.}~\bibnamefont {Watanabe}}, \bibinfo {author}
  {\bibfnamefont {T.}~\bibnamefont {Taniguchi}}, \bibinfo {author}
  {\bibfnamefont {J.}~\bibnamefont {Shan}}, \ and\ \bibinfo {author}
  {\bibfnamefont {K.~F.}\ \bibnamefont {Mak}},\ }\href {\doibase
  10.48550/ARXIV.2205.07354} {\enquote {\bibinfo {title} {Exciton density waves
  in coulomb-coupled dual moiré lattices},}\ } (\bibinfo {year}
  {2022})\BibitemShut {NoStop}%
\bibitem [{\citenamefont {Zhang}(2022)}]{YaHui2022}%
  \BibitemOpen
  \bibfield  {author} {\bibinfo {author} {\bibfnamefont {Y.-H.}\ \bibnamefont
  {Zhang}},\ }\href {\doibase 10.48550/ARXIV.2204.10937} {\enquote {\bibinfo
  {title} {Doping a mott insulator with excitons in moiré bilayer: fractional
  superfluid, neutral fermi surface and mott transition},}\ } (\bibinfo {year}
  {2022})\BibitemShut {NoStop}%
\bibitem [{\citenamefont {Camacho-Guardian}\ and\ \citenamefont
  {Cooper}(2022{\natexlab{a}})}]{Camacho-Guardian2021moire}%
  \BibitemOpen
  \bibfield  {author} {\bibinfo {author} {\bibfnamefont {A.}~\bibnamefont
  {Camacho-Guardian}}\ and\ \bibinfo {author} {\bibfnamefont {N.~R.}\
  \bibnamefont {Cooper}},\ }\href {\doibase 10.1103/PhysRevLett.128.207401}
  {\bibfield  {journal} {\bibinfo  {journal} {Phys. Rev. Lett.}\ }\textbf
  {\bibinfo {volume} {128}},\ \bibinfo {pages} {207401} (\bibinfo {year}
  {2022}{\natexlab{a}})}\BibitemShut {NoStop}%
\bibitem [{\citenamefont {Yu}\ \emph {et~al.}(2014)\citenamefont {Yu},
  \citenamefont {Liu}, \citenamefont {Gong}, \citenamefont {Xu},\ and\
  \citenamefont {Yao}}]{Yu2014b}%
  \BibitemOpen
  \bibfield  {author} {\bibinfo {author} {\bibfnamefont {H.}~\bibnamefont
  {Yu}}, \bibinfo {author} {\bibfnamefont {G.-B.}\ \bibnamefont {Liu}},
  \bibinfo {author} {\bibfnamefont {P.}~\bibnamefont {Gong}}, \bibinfo {author}
  {\bibfnamefont {X.}~\bibnamefont {Xu}}, \ and\ \bibinfo {author}
  {\bibfnamefont {W.}~\bibnamefont {Yao}},\ }\href {\doibase
  10.1038/ncomms4876} {\bibfield  {journal} {\bibinfo  {journal} {Nature
  Communications}\ }\textbf {\bibinfo {volume} {5}},\ \bibinfo {pages} {3876}
  (\bibinfo {year} {2014})}\BibitemShut {NoStop}%
\bibitem [{\citenamefont {Yu}\ and\ \citenamefont {Wu}(2014)}]{Yu2014}%
  \BibitemOpen
  \bibfield  {author} {\bibinfo {author} {\bibfnamefont {T.}~\bibnamefont
  {Yu}}\ and\ \bibinfo {author} {\bibfnamefont {M.~W.}\ \bibnamefont {Wu}},\
  }\href {\doibase 10.1103/PhysRevB.89.205303} {\bibfield  {journal} {\bibinfo
  {journal} {Phys. Rev. B}\ }\textbf {\bibinfo {volume} {89}},\ \bibinfo
  {pages} {205303} (\bibinfo {year} {2014})}\BibitemShut {NoStop}%
\bibitem [{\citenamefont {Wu}\ \emph {et~al.}(2015)\citenamefont {Wu},
  \citenamefont {Qu},\ and\ \citenamefont {MacDonald}}]{Wu2015b}%
  \BibitemOpen
  \bibfield  {author} {\bibinfo {author} {\bibfnamefont {F.}~\bibnamefont
  {Wu}}, \bibinfo {author} {\bibfnamefont {F.}~\bibnamefont {Qu}}, \ and\
  \bibinfo {author} {\bibfnamefont {A.~H.}\ \bibnamefont {MacDonald}},\ }\href
  {\doibase 10.1103/PhysRevB.91.075310} {\bibfield  {journal} {\bibinfo
  {journal} {Phys. Rev. B}\ }\textbf {\bibinfo {volume} {91}},\ \bibinfo
  {pages} {075310} (\bibinfo {year} {2015})}\BibitemShut {NoStop}%
\bibitem [{\citenamefont {Marzari}\ and\ \citenamefont
  {Vanderbilt}(1997)}]{Marzari1997}%
  \BibitemOpen
  \bibfield  {author} {\bibinfo {author} {\bibfnamefont {N.}~\bibnamefont
  {Marzari}}\ and\ \bibinfo {author} {\bibfnamefont {D.}~\bibnamefont
  {Vanderbilt}},\ }\href {\doibase 10.1103/PhysRevB.56.12847} {\bibfield
  {journal} {\bibinfo  {journal} {Phys. Rev. B}\ }\textbf {\bibinfo {volume}
  {56}},\ \bibinfo {pages} {12847} (\bibinfo {year} {1997})}\BibitemShut
  {NoStop}%
\bibitem [{\citenamefont {Remez}\ and\ \citenamefont
  {Cooper}(2021)}]{Remez2021}%
  \BibitemOpen
  \bibfield  {author} {\bibinfo {author} {\bibfnamefont {B.}~\bibnamefont
  {Remez}}\ and\ \bibinfo {author} {\bibfnamefont {N.~R.}\ \bibnamefont
  {Cooper}},\ }\href {\doibase 10.48550/ARXIV.2110.07628} {\enquote {\bibinfo
  {title} {Leaky exciton condensates in transition metal dichalcogenide moiré
  bilayers},}\ } (\bibinfo {year} {2021})\BibitemShut {NoStop}%
\bibitem [{\citenamefont {Zhang}\ \emph {et~al.}(2021)\citenamefont {Zhang},
  \citenamefont {Wu}, \citenamefont {Hou}, \citenamefont {Zhang}, \citenamefont
  {Chou}, \citenamefont {Watanabe}, \citenamefont {Taniguchi}, \citenamefont
  {Forrest},\ and\ \citenamefont {Deng}}]{Zhang2021}%
  \BibitemOpen
  \bibfield  {author} {\bibinfo {author} {\bibfnamefont {L.}~\bibnamefont
  {Zhang}}, \bibinfo {author} {\bibfnamefont {F.}~\bibnamefont {Wu}}, \bibinfo
  {author} {\bibfnamefont {S.}~\bibnamefont {Hou}}, \bibinfo {author}
  {\bibfnamefont {Z.}~\bibnamefont {Zhang}}, \bibinfo {author} {\bibfnamefont
  {Y.-H.}\ \bibnamefont {Chou}}, \bibinfo {author} {\bibfnamefont
  {K.}~\bibnamefont {Watanabe}}, \bibinfo {author} {\bibfnamefont
  {T.}~\bibnamefont {Taniguchi}}, \bibinfo {author} {\bibfnamefont {S.~R.}\
  \bibnamefont {Forrest}}, \ and\ \bibinfo {author} {\bibfnamefont
  {H.}~\bibnamefont {Deng}},\ }\href {\doibase 10.1038/s41586-021-03228-5}
  {\bibfield  {journal} {\bibinfo  {journal} {Nature}\ }\textbf {\bibinfo
  {volume} {591}},\ \bibinfo {pages} {61} (\bibinfo {year} {2021})}\BibitemShut
  {NoStop}%
\bibitem [{\citenamefont {Camacho-Guardian}\ and\ \citenamefont
  {Cooper}(2022{\natexlab{b}})}]{Camacho-Guardian2022moire}%
  \BibitemOpen
  \bibfield  {author} {\bibinfo {author} {\bibfnamefont {A.}~\bibnamefont
  {Camacho-Guardian}}\ and\ \bibinfo {author} {\bibfnamefont {N.~R.}\
  \bibnamefont {Cooper}},\ }\href {\doibase 10.48550/ARXIV.2206.06166}
  {\enquote {\bibinfo {title} {Optical non-linearities and translational
  symmetry breaking in driven-dissipative moiré exciton-polaritons},}\ }
  (\bibinfo {year} {2022}{\natexlab{b}})\BibitemShut {NoStop}%
\bibitem [{\citenamefont {Marcellina}\ \emph {et~al.}(2021)\citenamefont
  {Marcellina}, \citenamefont {Liu}, \citenamefont {Hu}, \citenamefont
  {Fieramosca}, \citenamefont {Huang}, \citenamefont {Du}, \citenamefont {Liu},
  \citenamefont {Zhao}, \citenamefont {Watanabe}, \citenamefont {Taniguchi},\
  and\ \citenamefont {Xiong}}]{Marcellina2021}%
  \BibitemOpen
  \bibfield  {author} {\bibinfo {author} {\bibfnamefont {E.}~\bibnamefont
  {Marcellina}}, \bibinfo {author} {\bibfnamefont {X.}~\bibnamefont {Liu}},
  \bibinfo {author} {\bibfnamefont {Z.}~\bibnamefont {Hu}}, \bibinfo {author}
  {\bibfnamefont {A.}~\bibnamefont {Fieramosca}}, \bibinfo {author}
  {\bibfnamefont {Y.}~\bibnamefont {Huang}}, \bibinfo {author} {\bibfnamefont
  {W.}~\bibnamefont {Du}}, \bibinfo {author} {\bibfnamefont {S.}~\bibnamefont
  {Liu}}, \bibinfo {author} {\bibfnamefont {J.}~\bibnamefont {Zhao}}, \bibinfo
  {author} {\bibfnamefont {K.}~\bibnamefont {Watanabe}}, \bibinfo {author}
  {\bibfnamefont {T.}~\bibnamefont {Taniguchi}}, \ and\ \bibinfo {author}
  {\bibfnamefont {Q.}~\bibnamefont {Xiong}},\ }\href {\doibase
  10.1021/acs.nanolett.1c01207} {\bibfield  {journal} {\bibinfo  {journal}
  {Nano Letters}\ }\textbf {\bibinfo {volume} {21}},\ \bibinfo {pages} {4461}
  (\bibinfo {year} {2021})}\BibitemShut {NoStop}%
\bibitem [{\citenamefont {Tang}\ \emph {et~al.}(2021)\citenamefont {Tang},
  \citenamefont {Gu}, \citenamefont {Liu}, \citenamefont {Watanabe},
  \citenamefont {Taniguchi}, \citenamefont {Hone}, \citenamefont {Mak},\ and\
  \citenamefont {Shan}}]{Tang2021}%
  \BibitemOpen
  \bibfield  {author} {\bibinfo {author} {\bibfnamefont {Y.}~\bibnamefont
  {Tang}}, \bibinfo {author} {\bibfnamefont {J.}~\bibnamefont {Gu}}, \bibinfo
  {author} {\bibfnamefont {S.}~\bibnamefont {Liu}}, \bibinfo {author}
  {\bibfnamefont {K.}~\bibnamefont {Watanabe}}, \bibinfo {author}
  {\bibfnamefont {T.}~\bibnamefont {Taniguchi}}, \bibinfo {author}
  {\bibfnamefont {J.}~\bibnamefont {Hone}}, \bibinfo {author} {\bibfnamefont
  {K.~F.}\ \bibnamefont {Mak}}, \ and\ \bibinfo {author} {\bibfnamefont
  {J.}~\bibnamefont {Shan}},\ }\href {\doibase 10.1038/s41565-020-00783-2}
  {\bibfield  {journal} {\bibinfo  {journal} {Nature Nanotechnology}\ }\textbf
  {\bibinfo {volume} {16}},\ \bibinfo {pages} {52} (\bibinfo {year}
  {2021})}\BibitemShut {NoStop}%
\bibitem [{\citenamefont {Wang}\ \emph
  {et~al.}(2021{\natexlab{b}})\citenamefont {Wang}, \citenamefont {Zhu},
  \citenamefont {Seyler}, \citenamefont {Rivera}, \citenamefont {Zheng},
  \citenamefont {Wang}, \citenamefont {He}, \citenamefont {Taniguchi},
  \citenamefont {Watanabe}, \citenamefont {Yan}, \citenamefont {Mandrus},
  \citenamefont {Gamelin}, \citenamefont {Yao},\ and\ \citenamefont
  {Xu}}]{Wang2021}%
  \BibitemOpen
  \bibfield  {author} {\bibinfo {author} {\bibfnamefont {X.}~\bibnamefont
  {Wang}}, \bibinfo {author} {\bibfnamefont {J.}~\bibnamefont {Zhu}}, \bibinfo
  {author} {\bibfnamefont {K.~L.}\ \bibnamefont {Seyler}}, \bibinfo {author}
  {\bibfnamefont {P.}~\bibnamefont {Rivera}}, \bibinfo {author} {\bibfnamefont
  {H.}~\bibnamefont {Zheng}}, \bibinfo {author} {\bibfnamefont
  {Y.}~\bibnamefont {Wang}}, \bibinfo {author} {\bibfnamefont {M.}~\bibnamefont
  {He}}, \bibinfo {author} {\bibfnamefont {T.}~\bibnamefont {Taniguchi}},
  \bibinfo {author} {\bibfnamefont {K.}~\bibnamefont {Watanabe}}, \bibinfo
  {author} {\bibfnamefont {J.}~\bibnamefont {Yan}}, \bibinfo {author}
  {\bibfnamefont {D.~G.}\ \bibnamefont {Mandrus}}, \bibinfo {author}
  {\bibfnamefont {D.~R.}\ \bibnamefont {Gamelin}}, \bibinfo {author}
  {\bibfnamefont {W.}~\bibnamefont {Yao}}, \ and\ \bibinfo {author}
  {\bibfnamefont {X.}~\bibnamefont {Xu}},\ }\href {\doibase
  10.1038/s41565-021-00969-2} {\bibfield  {journal} {\bibinfo  {journal}
  {Nature Nanotechnology}\ }\textbf {\bibinfo {volume} {16}},\ \bibinfo {pages}
  {1208} (\bibinfo {year} {2021}{\natexlab{b}})}\BibitemShut {NoStop}%
\bibitem [{\citenamefont {Yu}\ \emph {et~al.}(2015{\natexlab{b}})\citenamefont
  {Yu}, \citenamefont {Wang}, \citenamefont {Tong}, \citenamefont {Xu},\ and\
  \citenamefont {Yao}}]{Yu2015:2}%
  \BibitemOpen
  \bibfield  {author} {\bibinfo {author} {\bibfnamefont {H.}~\bibnamefont
  {Yu}}, \bibinfo {author} {\bibfnamefont {Y.}~\bibnamefont {Wang}}, \bibinfo
  {author} {\bibfnamefont {Q.}~\bibnamefont {Tong}}, \bibinfo {author}
  {\bibfnamefont {X.}~\bibnamefont {Xu}}, \ and\ \bibinfo {author}
  {\bibfnamefont {W.}~\bibnamefont {Yao}},\ }\href {\doibase
  10.1103/PhysRevLett.115.187002} {\bibfield  {journal} {\bibinfo  {journal}
  {Phys. Rev. Lett.}\ }\textbf {\bibinfo {volume} {115}},\ \bibinfo {pages}
  {187002} (\bibinfo {year} {2015}{\natexlab{b}})}\BibitemShut {NoStop}%
\bibitem [{\citenamefont {Rivera}\ \emph {et~al.}(2016)\citenamefont {Rivera},
  \citenamefont {Seyler}, \citenamefont {Yu}, \citenamefont {Schaibley},
  \citenamefont {Yan}, \citenamefont {Mandrus}, \citenamefont {Yao},\ and\
  \citenamefont {Xu}}]{Rivera2016}%
  \BibitemOpen
  \bibfield  {author} {\bibinfo {author} {\bibfnamefont {P.}~\bibnamefont
  {Rivera}}, \bibinfo {author} {\bibfnamefont {K.~L.}\ \bibnamefont {Seyler}},
  \bibinfo {author} {\bibfnamefont {H.}~\bibnamefont {Yu}}, \bibinfo {author}
  {\bibfnamefont {J.~R.}\ \bibnamefont {Schaibley}}, \bibinfo {author}
  {\bibfnamefont {J.}~\bibnamefont {Yan}}, \bibinfo {author} {\bibfnamefont
  {D.~G.}\ \bibnamefont {Mandrus}}, \bibinfo {author} {\bibfnamefont
  {W.}~\bibnamefont {Yao}}, \ and\ \bibinfo {author} {\bibfnamefont
  {X.}~\bibnamefont {Xu}},\ }\href {\doibase 10.1126/science.aac7820}
  {\bibfield  {journal} {\bibinfo  {journal} {Science}\ }\textbf {\bibinfo
  {volume} {351}},\ \bibinfo {pages} {688} (\bibinfo {year} {2016})},\ \Eprint
  {http://arxiv.org/abs/https://www.science.org/doi/pdf/10.1126/science.aac7820}
  {https://www.science.org/doi/pdf/10.1126/science.aac7820} \BibitemShut
  {NoStop}%
\bibitem [{\citenamefont {Leykam}\ \emph {et~al.}(2018)\citenamefont {Leykam},
  \citenamefont {Andreanov},\ and\ \citenamefont {Flach}}]{Leykam2018}%
  \BibitemOpen
  \bibfield  {author} {\bibinfo {author} {\bibfnamefont {D.}~\bibnamefont
  {Leykam}}, \bibinfo {author} {\bibfnamefont {A.}~\bibnamefont {Andreanov}}, \
  and\ \bibinfo {author} {\bibfnamefont {S.}~\bibnamefont {Flach}},\ }\href
  {\doibase 10.1080/23746149.2018.1473052} {\bibfield  {journal} {\bibinfo
  {journal} {Advances in Physics: X}\ }\textbf {\bibinfo {volume} {3}},\
  \bibinfo {pages} {1473052} (\bibinfo {year} {2018})},\ \Eprint
  {http://arxiv.org/abs/https://doi.org/10.1080/23746149.2018.1473052}
  {https://doi.org/10.1080/23746149.2018.1473052} \BibitemShut {NoStop}%
\bibitem [{\citenamefont {Bistritzer}\ and\ \citenamefont
  {MacDonald}(2011)}]{Bistritzer2010}%
  \BibitemOpen
  \bibfield  {author} {\bibinfo {author} {\bibfnamefont {R.}~\bibnamefont
  {Bistritzer}}\ and\ \bibinfo {author} {\bibfnamefont {A.~H.}\ \bibnamefont
  {MacDonald}},\ }\href {\doibase 10.1073/pnas.1108174108} {\bibfield
  {journal} {\bibinfo  {journal} {Proc. Natl. Acad. Sci. U.S.A.}\ }\textbf
  {\bibinfo {volume} {108}},\ \bibinfo {pages} {12233} (\bibinfo {year}
  {2011})}\BibitemShut {NoStop}%
\bibitem [{\citenamefont {Angeli}\ and\ \citenamefont
  {MacDonald}(2021)}]{Angeli2021}%
  \BibitemOpen
  \bibfield  {author} {\bibinfo {author} {\bibfnamefont {M.}~\bibnamefont
  {Angeli}}\ and\ \bibinfo {author} {\bibfnamefont {A.~H.}\ \bibnamefont
  {MacDonald}},\ }\href {\doibase 10.1073/pnas.2021826118} {\bibfield
  {journal} {\bibinfo  {journal} {Proceedings of the National Academy of
  Sciences}\ }\textbf {\bibinfo {volume} {118}},\ \bibinfo {pages}
  {e2021826118} (\bibinfo {year} {2021})},\ \Eprint
  {http://arxiv.org/abs/https://www.pnas.org/doi/pdf/10.1073/pnas.2021826118}
  {https://www.pnas.org/doi/pdf/10.1073/pnas.2021826118} \BibitemShut {NoStop}%
\bibitem [{\citenamefont {Castin}(2001)}]{castin:book}%
  \BibitemOpen
  \bibfield  {author} {\bibinfo {author} {\bibfnamefont {Y.}~\bibnamefont
  {Castin}},\ }in\ \href@noop {} {\emph {\bibinfo {booktitle} {Coherent Atomic
  Matter Waves}}},\ \bibinfo {editor} {edited by\ \bibinfo {editor}
  {\bibfnamefont {R.}~\bibnamefont {Caiser}}, \bibinfo {editor} {\bibfnamefont
  {C.}~\bibnamefont {Westbrook}}, \ and\ \bibinfo {editor} {\bibfnamefont
  {F.}~\bibnamefont {David}}}\ (\bibinfo  {publisher} {EDP Sciences and
  Springer-Verlag},\ \bibinfo {year} {2001})\BibitemShut {NoStop}%
\bibitem [{\citenamefont {Moskalenko}\ and\ \citenamefont
  {Snoke}(2000)}]{Moskalenko2000}%
  \BibitemOpen
  \bibfield  {author} {\bibinfo {author} {\bibfnamefont {S.~A.}\ \bibnamefont
  {Moskalenko}}\ and\ \bibinfo {author} {\bibfnamefont {D.~W.}\ \bibnamefont
  {Snoke}},\ }\href@noop {} {\emph {\bibinfo {title} {Bose-Einstein
  Condensation of Excitons and Biexcitons and Coherent Nonlinear Optics with
  Excitons.}}}\ (\bibinfo  {publisher} {Cambridge University Press},\ \bibinfo
  {year} {2000})\BibitemShut {NoStop}%
\end{thebibliography}
\end{document}

\title{Non-local interactions and supersolidity of \moire excitons}

\newcommand{\affiliationAarhus}{Center for Complex Quantum Systems, Department of Physics and Astronomy, Aarhus University, Ny Munkegade, DK-8000 Aarhus C, Denmark}

\author{Aleksi Julku}
\affiliation{\affiliationAarhus}

%Collaboration name if desired (requires use of superscriptaddress
%option in \documentclass). \noaffiliation is required (may also be
%used with the \author command).
%\collaboration can be followed by \email, \homepage, \thanks as well.
%\collaboration{}
%\noaffiliation

\date{\today}

%\begin{abstract}
%This is the SM.    
%\end{abstract}

\maketitle

\onecolumngrid

\section{Hybridized \moire exciton model}\label{Sec:hyb}

In this section we provide the details of the calculations related the hybridized \moire exciton continuum model of Refs.~\cite{Ruiz-Tijerina2019,Alexeev2019}. We first briefly go through the derivation of the model and show how to obtain the one-particle continuum \moire exciton Hamiltonian. Based on this, we derive the tight-binding model and discuss some additional details related to computing the exciton-exciton interaction terms.

We consider a TMDC heterobilayer system and label two layers as L1 and L2, with respective lattice constants being $a_1$ and $a_2$. We take L1 to be a MoSe$_2$ and L2 a WS$_2$ monolayer. TMDC monolayers have a hexagonal lattice structure and direct band gaps at the corners of their hexagonal Brillouin zone (BZ), namely in the $K$- and $K'$-valleys~\cite{Wang2018,Mueller2018}.  For two layers, the $K$-valleys are at $\bk = [4\pi/(3 a_1),0] \equiv \textbf{K}_1$ and $\bk = [4\pi/(3 a_2),0] \equiv \textbf{K}_2$. The $K$-points of two layers differ from each other by $\Delta \bkk = \bkk_2 - \bkk_1$. Due to finite lattice mismatch ($a_1 \neq a_2$) or possible relative twist angle $\theta$ between the layers, one has $\Delta \bkk \neq 0$. When the lattice mismatch and the twist angle is small, we have $|\Delta \bkk| << |\bkk_1|$. In this case, the inter-layer electron tunneling hybridizes  the low-energy electron states near the $K$- and $K'$-valleys. Correspondingly, the system acquires long-wavelength moire pattern with \moire periodicity $a_m$. Furthermore, due to large momentum mismatch between $K$ and $K'$-valleys, $K$ ($K'$)-valley electrons of L1 mix up only with $K$ ($K'$)-valley electrons of L2. Moreover, as the intrinsic spin-orbit coupling of TMDC monolayers leads to the spin-valley locking~\cite{Wang2018,Mueller2018}, one can, with a photon of a given helicity, optically create excitons in monolayer TMDCs only in a single valley~\cite{Zhang2019}. Furthermore, it turns out~\cite{Ruiz-Tijerina2019} that spin-down excitons have lower energy than the spin-up excitons.  We therefore consider only the  spin-down electron states in the $K$-valley, i.e. the ones near the $\bkk_1$ and $\bkk_2$ states.

%As our interest is excitons, we therefore consider only the spin-down electron states in the $K$-valley, i.e. the ones near the $\bkk_1$ and $\bkk_2$ states. Consequently, we therefore discard the valley and spin indices. 

\begin{figure}
  \centering
    \includegraphics[width=1.0\columnwidth]{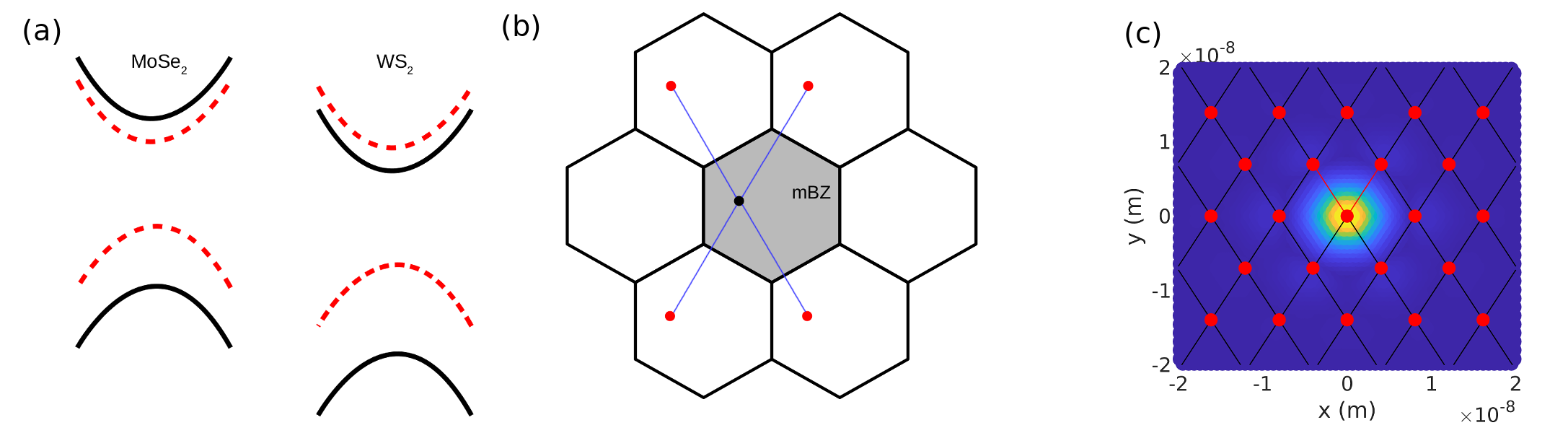}
    \caption{(a) Schematic of the valence and conduction band dispersions in the $K$-valley for both MoSe$_2$ and WS$_2$ monolayers. Red dashed (black solid) lines depicts spin-down (-up) electron bands. We consider only the spin down electrons. (b) Interlayer coupling Eq.~\eqref{t_c} couples a momentum $\bk$ (black dot) in the first mBZ (grey-shaded area) to other momenta $\bk \pm \bb^m_1$ and $\bk \pm \bb^m_2$ (red dots) that reside in adjacent mBZs. The coupling makes it possible to present the system Hamiltonian only with the momenta that reside in the first \moire Brillouin zone (mBZ). (c) Spatial profile of the excitonic Wannier function amplitude $|W_{x,\brr_i = 0} \rangle$ in case of $\theta = 0^\circ$. The red dots correspond to different lattice sites $\brr_i$ of the \moire lattice and red lines are the corresponding basis vectors of the \moire lattice. Black lines depict the boundaries of \moire unit cells. The lattice constant is the \moire periodicity $a_m$.}
   \label{Fig:S1}
\end{figure}

The schematic of the valence and conduction band dispersions  at the $K$-valley for decoupled  MoSe$_2$ and WS$_2$ monolayers are given in Fig.~\ref{Fig:S1}(a). To a good approximation, these band structures can be taken as parabolic bands. Furthermore, the valence and the conduction bands are separated by a large band gap. Then the Hamiltonian for two decoupled layers read
\begin{align}
&H^e_0 =  \sum_\bk \Big[\frac{k^2}{2 m_{c,1}} c^\dag_{c,1}(\bk) c_{c,1}(\bk) - \frac{k^2}{2 m_{v,1}} c^\dag_{v,1}(\bk) c_{v,1}(\bk)\Big]  + \sum_{\bk'} \Big[\frac{k'^2}{2 m_{c,2}} c^\dag_{c,2}(\bk') c_{c,2}(\bk') - \frac{k'^2}{2 m_{v,2}} c^\dag_{v,2}(\bk') c_{v,2}(\bk')\Big] \nonumber \\  
& + E_{gap,1}  + E_{gap,2}.
\end{align}
Here the first sum (second) describes the band dispersions of L1 (L2), $\bk$ ($\bk'$) is momentum measured from $\bkk_1$ ($\bkk_2$), and $c_{c,1}(\bk)$ [$c_{c,2}(\bk)$] are the annihilation operators for the conduction band electrons in L1 (L2). Similarly,  $c_{v,1}(\bk)$ and $c_{v,2}(\bk)$ are the operators for the valence band electrons. Moreover, $E_{gap,1}$ ($E_{gap,2}$) is the energy gap between the valence and conduction bands of L1 (L2). Finally, $m_{c,1}$, ($m_{c,2}$) and $m_{v,1}$ ($m_{v,2}$) are, respectively, the effective masses of conduction and valence band electrons in L1 (L2). 

Kinetic interlayer tunneling of electrons leads to the coupling between the electronic states of two layers and consequently to the long-wavelength \moire potential. To a good approximation, one can consider the coupling between two conduction electrons and between two valence electrons separately so that the \moire coupling Hamiltonian is 
\begin{align}
\label{tunneling_H}
H_t =  \sum_{\bk,\bk'} & \Big[ T^c(\bk,\bk') c^\dag_{c,1}(\bk)c_{c,2}(\bk') +  T^v(\bk,\bk') c^\dag_{v,1}(\bk)c_{v,2}(\bk') + h.c.  \Big]
\end{align}
where the first (second) term describes the tunneling of conduction (valence) electrons and $T^c(\bk,\bk')$ ($T^v$) gives the corresponding tunneling amplitude. For convenience, in $H_1$ \textit{both} $\bk$ and $\bk'$ are measured from $\bkk_1$.

It turns out \cite{Ruiz-Tijerina2019,Wang2017} that for large \moire periods ($|\Delta \bkk| << |\bkk_1|$) the tunneling terms can be written as (for details, see Ref.  \cite{Ruiz-Tijerina2019})
\begin{align}
\label{t_c}
T^c(\bk,\bk') = t_c \Big( \delta_{\bk,\bk'} + \delta_{\bk - \bk', \bb^m_{1}} + \delta_{\bk - \bk', \bb^m_{2}} \Big),    
\end{align}
where numerical value for $t_c$ is  obtained from experiments~\cite{Ruiz-Tijerina2019,Alexeev2019} and $\bb^m_{i}$ ($i=1,2)$ are the basis vectors of the momentum space of the \moire lattice. Similarly, hopping terms of valence band electrons are characterized by the hopping strength $t_v$. Following Ref.~\cite{Ruiz-Tijerina2019},  we use $t_c = 26$ meV and $t_v = 2t_c$ meV. From Eq.~\eqref{t_c} we see that the interlayer tunneling processes couple electrons of momenta $\bk$ in one layer with momenta $\bk \pm \bb^m_{1}$ and $\bk \pm \bb^m_{2}$ in the other layer. This is illustrated in  Fig.~\ref{Fig:S1}(b).

From the electronic interlayer hopping terms of Eq.~\eqref{t_c}, it is straightforward to obtain the \moire Hamiltonian for excitons.  We take into account both intra- and interlayer excitons, labelled as $| \text{X} \rangle$, $| \text{X}' \rangle$, $| \text{IX} \rangle$ and $|\text{IX}' \rangle$. Excitons $| \text{X} \rangle$ ($| \text{X}' \rangle$) and holes of $| \text{IX} \rangle$ ($|\text{IX}' \rangle$) reside in layer L1 (L2). The expressions for all four types of excitons read
\begin{align}
\label{Xscat1}
& | \text{IX} (\bqq) \rangle = \sum_{\bq} \phi_{\text{IX}}(\bq) c^\dag_{c,2}(x^{\text{IX}}_{e}\bqq + \bq) c_{v,1}(-x^{\text{IX}}_{h}\bqq + \bq) |0\rangle \nonumber \\
& | \text{IX}' (\bqq) \rangle =  \sum_{\bq} \phi_{\text{IX}'}(\bq) c^\dag_{c,1}(x^{\text{IX}'}_{e}\bqq + \bq) c_{v,2}(-x^{\text{IX}'}_{h}\bqq + \bq) |0\rangle  \nonumber \\
& | \text{X} (\bqq) \rangle =  \sum_{\bq} \phi_{\text{X}}(\bq) c^\dag_{c,1}(x^{\text{X}}_{e}\bqq + \bq) c_{v,1}(-x^{\text{X}}_{h}\bqq + \bq) |0\rangle \nonumber \\
& | \text{X}' (\bqq) \rangle = \sum_{\bq} \phi_{\text{X}'}(\bq) c^\dag_{c,2}(x^{\text{X}'}_{e}\bqq + \bq) c_{v,2}(-x^{\text{X}'}_{h}\bqq + \bq) |0\rangle
\end{align}
These expressions can be also found in Refs.~\cite{Yu2015:2,Rivera2016,Ruiz-Tijerina2019}. Here  $\bqq = \bk_e + \bk_h$ is the center of mass momentum with $\bk_e$ and $\bk_h$ being the electron and hole momenta, measured relative to their respective $K$-points. Furthermore, the coefficients $x^{t}_{e}$ and $x^{t}_{h}$ are the mass ratios of the electron and hole with respect to the total exciton mass, i.e. $x^{t}_{e} + x^{t}_{h} = 1$. Finally, the momentum sums run over the relative momenta, i.e. $\bq = x^{t}_{h}\bk_e - x^{t}_{e}\bk_h$ and $\phi_{t}(\bq) = \int d\br \frac{\exp(-i\bq \cdot \br)}{\sqrt{A}}\phi_t(\br)$ are the relative wavefunctions of excitons in the momentum space with $\phi_t(\br)$ being the corresponding wavefunctions in the real space. With these expressions, one can then write down the interlayer hopping terms for excitons~\cite{Ruiz-Tijerina2019}. For example, the coupling between $|\text{X}'\rangle$ and  $|\text{IX}'\rangle$ can be written as
\begin{align}
\label{Xscat2}
&\langle \text{IX}'(\bqq) | H_t | \text{X}'(\bqq') \rangle =  \sum_\bq \phi^*_{\text{IX}'}(\bq) \langle 0 | c^\dag_{v,2}(-x_h^{\text{IX}'}\bqq + \bq) c_{c,1}(x_e^{\text{IX}'}\bqq + \bq) | H_t \nonumber \\
& \times \sum_{\bq'} \phi_{\text{X}'}(\bq') c^\dag_{c,1}(x_e^{\text{X}'}\bqq' + \bq') c_{v,2}(-x_h^{\text{X}'}\bqq' + \bq')| 0 \rangle \nonumber \\
& =  \sum_{\bq,\bq'}  \phi^*_{\text{IX}'}(\bq)  \phi_{\text{X}'}(\bq')  \langle c_{c,1}(m_e^{\text{IX}'}\bqq + \bq)  |H_t| c^\dag_{c,1}(x_e^{\text{X}'}\bqq' + \bq') \rangle \delta_{-m_h^{\text{IX}'}\bqq + \bq, -x^{\text{X}'}_h \bqq' + \bq'} \nonumber \\
& =t_c\sum_{\bq'} \phi_{\text{IX}'}^*(\bq' + x^{\text{IX}'}_{h}\bqq - x^{\text{X}'}_{h}\bqq')\phi_{\text{X}'}(\bq') \sum_{i=0}^2 \delta_{\bqq - \bqq',C^i_3 \Delta \bkk}
\end{align}
where the last term follows from Eq.~\eqref{t_c} and the operator $C_3$ rotates a vector by $120$ degrees. In the same way one can derive the expressions for the remaining exciton-exciton scattering terms:
\begin{align}
\label{Xscat3}
&  \langle \text{IX}(\bqq) | H_t | \text{X}'(\bqq') \rangle =   t_v \sum_{\bq'} \phi_{\text{IX}}^*(\bq' - x^{\text{IX}}_{e}\bqq + x^{\text{X}'}_{e}\bqq')\phi_{\text{X}'}(\bq') \sum_{i=0}^2 \delta_{\bqq - \bqq',C^i_3 \Delta \bkk} \nonumber \\
& \langle \text{IX}'(\bqq) | H_t | \text{X}(\bqq') \rangle =  t_v \sum_{\bq'} \phi_{\text{IX}'}^*(\bq' - x^{\text{IX}'}_{e}\bqq + x^{\text{X}}_{e}\bqq')\phi_{\text{X}}(\bq') \sum_{i=0}^2 \delta_{\bqq - \bqq',C^i_3 \Delta \bkk} \nonumber \\
& \langle \text{IX}(\bqq) | H_t | \text{\text{X}}(\bqq') \rangle =   t_c \sum_{\bq'} \phi_{\text{IX}}^*(\bq' + x^{\text{IX}}_{h}\bqq - x^{\text{X}}_{h}\bqq')\phi_{\text{X}}(\bq') \sum_{i=0}^2 \delta_{\bqq - \bqq',C^i_3 \Delta \bkk}.
\end{align}
To calculate the exciton-exciton scattering elements, one needs to solve the relative wavefcuntions of excitons, i.e. $\phi_t(\bq)$. For  $t \in \{\text{X},\text{X}' \}$ ($t \in \{\text{IX},\text{IX}' \}$) this can be obtained from the two-body Hamiltonian of an electron and hole residing in the same (separate) layer(s). To a good approximation, the lowest bound state can be taken $s$-symmetric which is identified by the Bohr radius $a_{t,B}$ of the exciton. For our MoSe$_2$/WS$_2$ system, we use the Bohr radii values given in Ref.~\cite{Ruiz-Tijerina2019}. Consequently, one then has $\phi_t(\bq) = \sqrt{\frac{8\pi}{A a_{t,B}^4}}(q^2 + \frac{1}{a_{t,B}^2})^{-3/2}$ and it is possible to obtain analytic expressions for the scattering terms~\eqref{Xscat2} and \eqref{Xscat3} \cite{Ruiz-Tijerina2019}.

The single-particle dispersions of the excitons can be taken as
\begin{align}
\label{Xdisp}
E_{t}(\bqq) = E_{t}^0 + \frac{\hbar^2 \bqq^2}{2 m_t},    
\end{align}
where $t\in \{$X,X',IX,IX' $\}$, $m_t$ is the total exciton mass obtained from the electron and hole masses and the zero-momentum energies $E_{t}^0$ are determined by the experimental values~\cite{Ruiz-Tijerina2019}. With Eqs.~\eqref{Xscat2},\eqref{Xscat3} and \eqref{Xdisp}, one can write down the $K$-valley hybrid \moire exciton continuum Hamiltonian $H_H$ as
\begin{align}
\label{moire_ham_ex}
&H_H = \sum_{\bk \in \textrm{mBZ}} \Psi_{x}^\dag(\bk) \mathcal{H}_H (\bk) \Psi_x(\bk), \textrm{ with} \nonumber \\
& \Psi_x(\bk) = \begin{bmatrix} \tilde{x}_{\text{IX}}(\bk -\Delta \bkk) & \tilde{x}_{\text{IX}'}(\bk +\Delta \bkk) & \tilde{x}_{\text{X}}(\bk) & \tilde{x}_{\text{IX}'}(\bk)
\end{bmatrix}^T \nonumber \\
&\tilde{x}_{t}(\bk) = \begin{bmatrix}
x_{t,\bk} & x_{t,\bk + \bb^m_1} & x_{t,\bk - \bb^m_1} & x_{t,\bk + \bb^m_2}  & \dots &
\end{bmatrix} \nonumber \\
& \mathcal{H}_H(\bk) =  \begin{bmatrix}
 \mathcal{H}_{\text{IX}}(\bk - \Delta \bkk) & 0 & \mathcal{H}_{\text{IX},\text{X}}(\bk) & \mathcal{H}_{\text{IX},\text{X}'}(\bk) \\
 0 & \mathcal{H}_{\text{IX}'}(\bk + \Delta \bkk) & \mathcal{H}_{\text{IX}',\text{X}}(\bk) & \mathcal{H}_{\text{IX}',\text{X}'}(\bk) \\
 \mathcal{H}_{\text{IX},\text{X}}^\dag(\bk) & \mathcal{H}_{\text{IX}',\text{X}}^\dag(\bk) & \mathcal{H}_{\text{X}}(\bk) & 0 \\
 \mathcal{H}_{\text{IX},\text{X}'}^\dag(\bk) & \mathcal{H}_{\text{IX}',\text{X}'}^\dag(\bk) & 0 & \mathcal{H}_{\text{X}'}(\bk)
\end{bmatrix},
\end{align}
where the momentum sum is limited to the first \moire Brillouin zone (mBZ) and $x_{t,\bk}$  are the annihilation operators for excitons of the center of mass momentum $\bk$. Furthermore, the matrices $\mathcal{H}_{t}(\bk)$  are diagonal and contain the parabolic band dispersions of the excitons~\eqref{Xdisp}. Moreover, the matrices in the off-diagonal sector, i.e. $\mathcal{H}_{t,t'}(\bk)$, contain the \moire hopping terms. For example, the elements of $\mathcal{H}_{\text{IX}',\text{X}'}(\bk)$ are determined by Eq.~\eqref{Xscat2}. Due to the \moire coupling, the intralayer and interlayer excitons hybridize, giving a rise to hybridized \moire excitons that contain features of both the intra- and interlayer excitons \cite{Alexeev2019,Ruiz-Tijerina2019}, i.e. strong light-mattter coupling of the intralayer part and long lifetimes and strong interactions of the interlayer component.

By diagonalizing $\mathcal{H}_H(\bk)$ of Eq.~\eqref{moire_ham_ex}, one obtains:
\begin{align}
H_H = \sum_n\sum_{\bk\in \text{mBZ}} \epsilon_{\bk n} \gamma_{\bk n}^\dag \gamma_{\bk n}, 
\end{align}
where $n$ is the \moire band index and $\gamma_{\bk n}$ annihilates a \moire exciton at momentum $\bk$ and energy $\epsilon_{\bk n}$. In Fig. 1(c) of the main text the \moire band structure $\epsilon_{\bk n}$ for $\theta = 0.5^\circ$ of spin-down hybrid excitons is presented. The lowest exciton band $\epsilon_{\bk 1}$ becomes extremely flat and separated from other bands by a notable band gap once the twist angle is small enough.

\subsection{Tight-binding Hamiltonian for \moire excitons}\label{Sec:TB}

We show here that the lowest \moire exciton band can be described by a triangular tight-binding model. The Wannier functions for the states of the lowest band can be formed as
\begin{align}
& | W_{\brr_i} \rangle = \frac{1}{\sqrt{N}}\sum_{\bk \in \textrm{mBZ}} e^{-i \bk \cdot \brr_i} |\psi_{\bk 1} \rangle 
\end{align}
where $N$ is the number of \moire unit cells, $|\psi_{\bk,1} \rangle$ are the Bloch functions for the lowest \moire band and  $| W_{\brr_i} \rangle$ are the respective Wannier functions labelled by the \moire unit cells coordinates $\brr_i$ [see Fig. 1(d) of the main text]. As usual, the definition of the Wannier functions is not unique but instead depends on the gauge choice of the Bloch functions, i.e. the Wannier states are not gauge-invariant under the transformation $| \psi_{\bk 1} \rangle \rightarrow e^{i\phi(\bk)} | \psi_{\bk 1} \rangle$. The usual procedure to build a tight-binding model is to maximally localize the resulting spread of the Wannier functions~\cite{Marzari1997}. As we expand our Wannier functions by using only a single band and the Berry curvature of our \moire bands is zero (as our starting point was trivial parabolic bands of the continuum model), the choice of the gauge fields reduces to choosing a gauge in which the Bloch functions are smooth functions in the momentum space~\cite{Marzari1997}. The resulting Wannier state for excitons at the unit cell $\brr_i = 0$ is shown in Fig.~\ref{Fig:S1}(c) for the twist angle $\theta = 0^\circ$. We see that the Wannier functions are well localized and form a simple triangular lattice structure whose lattice site coordinates are characterized by the unit cell coordinates $\brr_i$.

As the Wannier states are well localized and form a triangular lattice, one can use a tight-binding approximation to describe the bands of interest. We can write the Bloch functions of excitons as $|\psi_{\bk 1} \rangle = \frac{1}{\sqrt{N}} \sum_{\brr_i} e^{i\bk \cdot \brr_i} | W_{\brr_i} \rangle$ and the corresponding annihilation operators are $\gamma_{\bk 1} = \frac{1}{\sqrt{N}} \sum_{\brr_i} e^{i\bk \cdot \brr_i} x_{i}$. One can then cast the tight-binding \moire Hamiltonian for the lowest \moire band excitons as
\begin{align}
& H_{\textrm{TB}} = \sum_\bk \epsilon_{\bk 1} \gamma^\dag_{\bk 1} \gamma_{\bk 1} = \sum_{a,b} t_{ab} x^\dag_{a} x_{b}, \textrm{ where} \nonumber \\
&\label{t_hop} t_{ab} = \frac{1}{N}\sum_{\bk \in \textrm{mBZ}} \epsilon_{\bk 1} e^{-i \bk \cdot (\brr_a -\brr_b)}.
\end{align}
Here $t_{ab}$ is the hopping term between the lattice sites $\brr_a$ and $\brr_b$ of the tight-binding model and $\epsilon_{\bk 1}$ is the energy of the lowest \moire exciton band obtained by diagonalizing the original continuum \moire Hamiltonian~\eqref{moire_ham_ex}. Similar procedure applies for the \moire electrons.

For small twist angles, it is sufficient to take into account only the nearest-neighbour (NN) hopping terms, $t_{\mathrm{NN}}$, to faithfully reproduce the original \moire band structure quantitatively. For $\theta \sim 1.8^\circ$, $|t_{\text{NN}}| \sim 0.36$ meV.  One should note that $H_H$ yields in general complex-valued hopping terms (see Fig. 1(d) of the main text) and the complex phase cannot be gauged away. Thus, in the present work we built our own cluster mean-field (CMF) model to probe the many-body physics of \moire excitons as earlier CMF studies have used only real-valued hopping parameters.

We use here a single band tight-binding model to describe the quasi-flat band of \moire excitons. This is because the flatness of the band and corresponding localization of the excitons arise from the quenched kinetic energy. Flat bands can, however, also arise from the destructive interference of the Bloch states~\cite{Leykam2018}. To treat such flat band systems, a single-band tight-binding description is not enough but one needs to build a many-band model by including sufficiently many bands to capture correctly the geometric properties of the Bloch states. Examples of such non-trivial flat bands are the ones of twisted bilayer graphene~\cite{Bistritzer2010}, twisted homobilayer TMDC systems~\cite{Angeli2021} and various standard lattice models such as Lieb or kagome lattice~\cite{Leykam2018}.

\subsection{Exciton-exciton interactions}\label{sec_int}

In this subsection we show how to obtain the direct interaction vertex between two excitons, i.e. $g^{dir}_{t\tilde{t}}(\bq)$. To this end, we need to first write down the momentum-space Coulomb interaction $V_C(\bq)$ between two particles (either electrons or holes) of charge $q_1$ and $q_2$. For convenience, we write $V_C(\bq) = \frac{q_1 q_1}{2q \epsilon_{\text{intra},l}(q)}$ when two particles reside in the same layer $l$ and $V_C(\bq) = \frac{q_1 q_1}{2q \epsilon_{\text{inter}}(q)}$ when they reside in different layers. The momentum-dependent dielectric functions $\epsilon_{\text{inter}}(q)$ and $\epsilon_{\text{intra},l}(q)$ for our two-layer geometry is derived in Sec.~\ref{keldysh_sec}. Once $V_C(\bq)$ is known, one can calculate for example the IX-IX exciton interaction vertex as
\begin{align}
\label{int}
g^{dir}_{\text{IX},\text{IX}}(\bq,\bk,\bk') = \int d\br_{e} \int d\br'_e \int d\br_{h} \int d\br'_h \langle  \text{IX} (\bk + \bq) | \br_e \br_h \rangle \langle  \text{IX} (\bk' - \bq) | \br'_e \br'_h \rangle V(\br_e,\br_h,\br'_e,\br'_h) \langle \br'_e \br'_h | \text{IX} (\bk')  \rangle \langle \br_e \br_h | \text{IX} (\bk)  \rangle
\end{align}
with 
\begin{align}
V(\br_e,\br_h,\br'_e,\br'_h) = V_C(\br_e,\br'_e) + V_C(\br_h,\br'_h) + 2V_C(\br_e,\br'_h).    
\end{align}
Here $V_C(\br_1,\br_2)$ is the Coulomb interaction in the real space and $\langle \br_e \br_h | \text{IX} (\bk)  \rangle$ is the total exciton wavefunction which can be cast as~\cite{Yu2015:2,Rivera2016,Ruiz-Tijerina2019}
\begin{align}
\langle \br_e \br_h | \text{IX} (\bk)  \rangle = \frac{e^{i \bk \cdot \br_{\text{CM}}}}{A} \sum_{\bq'} \phi_{\text{IX}}(\bq') e^{i\bq' \cdot \br_{\text{rel}}} \langle \br_e | u^c_1(x_e^{\text{IX}}\bk + \bq') \rangle   \langle u^v_2(-x_h^{\text{IX}}\bk + \bq') | \br_h \rangle.  
\end{align}
Here $\br_e$ ($\br_h$) is the spatial coordinate of the electron (hole), $\br_{\text{CM}} = x_e^{\text{IX}}\br_e + x_h^{\text{IX}}\br_e$ is the center-of-mass coordinate, and $\br_{\text{rel}} = \br_e - \br_h$ is the relative coordinate. Moreover, $| u^c_L(\bk) \rangle$ [$| u^v_L(\bk) \rangle$] is the periodic part of the electronic Bloch function  of the lowest conduction (highest valence) band in monolayer $L$. %For our purposes, we can approximate $u_{x_e^{IX}\bk + \bq',1}^c(\br_e) \approx u_{-x_h^{IX}\bk + \bq',2}^v(\br_h) \approx 1$. Physically, this assumption means that excitons are localized in the momentum space around their respective K-points. By using the Bloch functions obtained from the faithful tight-binding models for TMDC monolayers REF, we have numerically confirmed that this approximation does not change the resulting interaction vertices considerably. We can then write the exciton wa

From Eq.~\eqref{int} one can with a straightforward but lengthy algebra obtain
\begin{align}
& g^{dir}_{\text{IX},\text{IX}}(\bq,\bk,\bk') = \frac{e^2}{2 A q} \Big[ \frac{F(\bk,\bq) F(\bk',-\bq)}{\epsilon_{\text{intra},1}(q)} + \frac{G(\bk,\bq)G(\bk',-\bq)}{\epsilon_{\text{intra},2}(q)} -\frac{2F(\bk,\bq) G(\bk',-\bq)}{\epsilon_{\text{inter}}(q)} \Big]     
\end{align}
where
\begin{align}
& F(\bk,\bq) = \sum_{\bq'} \phi_{\text{IX}}(\bq')\phi_{\text{IX}}(\bq' - x_h^{\text{IX}}\bq) \langle u^v(x_e^{\text{IX}}(\bk + \bq) + \bq') | u^v(x_e^{\text{IX}}\bk - x_h^{\text{IX}}\bq +\bq') \rangle \\
& G(\bk,\bq) = \sum_{\bq'} \phi_{\text{IX}}(\bq')\phi_{\text{IX}}(\bq' + x_e^{\text{IX}}\bq) \langle u^c(-x_h^{\text{IX}}\bk + x_e^{\text{IX}}\bq + \bq') | u^c(-x_h^{\text{IX}}(\bk + \bq) +\bq') \rangle.
\end{align}
To continue, we assume that the overlaps between the Bloch functions inside $F$ and $G$ are close to unity. Physically, this assumption means that only the scattering terms near the $K$-valley matter which is a plausible approximation. %Furthermore, by using the Bloch functions obtained from the faithful tight-binding model for TMDC monolayers REF, we have numerically confirmed that this approximation does not change the resulting interaction vertices considerably. 
A similar approximation was also used in Ref.~\cite{Yu2015:2}. One thus obtains
\begin{align}
g^{dir}_{\text{IX},\text{IX}}(\bq,\bk,\bk') \approx g^{dir}_{\text{IX},\text{IX}}(\bq) = \frac{e^2}{2 q A} \Bigg\{ \frac{f_{\text{IX}}(x^{\text{IX}}_{h}\bq)^2}{\epsilon_{\textrm{intra},2}(q)}+ \frac{f_{\text{IX}}(x^{\text{IX}}_{e}\bq)^2}{\epsilon_{\textrm{intra},1}(q)}  -\frac{2 f_{\text{IX}}(x_{e}^{\text{IX}}\bq) f_{\text{IX}}(x_{h}^{\text{IX}}\bq)}{\epsilon_{\textrm{inter}}(q)} \Bigg\} 
\end{align}
with $f_{\text{IX}}(\bk) = \sum_{\tilde{\bq}} \phi^*_{\text{IX}} (\tilde{\bq}) \phi_{\text{IX}}(\tilde{\bq} + \bk)$. One can in the same way also derive the IX-IX' and IX'-IX' interaction terms:
\begin{align}
 g^{dir}_{\text{IX}',\text{IX}'}(\bq) = \frac{e^2}{2 q A} \Bigg\{ & \frac{f_{\text{IX}'}(x^{\text{IX}'}_{h}\bq)^2}{\epsilon_{\textrm{intra},1}(q)}+ \frac{f_{\text{IX}'}(x^{\text{IX}'}_{e}\bq)^2}{\epsilon_{\textrm{intra},2}(q)}  -\frac{2 f_{\text{IX}'}(x_{e}^{\text{IX}'}\bq) f_{\text{IX}'}(x_{h}^{\text{IX}'}\bq)}{\epsilon_{\textrm{inter}}(q)} \Bigg\}  \\
g^{dir}_{\text{IX},\text{IX}'}(\bq) =  \frac{e^2}{2 q A} \Bigg\{ & \frac{f_{\text{IX}}(x^{\text{IX}}_{h}\bq) f_{\text{IX}'}(x^{\text{IX}'}_{h}\bq)}{\epsilon_{\textrm{inter}}(q)} + \frac{f_{\text{IX}}(x^{\text{IX}}_{e}\bq) f_{\text{IX}'}(x^{\text{IX}'}_{e}\bq)}{\epsilon_{\textrm{inter}}(q)}  - \frac{f_{\text{IX}}(x^{\text{IX}}_{e}\bq) f_{\text{IX}'}(x^{\text{IX}'}_{h}\bq)}{\epsilon_{\textrm{intra},1}(q)} - \frac{f_{\text{IX}}(x^{\text{IX}}_{h}\bq) f_{\text{IX}'}(x^{\text{IX}'}_{e}\bq)}{\epsilon_{\textrm{intra},2}(q)} \Bigg\} 
\end{align}

\section{Effective \moire potential model}

\begin{figure}
  \centering
    \includegraphics[width=0.85\columnwidth]{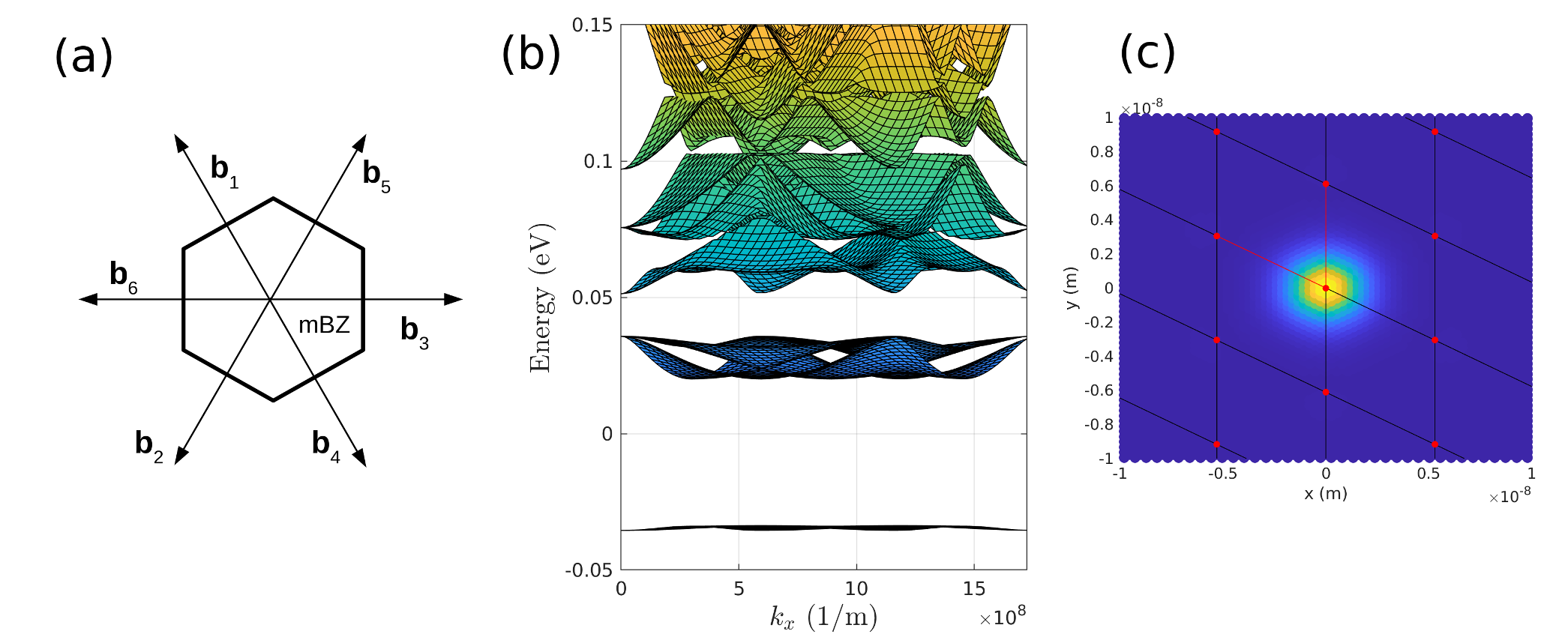}
    \caption{ Effective \moire potential model. (a) Vectors $\bb_j$ connecting adjacent \moire BZs. (b) \Moire exciton band structure obtained from the effective \moire potential Hamiltonian $H_E$~\eqref{cont} for the twist angle $\theta = 3^\circ$. (c) Spatial profile of the excitonic Wannier function amplitude $|W_{x,\brr_i = 0} \rangle$ in case of $\theta = 3^\circ$. The red dots correspond to different lattice sites $\brr_i$ of the \moire lattice and red lines are the corresponding basis vectors of the \moire lattice. Black lines depict the boundaries of \moire unit cells.}
   \label{Fig:S_cont}
\end{figure}

In Refs.~\cite{Yu2017,Gotting2022,Wu2018a,Tran2019} the properties of \moire interlayer excitons were studied by using a simplified \moire Hamiltonian that treats the effects of the \moire interlayer coupling as an effective single-particle  \moire potential $\Delta(\br)$ for excitons. In this section we go through our analysis by using this alternative model which we call the effective \moire potential model $H_E$. To this end, we write the one-particle Hamiltonian for the \moire interlayer excitons as
\begin{align}
\label{cont}
H_E = -\frac{\hbar^2 \nabla^2}{2\mu} + \Delta(\br),    
\end{align}
where $\mu = m_e m_h/(m_e + m_h)$ is the reduced mass of the exciton and the potential reads
\begin{align}
\Delta(\br) = V \sum_{j=1,2,3} \cos(\bb_j \cdot \br + \psi),    
\end{align}
where the vectors $\bb_j$ are depicted in Fig.~\ref{Fig:S_cont}(a). As $\bb_j$ connect adjacent \moire Brillouin zones, it is clear that their magnitude and thus the periodicity of $\Delta(\br)$ depends on the twist angle $\theta$. The values for $V$ are usually obtained from experiments or by numerically fitting the band structure to the DFT calculations.

In an experimental study of Ref.~\cite{Tran2019} the offset angle was set to $\phi = \pi$ which yields 
\begin{align}
\Delta(\br) = -V \sum_{j=1}^6 \exp(i \bb_j \cdot \br).    
\end{align}
Moreover, in Ref.~\cite{Tran2019} the strength of the \moire potential was set to $V= 18$ meV to match the experiments in case of a MoSe$_2$/WSe$_2$. We take this value and keep it constant for different twist angles.

One can easily diagonalize Eq.~\eqref{cont} in the basis of plane wave functions $\langle \br |\bk \rangle = \frac{1}{\sqrt{A}} e^{i\bk \cdot \br}$. This yields the eigenvalue problem
\begin{align}
\label{cont2}
\frac{\hbar^2 k^2}{2\mu} \psi_\bk - V \sum_j \psi_{\bk + \bb_j} = E\psi_\bk,   
\end{align}
with $|\psi \rangle = \sum_\bk \psi_\bk | \bk \rangle$ being the eigenstates of $H$. We see that the \moire potential couples the state of momentum $\bk$ with states of momenta $\bk + \bb_j$. One can thus express the Hamiltonian only with the momenta within the first mBZ such that $H_E \equiv \sum_{\bk \in \text{mBZ}} H_E(\bk)$. By diagonalizing $H_E(\bk)$ for each $\bk$ in the mBZ, one obtains the band structure and eigenstates of \moire excitons. As an example, in Fig.~\ref{Fig:S_cont}(b) the \moire exciton band structure is shown for $\theta = 3^\circ$. The lowest \moire exciton band is well isolated from the higher bands and becomes extremely flat as a function of decreasing $\theta$, in the same way as in case of $H_H$ [see Fig. 1(c) of the main text].

A difference between continuum models $H_H$ of Sec.~\ref{Sec:hyb} and $H_E$ is that the starting point of $H_H$ is the microscopic interlayer tunneling term~\eqref{tunneling_H} and the emergence of hybridized \moire excitons naturally arises from the tunneling Hamiltonian, allowing systematical studies of hybridized \moire excitons. Therefore, $H_H$ is more suitable than $H_E$ when \moire excitons consist of both intra- and interlayer excitons. In contrast, $H_E$ includes only the contribution of interlayer excitons and introduces \moire effects somewhat more heuristically via the effective potential $\Delta(\br)$. The advantage of $H_E$ is its simplicity which has made it a popular tool to study \moire effects of both electrons and excitons~\cite{Wu2019a,Wu2018,Pan2020,Morales-Duran2021,Yu2017,Wu2017,Gotting2022,Wu2018a,Tran2019}.

As the lowest \moire exciton band is well isolated from the higher bands, we can once again construct the localized Wannier functions for the lowest band, in the same way as was done in Sec.~\ref{Sec:TB} for $H_H$. As before, the zero Berry curvature means that the problem reduces to finding a smooth gauge for the Bloch functions of the lowest band. The resulting spatial profile of the Wannier state at $\brr_i = 0$ is shown in Fig.~\ref{Fig:S_cont}(c) for $\theta = 3^\circ$. We see that the state is well localized and the resulting tight-binding model requires only the inclusion of the nearest-neighbour hopping term $t_{\text{NN}}$. For $\theta = 3^\circ$, $t_{\text{NN}} \sim 0.2$ meV.

The exciton-exciton interaction can be computed in the same way as done in the main text for $H_H$. We start with the direct Coulomb interaction
\begin{align}
\label{cont_int}
H_{int} = \frac{1}{2A} \sum_{\bk,\bk',\bq} g^{dir}(\bq) x^\dag(\bk+\bq) x^\dag(\bk'-\bq) x(\bk') x(\bk),    
\end{align}
where $g^{dir}(\bq)$ is the bare Coulomb interaction between two interlayer excitons and $x(\bk)$ is the annihilaton operator for a interlayer exciton of momentum $\bk$. Note that $H_E$ includes only one kind of interlayer excitons and thus Eq.~\eqref{cont_int} does not include index $t$ accounting for different types of excitons, in contrast to the interaction term of $H_H$ (see Eq. 2 in the main text).  One then follows exactly the same procedure as for $H_H$, i.e. one expands the bare exciton operators in the basis of \moire excitons and keeps only the lowest \moire band degrees of freedom, to eventually obtain the interaction term for the tight binding model as
\begin{align}
H_{int} \sum_{abcd} \approx g_{abcd} x^\dag_a x^\dag_b x_c x_d,    
\end{align}
where $x_i$ annihilates an exciton of the lowest \moire exciton band at lattice site $i$. As in case of $H_H$, non-direct interaction terms are negligible and we are left only with the direct terms to obtain the interacting tight-binding Hamiltonian in the form of Eq. (1) of the main text. The bare interaction term $g^{dir}$ requires the knowledge of the exciton wavefunction, or if assuming the s-wave symmetry, the Bohr radius $a_B$ of the exciton. By numerically solving the two-body Hamiltonian for an electron and hole located in separate layers, and using the dielectric functions derived in Sec.~\ref{keldysh_sec}, one finds that $a_B \sim 3$ nm in case of a MoSe$_2$/WSe$_2$ stack. We use here the relative permittivity of $\epsilon_r \sim 4$ as in experiments of Ref.~\cite{Tran2019} the MoSe$_2$/WSe$_2$ system was encapsulated in boron nitride. 

In Fig. 2(b) of the main text, we show the calculated local interaction $U_0 \equiv g_{aaaa}$, direct nearest-neighbour and next-nearest-neighbour interaction terms¸ $U_{\text{NN}}$ and $U_{\text{NNN}}$, as a function of the twist angle $\theta$ for $H_E$. As mentioned already in the main text, the nearest-neighbour interaction $U_{\text{NN}}$ is much larger compared to the hopping $t_{\text{NN}}$ than in case of $H_H$. For example, in case of $\theta \sim 3^\circ$, $U_{\text{NN}}/t_{\text{NN}} \sim 10$. As a result, one has to use relatively large twist angle $\theta$ in order to observe supersolidity, as otherwise strong $U_{\text{NN}}$ drives the system to the insulating state. 

\begin{figure}
  \centering
    \includegraphics[width=0.6\columnwidth]{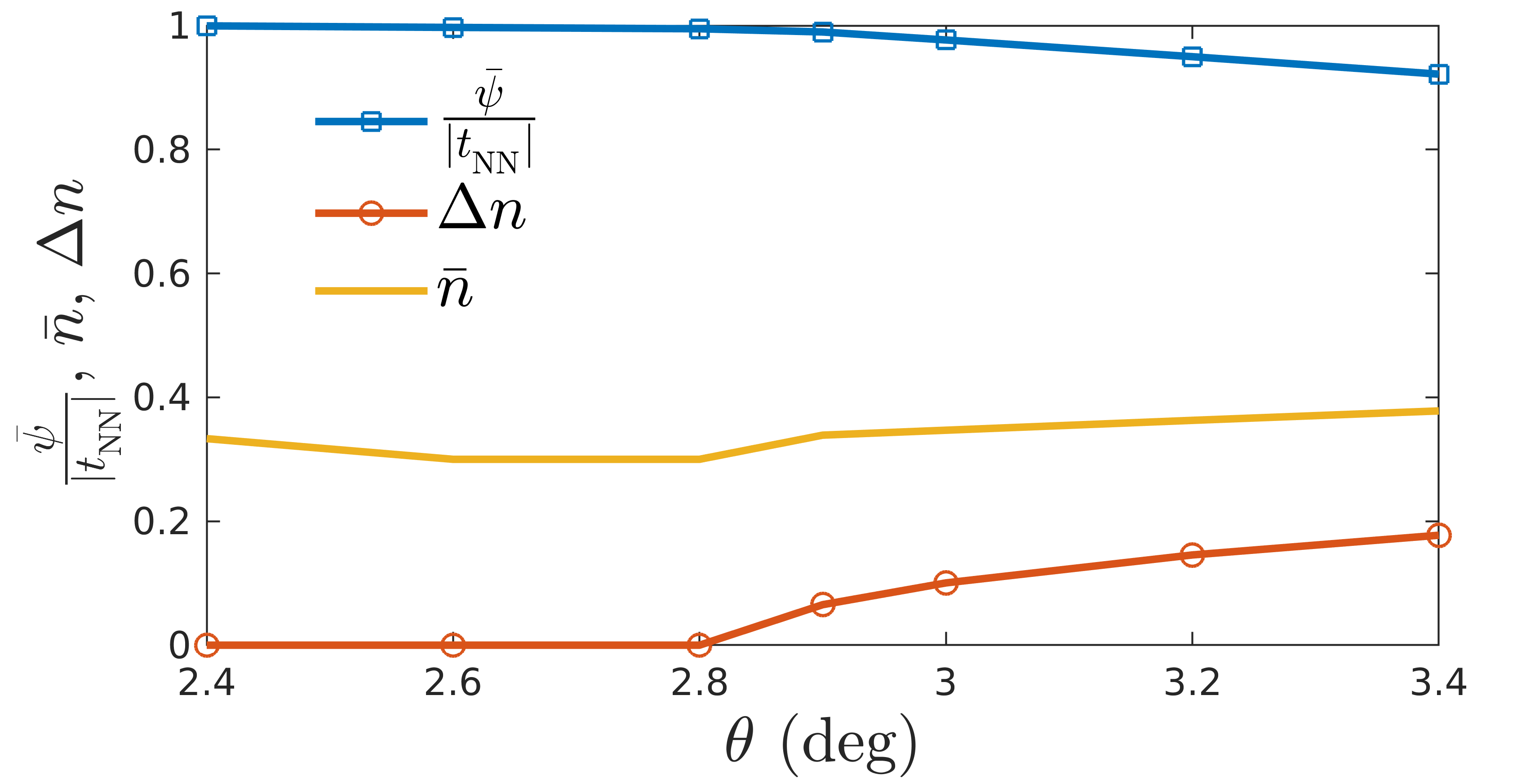}
    \caption{Effective \moire potential model: average superfluid order parameter $\bar{\psi}$, the staggering parameter $\Delta n$ and average density $\bar{n}$ obtained from the CMF as a function of $\theta$ at $\mu/U_{\text{NN}} = 5.7$ and $T=0$. }
   \label{Fig:S_cont2}
\end{figure}

In the main text we performed finite temperature calculations for the twist angle of $\theta = 3^\circ$ and we found $T_c \sim 1$ K. As can be seen from Fig.~\ref{Fig:S_cont2}, larger twist angles yield larger $\bar{\psi}$ and thus should also give higher $T_c$. However, the issue with larger twist angles is that the area of the \moire unit cell, $A_{\text{uc}}$ becomes smaller and thus the exciton density becomes larger. As was shown in experimental study of Ref.~\cite{Wang2019b}, above the critical density of $n_c \sim 1.6-3\times 10^{-12}$  cm$^{-2}$, \moire excitons dissociate to free electron-hole plasma and the picture of well-defined excitons breaks down. For $\theta = 3^\circ$ and $\mu/U_{\text{NN}} = 5.8$, the calculated exciton density is roughly $\bar{n} \sim 1.05 \times 10^{12}$ cm$^{-2}$ which is on the verge of critical density $n_c$. Therefore, one can anticipate that our results for larger twist angles are not necessarily experimentally applicable.

The reason why one needs such high twist angles in case of $H_E$ is the large value of the ratio $U_{\text{NN}}/t_{NN}$. This, on the other hand, follows from a high \moire potential amplitude $V$ in Eq.~\eqref{cont2}. In this work we used $V = 18$ meV which is the same value as the one used in the experimental work of Ref.~\cite{Tran2019}. However, one can also tune $V$ with an external electric field or pressure~\cite{Morales-Duran2021,Yu2017}, using different embedding material~\cite{Gotting2022} or different stacking order~\cite{Wu2018a}. Therefore, it is expected that $V$ can be reduced which would in turn lead to supersolidity to emerge at smaller angles and exciton densities. One can thus circumvent the problem of high exciton densities by tuning the system parameters, highlighting the flexibility of \moire  systems.

%$n_c \sim 1.6-3\times 10^{-12}$  cm$^{-2}$ ~\cite{Wang2019b}

\section{Dielectric function}\label{keldysh_sec}

In this section we derive the momentum-dependent dielectric functions $\epsilon_{\text{intra}, l}(\bq)$ and  $\epsilon_{\text{inter}}(\bq)$ for our bilayer system, following the general multilayer derivation of Ref.~\cite{Danovich2018}. The geometry we consider has layer 1 and 2 embedded in a dielectric medium of relative permittivity $\epsilon_r$ and separated with distance $d_l$. Layers are taken to be parallel to the (x,y)-plane. We start with the relation for the electric displacement field $\textbf{D}(\br,z) = \epsilon_0 \epsilon_r \textbf{E}(\br,z) + \textbf{P}(\br,z)$, where $\textbf{E}$ is the electric field, $\textbf{P}$ is the polarization, $\epsilon_0$ is the vacuum permittivity, and $\br$ denotes the spatial coordinates in the $(x,y)$-plane. In our case the polarization is finite only in the monolayers so we write $\textbf{P}(\br,z) = \textbf{P}_{2D}(\br,d_j)\delta(z-d_j)$, with $d_1 =0$ and $d_2 = d_l$ being the z-coordinates for layer 1 and 2 and $\textbf{P}_{2D}(\br,d_j)$ is the two-dimensional polarization. As usual, the polarization can be taken to be proportional to the electric field, i.e. we write $\textbf{P}_{2D}(\br,d_j) = \epsilon_0 \kappa_j \textbf{E}(\br,d_j)$, where  $\kappa_j$ is the in-plane susceptibility for layer $j$ and has the units of length.

To obtain the dielectric function, one needs to compute the electric potential $\phi(\br,z)$ caused by a free charge distribution $\rho_F(\br,z) = \rho_F(\br)\delta(z -d_1)$ placed in layer 1. By using the relations $\nabla \cdot \textbf{D}(\br,z) = \rho_F(\br,z)$ and $\textbf{E}(\br,z) = - \nabla \phi(\br,z)$, where $\phi(\br,z)$ is the scalar potential, we can cast $\textbf{D}(\br,z) = \epsilon_0 \epsilon_r \textbf{E}(\br,z) + \textbf{P}(\br,z)$ as
\begin{align}
\rho_F(\br) \delta(z-d_1) = -\epsilon_0 \epsilon_r \nabla^2 \phi(\br,z) - \epsilon_0 \sum_{i=1}^2 \kappa_j \nabla_{||}^2\phi(\br,d_i) \delta(z-d_i).    
\end{align}
Here $\nabla_{||}$ is the gradient with respect to the in-plane coordinate $\br$. By Fourier transforming to the momentum space [with $\bq \equiv (q_x,q_y)$ and $k \equiv q_z$], we have
\begin{align}
\rho_F(\bq) e^{-ik d_1} = \epsilon_0 \epsilon_r (q^2 + k^2) \phi(\bq,k) + q^2 \epsilon_0 \sum_{i=1}^2 \kappa_i \phi(\bq,d_i)e^{-ik d_i}.    
\end{align}
By further inverse Fourier transforming back to the $z$-coordinate space and using the identity
\begin{align}
\int_{-\infty}^\infty \frac{dk e^{ik z}}{q^2 + k^2} = \frac{\pi e^{-k|z|}}{q}    
\end{align}
we get
\begin{align}
\label{keldysh1}
\rho_F(\bq) e^{-q|z-d_1|} = 2\epsilon_0 \epsilon_r q \phi(\bq,z) + q^2\epsilon_0\sum_{i=1}^2 e^{-q|z-d_i|}\kappa_i \phi(\bq,d_i).    
\end{align}
We are interested in the potential function caused by the point-like particles of charge $q_F$, i.e. we write $\rho_F(\br) = q_F\delta(\br)$ which yields $\rho_F(\bq) = q_F$. Equation~\eqref{keldysh1} can then be easily solved for the potentials $\phi(\bq,d_1)$ and $\phi(\bq,d_2)$, giving
\begin{align}
\label{keldysh2}
& \begin{bmatrix} \phi(\bq,d_1) \\ \phi(\bq,d_2) 
\end{bmatrix}
= \frac{q_F}{2\epsilon_0 \epsilon_r q [ (1 + r^*_1 q)( 1+ r^*_2 q) - q^2r^*_1 r^*_2 e^{-2qd_l}]} \begin{bmatrix}
1 + r^*_2q - r^*_2 q e^{-2qd_l}
\\
e^{-qd_l}
\end{bmatrix},    
\end{align}
where we have defined the screening lengths $r_j^* = \kappa_j/(2\epsilon_r) = r^*_{j,0}/\epsilon_r$, where $r^*_{j,0}$ is the screening length of monolayer embedded in vacuum. For $r^*_{j,0}$ we use the values reported in Ref.~\cite{Ruiz-Tijerina2019}. We can now identify the dielectric functions from Eq.~\eqref{keldysh2} as
%\begin{align}
%&\frac{1}{\epsilon_{intra,1}(q)} = \frac{1 + r_2 q - r_2q e^{-2qd_l}}{\epsilon_0 \epsilon_r [(1 + rq)^2 - r^2 q^2 e^{-2qd_l}]} % \nonumber \\ 
%&\frac{1}{\epsilon_{inter}(q)} = \frac{e^{-qd_l}}{\epsilon_0 \epsilon_r [(1 + rq)^2 - r^2 q^2 e^{-2qd_l}]}.
%\end{align}
\begin{align}
&\frac{1}{\epsilon_{\mathrm{intra},1}(q)} = \frac{1 + r^*_2 q - r^*_2q e^{-2qd_l}}{\epsilon_0 \epsilon_r [(1 + r^*_1 q)(1 + r^*_2 q) - r^*_1 r^*_2 q^2 e^{-2qd_l}]}  \nonumber \\ 
&\frac{1}{\epsilon_{\mathrm{inter}}(q)} = \frac{e^{-qd_l}}{\epsilon_0 \epsilon_r [(1 + r^*_1 q)(1 + r^*_2 q) - r^*_1 r^*_2 q^2 e^{-2qd_l}]}
\end{align}
The intralayer dielectric function for layer $2$, $\epsilon_{\mathrm{intra},2}$ , is obtained from  $\epsilon_{\mathrm{intra},1}$ by interchanging $r_1^*$ and $r_2^*$. In our MoSe$_2$/WS$_2$ calculations we used $\epsilon_r = 2.45$~\cite{Ruiz-Tijerina2019} and in case of MoSe$_2$/WSe$_2$ $\epsilon_r = 4$~\cite{Tran2019}.

%Here $d_l$ is the interlayer distance, $\epsilon_0$ is the vacuum permittivity, $\epsilon_r$ is the (averaged) relative permittivity of the embedding medium and $r$ is the screening length within a TMDC monolayer. To a good approximation, $r$ can be assumed to be the same for both the TMDC monolayers~\cite{Ruiz-Tijerina2019}. In our MoSe$_2$/WS$_2$ calculations we use $r = 3.8$ nm and $\epsilon_r = 2.45$~\cite{Ruiz-Tijerina2019}.

%In the main text we have taken $r^*_1 \approx r^*_2$ which for a MoSe$_2$/WS$_2$ system is a fine approximation~\cite{Ruiz-Tijerina2019}. By identifying $\phi(\bq,1)$ and $\phi(\bq,2)$ in Eq.~\eqref{keldysh2} as intra and interlayer potentials, we obtain the expressions for $\epsilon_{\text{intra}}(q)$ and $\epsilon_{\text{inter}}(q)$ used in the main text.

\section{Treating \moire excitons as bosons}

In the main text, the many-body phases of \moire excitons are studied by assuming the bosonic commutation relations between \moire excitons $x_i$. One should, however, note that excitons are not pure bosons but bound electron-hole pairs and therefore they do not strictly follow the bosonic commutation relations. Namely, if the exciton density is high enough compared to the Bohr radius of the exciton, the non-bosonic nature arising from the fermionic statistics of electrons and holes should play a role~\cite{Moskalenko2000}.

Basically, the exciton operators $x_i$ fulfill the following commutation relations:
\begin{align}
& [x_i,x_j] = 0 \label{comm1} \\
& [x_i,x_j^\dag] = \delta_{i,j} + \mathcal{O}(\bar{n} (a_B/a_m)^2) \label{comm2}
\end{align}
where $a_B$ is the Bohr radius of excitons, $a_m$ is the \moire  periodicity and $\bar{n}$ is the filling fraction of the \moire lattice, i.e. exciton occupation number per \moire unit cell. The first set of commutation relations, Eq.~\eqref{comm1}, are the same as for ideal bosons. The next set,  Eq.~\eqref{comm2}, are otherwise the same as for the ideal bosons but the correction term $\mathcal{O}((\bar{n} a_B/a_m)^2)$ is needed to account the finite spread of the exciton wavefunction and resulting fermionic statistics stemming from the electron and hole parts of the exciton. As we see, this correction term is negligible when the Borh radius of the exciton is small compared to the density $\bar{n}/a_m^2$. In the calculations of the main text, this is the case in relevant parameter regime. For example, Fig 4(a) of the main text, for $\theta =1.4^\circ$ we have $\bar{n} \sim 0.5$, $a_m \sim 6.8$ nm and $a_B \sim 2$ nm, yielding $\bar{n} (a_B/a_m)^2 \sim 0.042$. Correspondingly, in Fig 4(b) we have $\bar{n} \sim 0.35$, $a_m \sim 6$ nm and $a_B \sim 3$ nm, yielding  $\bar{n} (a_B/a_m)^2 \sim 0.085$. In other words, our density regime is such that the \moire excitons do not overlap considerably and therefore it is a good approximation to treat them as bosons. 

One should furthermore note that we are assuming the hard-core boson constraint which is  a feasible assumption as explained in the main text, and thus two excitons cannot reside within a same lattice site. This furthermore makes it physically feasible to treat \moire excitons as ideal bosons. If the exciton density was very high, as is for example assumed in Ref.~\cite{Camacho-Guardian2021moire}, one should properly take into account the non-bosonic nature of excitons. For example in Ref.\cite{Camacho-Guardian2021moire}, this is done by including the so-called saturation effects in the light-matter coupling Hamiltonian.

The commutation relations Eqs.~\eqref{comm1}-\eqref{comm2} has been derived for example in Ref.~\cite{Moskalenko2000}. Here the derivation in case of \moire excitons is shown by using, for simplicity,  the effective \moire potential model model $H_E$. The derivation for hybridized \moire exciton model, $H_H$, is similar. We start by writing down the explicit form for \moire excitons $x_i$:
\begin{align}
\label{xi}
x_i = \frac{1}{\sqrt{N}} \sum_{\bk \in \text{mBZ}} e^{-i\bk \cdot \brr_i} \gamma_{\bk1},    
\end{align}
where $\gamma_{\bk1}$ annihilates a \moire exciton at momentum $\bk$ in the lowest \moire band and $\brr_i$ is the spatial coordinate of the $i$th \moire lattice site. By writing $\gamma_{\bk 1}$ operators in the original interlayer exciton basis, i.e. $\gamma_{\bk1} = \sum_\alpha u_{\alpha 1}(\bk) x(\bk + \bG_\alpha)$, where $u_{\alpha 1}(\bk)$ are the periodic Bloch functions for \moire excitons, one can write the commutation relations of lattice \moire excitons $x_i$ as 
\begin{align}
& [x_i,x_j] = \frac{1}{N}\sum_{\bk,\bk' \in \text{mBZ}} e^{-i \bk \cdot \brr_i} e^{-i \bk' \cdot \brr_j} \sum_{\alpha\beta} u_{\alpha1}(\bk)u_{\beta1}(\bk')[x(\bk+\bG_\alpha),x(\bk'+\bG_\beta)] \label{comm_full1} \\
& [x_i,x_j^\dag] = \frac{1}{N}\sum_{\bk,\bk' \in \text{mBZ}} e^{-i \bk \cdot \brr_i} e^{i \bk' \cdot \brr_j} \sum_{\alpha\beta} u_{\alpha1}(\bk)u^*_{\beta1}(\bk')[x(\bk+\bG_\alpha),x^\dag(\bk'+\bG_\beta)]. \label{comm_full2}
\end{align}
To proceed, we note that $x(\bqq) = \sum_{\bq} \phi(\bq) c^\dag_c(x_e \bqq + \bq) c_v(-x_h\bqq + \bq)$, where $\phi(\bq)$ is the relative wavefunction of the exciton. Importantly, due to fermionic commutation relations of conduction band electrons $c_c(\bk)$ and valence band electrons $c_v(\bk)$, we have $[x(\bqq),x(\bqq')] = 0$ for any $\bqq$ and $\bqq'$. This means that, due to Eq.~\eqref{comm_full1}, we automatically have $[x_i,x_j] = 0$.

To obtain Eq.~\eqref{comm2}, one can show~\cite{Moskalenko2000} that $[x(\bqq),x^\dag(\bqq')] = \delta_{\bqq,\bqq'} + \mathcal{O}(n_x a_B^2)$, where $n_x$ is the exciton density. By plugging this into Eq.~\eqref{comm_full2}, one obtains
\begin{align}
[x_i,x_j^\dag] = \delta_{i,j} + \frac{\mathcal{O}(n_x a_B^2)}{N}\sum_{\bk,\bk' \in \text{mBZ}} e^{-i \bk \cdot \brr_i} e^{i \bk' \cdot \brr_j} \langle u_1(\bk) | u_1(\bk') \rangle = \delta_{i,j} + \mathcal{O}(n_x a_B^2)     
\end{align}
as $|\langle u_1(\bk) | u_1(\bk') \rangle| \leq 1$ and the vectors $|u_1(\bk)\rangle$ contain the coefficients $u_{\alpha 1}(\bk)$. By noting that $n_x \sim \bar{n}/a_m^2$, we arrive to Eq.~\eqref{comm2}.

\section{Details on cluster mean-field theory}

\begin{figure}
  \centering
    \includegraphics[width=1.0\columnwidth]{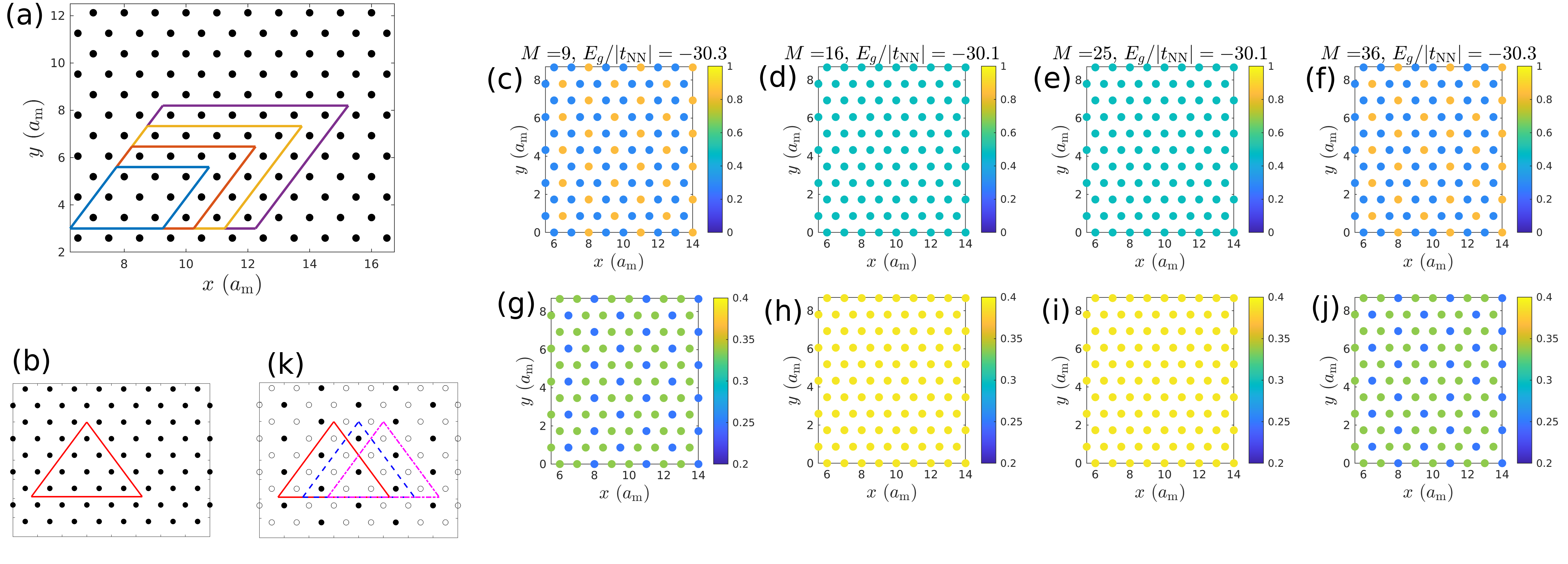}
    \caption{(a) Different choices of the unit cell for the CMF calculations in case of $M=9$ (blue parallelogram), $M=16$ (red), $M=25$ (yellow) and $M=36$ (purple). (b) 10-site cluster geometry used in all the  calculations (c)-(f) Zero-temperature density profiles $n_i$ for $M=9$, $M=16$, $M=25$ and $M=36$ at $\theta = 0.7^\circ$ and $\mu/U_{\text{NN}} = 5.8$ by using the hopping and interaction parameters obtained from $H_H$. (f)-(i) Corresponding superfluid order parameter $\psi_i$. Also the ground state energy per lattice site, $E_g$ is given. (k) Division of the \moire lattice to two sublattice degrees of freedom. Three triangles show all possible ways to embed a 10-site cluster to the underlying sublattice structure.}
   \label{Fig:CMF_N}
\end{figure}

In this section we provide some additional details on our cluster mean-field calculations. We consider $\sqrt{M} \times \sqrt{M}$ inequivalent lattice sites, labelled as $i \in\{1,2,\dots M\}$, and assume periodic boundary conditions. In Fig.~\ref{Fig:CMF_N}(a) we show the examples of $M=9$,  $M=16$,  $M=25$ and $M=36$. We then define $M$ clusters, labelled by index $C  \in\{1,2,\dots M\}$, such that the center of cluster $C =i$ is lattice site $i$. In all our calculations, we use a 10-site cluster depicted in Fig.~\ref{Fig:CMF_N}(b). The calculations are then performed by first choosing an initial ansatz for the mean-fields $\psi_i$ and $n_i$. Based on these values, $M$ cluster problems are solved independently by using $\psi_i$ and $n_i$. Once the cluster problems are exactly diagonalized, one can calculate new values for the mean fields of each site $i$ as
\begin{align}
\psi_i = \frac{1}{M_i} \sum_{C} \langle x_i \rangle \\
n_i = \frac{1}{M_i} \sum_{C} \langle x^\dag_i x_i \rangle
\end{align}
where the sums include only the clusters that contain site $i$. Correspondingly, $M_i$ denotes the number of clusters that include site $i$. Thermal average $\langle \dots \rangle$ for cluster $C$ is computed as explained in the main text. Iterative process is continued till $\psi_i$ and $n_i$ converge to a stable solution for all $i$ .

In Figs.~\ref{Fig:CMF_N}(c)-(j) we provide the obtained density $n_i$ [panels (c)-(f)] and superfluid order parameter $\psi_i$ [panels (g)-(j)] profiles for each $M$ computed at $\theta = 0.7^\circ$ in case of the hybrid exciton model $H_H$. To see which configuration is the most feasible, we moreover provide the corresponding ground state energy per lattice site. One can see that the energy is minimized for $M=9$ and $M=36$ and that these two cases yield the same two-sublattice structure. Two other cases of higher ground state energy, i.e. $M=16$ and $M=25$, are incommensurate with this two-sublattice pattern and thus yield a homogeneous superfluid phase. We can therefore conclude that it is sufficient to perform the CMF computations with $M=9$ which can equally well describe uniform solutions and the phase of broken translational symmetry. In fact, to capture the two-sublattice pattern, one is required to consider only three inequivalent cluster problems, as shown in Fig.~\ref{Fig:CMF_N}(k). This trick was used in the calculations of the main text and also in Ref.~\cite{Yamamoto2012} where CMF method was applied to a study hard-core bosons in a triangular lattice.

\section{Gross-Pitaevskii mean-field and Bogoliubov theory}

In the parameter regime studied in Figs. 3(a)-(b) of the main text, the superfluid gap for the supersolid state is maximized for twist angle of $\theta = 1.5^\circ$ which corresponds to the ratio of $U_{\text{NN}}/|t_{\text{NN}}| \sim 1.2$. Such moderate interaction strength implies that one could gain at least some qualitative insight by using a simple mean-field weak-coupling Gross-Pitaevskii equation (GPE)  accompanied with the Bogoliubov theory to study the effect of fluctuations. In this section we show that GPE and Bogoliubov theories indeed yield qualitatively similar supersolid solutions as more advanced cluster mean field theory used in the main text.

\begin{figure}
  \centering
    \includegraphics[width=0.75\columnwidth]{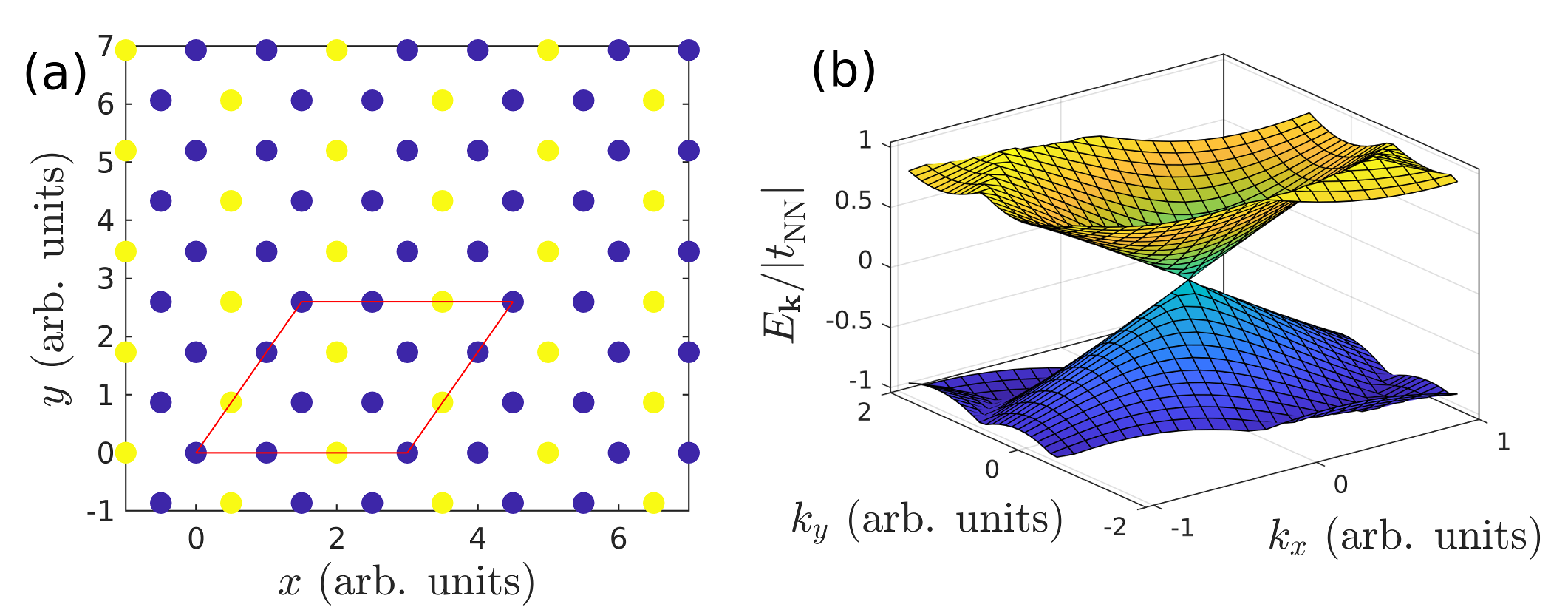}
    \caption{(a) Density profile of the superfluid obtained from the GPE in case of $\theta = 0.7^\circ$ which corresponds to $U_{\text{NN}}/|t_{\text{NN}}| = 1.4$ in case of $H_H$. Bright (dark) color corresponds to the occupation number $1.08$ ($0.06$) and the red parallelogram indicates the unit cell used in the Bogoliubov analysis. (b) Corresponding lowest quasi-particle and quasi-hole Bogoliubov excitation energy spectrum in the momentum space. }
   \label{Fig:GPE}
\end{figure}

We start by writing down the Heisenberg equation of motion for \moire exciton operators $x_i(t)$ in presence of the hard-core boson constraint as
\begin{align}
i\hbar \frac{\partial x_i(t)}{\partial t} = \sum_{j} t_{ij} x_j(t) + 2\sum_{j}U_{ij} x_i(t) x_j^\dag(t) x_j(t)  
\end{align}
By writing $x_i(t) = x^0_i(t) + \delta x_i(t)$ with $x^0_i \equiv \langle x_i \rangle$ describing the superfluid component, we obtain
\begin{align}
\label{eGPE}
i\hbar \frac{\partial x_i^0(t)}{\partial t} &= \sum_{j} t_{ij}  x^0_j(t) + 2\sum_{ij}U_{ij}\Big[ |x_j^0 (t)|^2 x^0_i(t) + \langle \delta x_j^\dag(t) \delta x_j(t) \rangle x^0_i(t) + \langle \delta x_i(t) \delta x_j(t) \rangle [x^0_j(t)]^* \nonumber \\
&+ \langle \delta x_i(t) \delta x_j^\dag(t) \rangle x^0_j(t) + \langle \delta x_i(t) \delta x_j^\dag(t) \delta x_j(t) \rangle \Big].
\end{align}
By ignoring the fluctuation terms and writing for the condensate part $x^0_i(t) = e^{-\mu t/\hbar} x_i^0$ with $\mu$ being the chemical potential, one gets the GPE:
\begin{align}
\label{GPE}
\mu x_i^0 =   \sum_{j} t_{ij}  x^0_j + 2\sum_j U_{ij} |x_j^0|^2 x^0_i.
\end{align}
By solving the GPE self-consistently, one obtains the density profile $x^0_i$ for the superfluid. In Fig.~\ref{Fig:GPE}(a) we show a density distribution obtained from the GPE for the twist angle $\theta = 0.7^\circ$. We see that the GPE yields a supersolid solution with the same two-sublattice pattern and periodicity as more advanced CMF. %Therefore, the sublattice division used in the CMF calculations is justified also from the mean-field point of view.

To confirm that the obtained supersolid phase is a stable solution, one can compute the Bogoliubov spectrum.
 To this end, we define a unit cell for the system as the red parallelogram shown in Fig.~\ref{Fig:GPE}(a), i.e. the unit cell contains nine lattice sites. We therefore reformulate our Hamiltonian as
 \begin{align}
 H = \sum_{i\alpha j \beta} t_{i\alpha,j\beta} x^\dag_{i\alpha}x_{j\beta} + \sum_{i\alpha j\beta} U_{i\alpha,j\beta} x^\dag_{i\alpha} x^\dag_{j\beta} x_{j\beta} x_{i\alpha},
 \end{align}
where $x_{i\alpha}$ denotes now the bosonic annhilation operator for a lattice site residing in the $i$th unit cell in the sublattice $\alpha$. For our choice of the unit cell, we have 9 sublattices, i.e. $\alpha \in [1,9]$. To formulate the Bogoliubov theory, one then writes $x_{i\alpha} = x^0_\alpha + \delta x_{i\alpha}$, where  the superfluid part $x^0_\alpha$  does not depend on the unit cell index $i$ due to our choice for the unit cell. One can proceed by performing the Fourier transformation as $x_{i\alpha} = \frac{1}{\sqrt{N}}\sum_\bk e^{\bk \cdot \br_i} x_{\bk\alpha}$, where $N$ is the number of unit cells and $\br_i$ the spatial location of the $i$th unit cell. By keeping the fluctuation terms up to the quadratic order in the Hamiltonian $H$, one then obtains
\begin{align}
H \approx H_C + H_B,    
\end{align}
where $H_C$ is constant describing the ground state energy and the Bogoliubov Hamiltonian $H_B$ reads
\begin{align}
&H_B = \frac{1}{2}\sum_{\bk  \neq 0} \Psi^\dag_\bk \mathcal{H}_B(\bk) \Psi_\bk  \textrm{ with} \\
& \Psi_\bk = \begin{bmatrix} \delta \tilde{x}_\bk & \delta \tilde{x}^\dag_{-\bk} \end{bmatrix} \\
& \delta \tilde{x}_\bk = \begin{bmatrix}
\delta x_{\bk,1} & \delta x_{\bk,2} & \cdots \delta x_{\bk,9}
\end{bmatrix} \\
& \mathcal{H}_B(\bk) = \begin{bmatrix}
 \mathcal{H}(\bk) - \mu + A(\bk) & B(\bk) \\
  B^\dag(\bk) & \mathcal{H}^*(-\bk) - \mu + A^*(-\bk) 
  \end{bmatrix}  \\
& A_{\alpha\beta}(\bk)= 2n_0 U_{\alpha\beta}(\bk) x^0_\alpha [x^0_\beta]^* + 2n_0 \Bigg( \sum_\gamma |x^0_\gamma|^2 U_{\gamma\beta}(0) \Bigg) \delta_{\alpha,\beta}    \\
& B_{\alpha\beta}(\bk) =  2n_0 U_{\alpha\beta}(\bk) x^0_\alpha x^0_\beta 
\end{align}
Here $n_0$ is the condensate density,  $\delta x_{\bk \alpha} = \frac{1}{\sqrt{N}}\sum_i e^{\bk \cdot \br_i} \delta x_{\i\alpha}$, matrix $\mathcal{H}(\bk)$ [$U(\bk)$] includes the Fourier-transformed hopping (interaction) terms and the momentum sum runs over all the momenta in the first Brillouin zone but excludes the superfluid state $\bk=0$. By diagonalizing $L_\bk = \sigma_z \mathcal{H}_B(\bk)$, with $\sigma_z$ being the Pauli matrix in the particle-hole space, one obtains the Bogoliubov energy bands $E_9(\bk) \geq ... E_2(\bk) \geq E_1(\bk) \geq 0 \geq -E_1(-\bk) \geq ... -E_9(-\bk)$ for each $\bk$~\cite{castin:book}. Here positive (negative) energies describe quasi-particle (-hole) excitations. In Fig.~\ref{Fig:GPE}(b) we show the lowest quasi-particle and highest quasi-hole Bogoliubov dispersions, $E_1(\bk)$ and $-E_1(-\bk)$, respectively, in case of $\theta = 0.7^\circ$. We see that the there exist a gapless Goldstone mode at $\bk \rightarrow 0$ and that the dispersion of the excitations never becomes zero outside $\bk=0$. This implies that our solution for the GPE is thermodynamically stable.

One should keep in mind the limitations of the GPE and the Bogoliubov theory. As they are weak-coupling approaches, they cannot capture the phase transition from the superfluid and supersolid states to the insulating states as a function of decreasing twist angle. They can be only used to gain some qualitative intuition in the weak-coupling regime, i.e. at large twist angles. Moreover, the superfluid order parameter $\psi_i$ obtained from the GPE follows the density profile of the condensate as $\psi_i = x^0_i$. This is in contrast to supersolid phases found in the CMF, where the variation of the superfluid order parameter  is much smaller than that of the density profile and it is furthermore  maximized in the sites where the density is minimized, as shown for example in Fig.~\ref{Fig:CMF_N}(g). To improve the GPE description, one could compute the fluctuation terms $\langle \delta x_j^\dag \delta x_i \rangle$ and $\langle \delta x_i \delta x_j \rangle$ from the Bogoliubov theory and use Eq.~\eqref{eGPE} to obtain the extended GPE. We have checked numerically that this does not qualitatively change the density profiles obtained from the original GPE of Eq.~\eqref{GPE}.

%\bibliography{bib_file}
%merlin.mbs apsrev4-1.bst 2010-07-25 4.21a (PWD, AO, DPC) hacked
%Control: key (0)
%Control: author (8) initials jnrlst
%Control: editor formatted (1) identically to author
%Control: production of article title (-1) disabled
%Control: page (0) single
%Control: year (1) truncated
%Control: production of eprint (0) enabled
%